\documentclass[useAMS,useAMSmath, usenatbib, usegraphicx, referee]{mn2e}

\usepackage{amsbsy}

\title[Electron-positron energy deposition near the rotation axis ]{Electron-positron energy deposition rate from neutrino pair annihilation on the rotation axis of neutron and quark stars}
\author[Kov\'{a}cs, Cheng and Harko]{Z. Kov\'{a}cs\thanks{E-mail:
zkovacs@mpifr-bonn.mpg.de}, K. S. Cheng\thanks{E-mail:
hrspksc@hkucc.hku.hk} and T. Harko\thanks{E-mail:
harko@hkucc.hku.hk}
\\
Department of Physics and Center for Theoretical and Computational Physics, The University of Hong Kong, \\
Pok Fu Lam Road, Hong Kong, Hong Kong SAR, P. R. China
}
\begin{document}


\pagerange{\pageref{firstpage}--\pageref{lastpage}} \pubyear{2002}

\maketitle


\def\apj{Astrophys. J.}
\def\mnras{Month. Not. Roy. Astr. Soc.}
\def\prd{Phys. Rev. D}

\date{Accepted 1988 December 15. Received 1988 December 14; in original form 1988 October 11}

\pagerange{\pageref{firstpage}--\pageref{lastpage}} \pubyear{2002}

\maketitle

\label{firstpage}

\begin{abstract}
We investigate the deposition of energy due to the annihilations of neutrinos and antineutrinos on the rotation axis of rotating neutron and quark stars, respectively. The source of the neutrinos is assumed to be a neutrino-cooled accretion disk around the compact object.  Under the  assumption of the separability of the neutrino null geodesic equation of motion we obtain the general relativistic expression of the energy deposition rate for arbitrary stationary and axisymmetric space-times. The neutrino trajectories are obtained by using a ray tracing algorithm, based on numerically solving the Hamilton-Jacobi equation for neutrinos by reversing the proper time evolution. We obtain the energy deposition rates for several classes of rotating neutron stars, described by different equations of state of the neutron matter, and for quark stars, described by the MIT bag model equation of state and in the CFL (Color-Flavor-Locked) phase, respectively. The electron-positron energy deposition rate on the rotation axis of rotating neutron and quark stars is studied for two accretion disk models (isothermal disk and accretion disk in thermodynamical equilibrium). Rotation and general relativistic effects modify the total annihilation rate of the neutrino-antineutrino pairs on the rotation axis of compact stellar, as measured by an observer at infinity. The differences in the equations of state for neutron and quark matter also have important effects on the spatial distribution of the energy deposition rate by neutrino-antineutrino annihilation.
\end{abstract}


\section{Introduction}

The neutrino-antineutrino annihilation into electrons and positrons is an important candidate to explain the energy source of the gamma ray bursts (GRBs) \cite{Pa90, MeRe92,RuJa98, RuJa99, AsIw02}. The study of the electron energy deposition rate from the $\nu +\bar{\nu}\rightarrow e^{+}+e^{-}$ neutrino annihilation reaction was initiated by Cooperstein et al. (1986), Cooperstein et al. (1987),  and Goodman et al. (1987), respectively, and this process has been intensively investigated in the physical and astrophysical literature. Neutrino-antineutrino annihilation into electrons and positrons can deposit more than $10^{51}$ ergs above the neutrino-sphere of a type II supernova \citep{Go87}. For a full understanding of the effects of the neutrino annihilation in strong gravitational fields, general relativistic effects must be taken into account \citep{SaWi99}. In a Schwarzschild geometry, the efficiency of the $\nu +\bar{\nu}\rightarrow e^{+}+e^{-}$ process is enhanced over the Newtonian values up to a factor of more than $4$, in the regime applicable to Type II supernovae, and by up to a factor of $30$ for collapsing neutron stars \citep{SaWi99}. The neutrino pair annihilation rate into electron pairs between two neutron stars in a binary system was calculated by Salmonson \& Wilson (2001).

The gravitational effects on neutrino pair annihilation near the neutrinosphere and around the thin accretion disk were considered in Asano \& Fukuyama (2000), by assuming that the accretion disk is isothermal, and that the gravitational field is dominated by the Schwarzschild black hole. General relativistic effects were studied only near the rotation axis. Using idealized models of the accretion disk, Asano \& Fukuyama (2001) investigated the relativistic effects on the energy deposition rate via neutrino pair annihilation near the rotation axis of a Kerr black hole, by assuming that the neutrinos are emitted from the accretion disk. The Kerr parameter, $a$, affects not only the behavior of the neutrinos, but also the inner radius of the accretion disk. When the deposition energy is mainly contributed by the neutrinos coming from the central part, the redshift effect becomes dominant as $a$ becomes large, and the energy deposition rate is reduced. On the other hand, for a small $a$, the bending effect becomes dominant and makes the energy increase by factor of 2, as compared with the case that neglects the relativistic effects \citep{AsFu01}.

General relativity and rotation cause important differences in the spatial distribution of the energy deposition rate by neutrino $\nu $ and antineutrino $\bar{\nu }$-annihilation \citep{Bi07}. The energy-momentum deposition rate (MDR) from the $\nu -\bar{\nu}$ collisions above a rotating black hole/thin accretion disk system was calculated by Miller et al. (2003), by imaging the accretion disk at a specified observer using the full geodesic equations, and calculating the cumulative MDR from the scattering of all pairs of neutrinos and antineutrinos arriving at the observer. The dominant contribution to the MDR comes from near the surface of the disk with a tilt of approximately $\pi/4$ in the direction of the disk's rotation. The MDR at large radii is directed outward in a conic section centered around the symmetry axis and is larger by a factor of 10-20 than the on-axis values. There is also a linear dependence of the MDR on the black hole angular momentum.
The deposition of energy and momentum, due to the annihilation of neutrinos $\nu $ and
antineutrinos $\bar{\nu }$ in the vicinity of steady, axisymmetric accretion tori around stellar-mass black holes was investigated in Birkl et al. (2007). The influence of general relativistic effects were analyzed in combination
with different neutrinosphere properties and spatial distribution of the
energy deposition rate. Assuming axial symmetry, the annihilation rate 4-vector was numerically computed.
 As compared to Newtonian calculations,
general relativistic effects increase the total annihilation rate measured by an observer at infinity by a factor of two when the neutrinosphere is a thin disk, but the increase is only 25\% for toroidal and spherical neutrinospheres. Thin disk models yield the highest energy deposition rates for neutrino-antineutrino annihilation, and spherical neutrinospheres the lowest ones, independently of whether general relativistic effects are included or not.

The study of the structure of neutron star disks based on the two-region (i.e., inner and outer) disk scenario was performed by Zhang \& Dai (2009), who calculated the neutrino annihilation luminosity from the disk in various cases.  As compared with the black hole disk, the neutrino annihilation luminosity above the neutron star disk is higher.  The neutron star disk with the advection-dominated inner disk could produce the highest neutrino luminosity, while the disk with an outflow has the lowest\citep{ZhDa09}. A detailed general relativistic calculation of the neutrino path for a general metric describing a rotating star was studied in Mallick \& Majumder (2009). The minimum photosphere radius was calculated for stars with two different equations of state, each rotating with two different velocities. The results show that the minimum photosphere radius for the hadronic star is greater than for the quark star, and that the minimum photosphere radius increases as the rotational velocity of the star decreases. The minimum photosphere radius along the polar plane is larger than that along the equatorial plane. The estimate of the energy deposition rate for neutrino pair annihilation for the neutrinos coming from the equatorial plane of a rotating neutron star was calculated along the rotation axis in Bhattacharyya et al. (2009), by using the Cook-Shapiro-Teukolsky metric. The neutrino trajectories, and hence the neutrino emitted from the disk, are affected by the redshift due to the disk rotation and gravitation. The energy deposition rate is very sensitive to the value of the temperature, and its variation along the disk. The rotation of the star has a negative effect on the energy deposition rate, it decreases with increase in rotational velocity. A phase transition during a supernova explosion can induce stellar collapse and result in large amplitude stellar oscillations \cite{Cha09}. Extremely intense pulsating neutrino fluxes, with submillisecond period and with neutrino energy (greater than 30 MeV), can be emitted because the oscillations of the temperature and density are out of phase almost $180^{\circ}$. The dynamical evolution of a phase-transition-induced collapse of a neutron star to a hybrid star, which consists of a mixture of hadronic matter and strange quark matter, was studied in Cheng et al. (2009). It was found that both the temperature and the density at the neutrinosphere are oscillating with acoustic frequency. Consequently, extremely intense, pulsating neutrino/antineutrino fluxes will be emitted periodically. Since the energy and density of neutrinos at the peaks of the pulsating fluxes are much higher than the non-oscillating case, the electron/positron pair creation rate can be enhanced significantly. Some mass layers on the stellar surface can be ejected, by absorbing energy of neutrinos and pairs. These mass ejecta can be further accelerated to relativistic speeds by absorbing electron/positron pairs, created by the neutrino and antineutrino annihilation outside the stellar surface.

It is the purpose of the present paper to consider a comparative systematic study of the neutrino-antineutrino annihilation process on the rotation axis of rotating neutron and strange stars, respectively, and to obtain the basic physical parameters characterizing this process (the electron-positron energy deposition rate per unit volume and unit time), by taking into account the full general relativistic corrections. The source of the neutrinos is assumed to be the accretion disk that can be formed around neutron and quark stars \citep{ZhDa09}. Neutrino-cooled accretion disks around stellar mass black holes or neutron stars are plausible candidates for the central engine of gamma-ray bursts \citep{ZhDa09}.  As a first step in obtaining the electron-positron energy deposition rate on the rotation axis of different types of neutron and quark stars we generalize the relativistic description of the neutrino-antineutrino annihilation process to the case of arbitrary stationary and axisymmetric geometries. By assuming that the neutrinos move on null geodesics, the main kinematic parameter of the collision process (the scalar product of the four momenta of the colliding $\nu \bar{\nu}$ pairs) are derived under the assumption that, similarly to the case of the Kerr black holes, the geometric variable in the geodesic equation of motion for the neutrinos can be separated. The domain of validity of this assumption is carefully investigated, and it is shown that in the case of the low mass stellar models with masses around $M=1.8 M_{\odot}$, the error is under 0.5\% for angular velocities of the order of $2\times10^3 {\rm s}^{-1}$, while it is more the 1\% for angular velocities higher than $4\times10^3 {\rm s}^{-1}$. The electron energy deposition rate is calculated for a general axisymmetric metric under the separability assumption. The neutrino trajectories are studied by using a ray tracing algorithm, based on numerically solving the Hamilton-Jacobi equation for neutrinos by reversing the proper time evolution. The electron-positron energy deposition rate on the rotation axis of rotating neutron and quark stars is studied for two accretion disk models (isothermal disk and accretion disk in thermodynamical equilibrium). In all these cases the electron energy deposition rate is obtained for different equations of state of the neutron and quark matter, as well as for different physical parameters of the stars.

In order to compute the electron-positron energy deposition rate, the metric outside the rotating general relativistic stars must be determined. In the present study we obtain the equilibrium configurations of the rotating neutron and quark stars by using the RNS code, as introduced in \cite{SteFr95}, and discussed in detail in \cite{Sterev}. This code was used for the study of different models of rotating neutron stars in \cite{No98} and for the study of the rapidly rotating strange stars \cite{Ste99}. The software provides the metric potentials for various types of compact rotating general relativistic objects, which can be used to obtain the electron-positron energy deposition rate on the rotation axis of rotating neutron and quark stars.

The present paper is organized as follows. In Section \ref{2} we present the basic formalism for the calculation of the kinematic parameters (four-momenta) of the neutrino-antineutrino annihilation. The general relativistic energy deposition rate is obtained in Section \ref{3}. The equations of state of dense neutron and quark matter used in the present study are presented in Section \ref{eos}. In Section \ref{5} we obtain the electron-positron energy deposition rates on the rotation axis of the considered classes of neutron and quark stars by assuming that the disk is isothermal. The case of the disk in thermodynamic equilibrium is discussed in Section \ref{6}. We discuss and conclude our results in Section \ref{7}.

\section{Four-momenta of massless particles at the rotational axis in stationary and axisymmetric spacetimes}\label{2}

Let us consider in the coordinate system $(t,r,\theta,\phi)$ an arbitrary stationary and axially symmetric spacetime, given by the metric
\begin{equation}
ds^{2}=c^2g_{tt}dt^{2}+cg_{t\phi }dtd\phi +g_{rr}dr^{2}+g_{\theta \theta
}d\theta ^{2}+g_{\phi \phi }d\phi ^{2}\;,  \label{ds2rcoappr}
\end{equation}%
and adapted to the symmetries of the geometry,
with the time- and space-like Killing vector fields $(\partial /\partial t)$ and $(\partial /\partial \phi)$, respectively. In the adapted coordinate system the metric coefficients depend only on the coordinates $r$ and $\theta$, measuring the curvature, isotropic, quasi-isotropic etc. radial distances, and the poloidal angle, respectively.

For massless particles propagating along null geodesics in the spacetime (\ref{ds2rcoappr}),
the geodesic equations containing the time-derivatives $\dot{r}=g^{rr}p_{r}$
and $\dot{\theta}=g^{\theta\theta}p_{\theta}$ can be written as
\begin{equation}
g^{rr}p_{r}^{2}+g^{\theta\theta}p_{\theta}^{2}=\frac{g_{\phi\phi}(\omega_{0}/c)^{2}+2g_{t\phi}(\omega_{0}/c)L+g_{tt}L^{2}}{g_{t\phi}^{2}-g_{tt}g_{\phi\phi}},
\label{eq:geodeq1}
\end{equation}
with the constants of motion $p_{t}=-\omega_{0}/c$ and $p_{\phi}=L$. Here $p_t,p_{r},p_{\theta}$
and $p_{\phi}$ are the covariant components of the 4-momentum ${\boldsymbol p}$ of
the light like particles. By applying the formal representations
\begin{eqnarray}
g^{rr}g_{\theta\theta} & = & g_{\theta\theta}/g_{rr}=f(r)\:,
\label{grrgthth}\\
\frac{g_{\phi\phi}g_{\theta\theta}}{g_{t\phi}^{2}-g_{tt}g_{\phi\phi}} & = & \frac{g^{2}(r)}{f(r)}-a^{2}\sin^{2}\theta\:,\label{gppgththg2}\\
\frac{g_{t\phi}g_{\theta\theta}}{g_{t\phi}^{2}-g_{tt}g_{\phi\phi}} & = & a-\frac{g(r)h(r,\theta)}{f(r)}\:,\label{gtpgththg2}\\
\frac{g_{tt}g_{\theta\theta}}{g_{t\phi}^{2}-g_{tt}g_{\phi\phi}} & = & \frac{h^{2}(r,\theta)}{f(r)}-\frac{1}{\sin^{2}\theta},\label{gttgththg2}
\end{eqnarray}
Eq.~(\ref{eq:geodeq1}) can be written as
\begin{equation}
fp_{r}^{2}+p_{\theta}^{2}=\frac{1}{f}\left(g\frac{\omega_{0}}{c}-hL\right)^{2}-\left(a\frac{\omega_{0}}{c}\sin\theta-\frac{L}{\sin\theta}\right)^{2}\:.
\label{eq:geodeqs2}
\end{equation}
In Eqs.~(\ref{grrgthth})-(\ref{gttgththg2})  we have introduced a constant $a$, and the functions
$f=f(r),$ $g=g(r)$ and $h=h(r,\theta)$, and we have assumed that
$h$ depends on both $r$ and $\theta$, but $f$ and $g$ are only
functions of the radial coordinate. This is a rather strong assumption,
since these functions are constructed from the metric coefficients
via Eqs.~(\ref{grrgthth})-(\ref{gttgththg2}), which for a general stationary and axially symmetric
spacetime depend on both $r$ and $\theta$,
\begin{eqnarray}
g_{\pm} & = & -\frac{1}{g_{tt}}\left[g_{t\phi}h\pm\sqrt{\frac{f}{\sin^{2}\theta}(g_{t\phi}^{2}-g_{tt}g_{\phi\phi})}\right]\:,\label{eq:gpm}\\
h^{2} & = & \frac{g_{\theta\theta}}{g_{rr}}\left(\frac{g_{tt}g_{\theta\theta}}{g_{t\phi}^{2}-g_{tt}g_{\phi\phi}}+\frac{1}{\sin^{2}\theta}\right)\:.
\label{eq:h2}
\end{eqnarray}
We will examine the validity of these assumptions for the neutron and
quark star spacetimes under our study in Appendix A. In the following we assume that $g(r)\approx g_{+}(r,0)$ is a good approximation for the function $g$. By virtue of the pure radial dependence of $f$ and $h$, the geodesic equation (\ref{eq:geodeqs2}) for $L=0$ can be separated with respect to the coordinates $r$ and $\theta$:
\begin{eqnarray}
p_{r}^{2} & = & \left(\frac{g_{+}}{f}\frac{\omega_{0}}{c}\right)^{2}-\frac{J^{2}}{f}\:,\label{eq:p2r}\\
p_{\theta}^{2} & = & -\left(a\frac{\omega_{0}}{c}\sin\theta\right)^{2}+J^{2},\label{eq:p2th}\end{eqnarray}
where $J$ is a separation constant. In the following we use the positive branch
$g_{+}$ of Eq.~(\ref{eq:gpm}), which gives $r^{2}+a^{2}$ for the Kerr
metric, where $a$ becomes the rotational parameter, and leads to consistent
identities in the decompositions in Eqs.~(\ref{gppgththg2}) and (\ref{gtpgththg2}).
If we insert the expressions for $p_r$ and $p_{\theta}$ given by Eqs.~(\ref{eq:p2r}) and (\ref{eq:p2th}) into the scalar product of the 4-momenta of a scattering $\nu\bar{\nu}$ pair for $L=0$ we obtain
\begin{eqnarray*}
\boldsymbol{p}_{\nu}\cdot\boldsymbol{p}_{\bar{\nu}} & = & g^{tt}\frac{\omega_{0\nu}}{c}\frac{\omega_{0\bar{\nu}}}{c}+g^{rr}p_{\nu r}p_{\bar{\nu}r}+g^{\theta\theta}p_{\nu\theta}p_{\bar{\nu}\theta}\\
 & = & \frac{\omega_{0\nu}}{c}\frac{\omega_{0\bar{\nu}}}{c}g^{tt}+g^{rr}\frac{g_{+}^{2}}{f^{2}}\frac{\omega_{0\nu}}{c}\frac{\omega_{0\bar{\nu}}}{c}\sqrt{1-\frac{f}{g_{+}^{2}}\frac{cJ_{\nu}^{2}}{\omega_{0\nu}}}\sqrt{1-\frac{f}{g_{+}^{2}}\frac{cJ_{\bar{\nu}}^{2}}{\omega_{0\bar{\nu}}}}\\
 &  & +g^{\theta\theta}\frac{\omega_{0\nu}}{c}\frac{\omega_{0\bar{\nu}}}{c}\sqrt{\frac{c^{2}J_{\nu}^{2}}{\omega_{0\nu}^{2}}-a^{2}\sin^{2}\theta}\sqrt{\frac{c^{2}J_{\bar{\nu}}^{2}}{\omega_{0\bar{\nu}}^{2}}-a^2\sin^2\theta}.
 \end{eqnarray*}
 By introducing the variable $\rho_{\nu}$, and the horizontal collision angle $\theta_{\nu}$
\[
\rho_{\nu}=\frac{cJ_{\nu}}{\omega_{0\nu}}\:,\qquad\sin\theta_{\nu}=\frac{\sqrt{f}}{g_{+}}\rho_{\nu}\:,\]
we can cast the result in the form
\begin{eqnarray*}
\boldsymbol{p}_{\nu}\cdot\boldsymbol{p}_{\bar{\nu}} & = & \frac{\omega_{0\nu}\omega_{0\bar{\nu}}}{c^{2}}\left(g^{tt}+g^{\theta\theta}\frac{g_{+}^{2}}{f}\cos\theta_{\nu}\cos\theta_{\bar{\nu}}+g^{\theta\theta}\sqrt{\frac{g_{+}^{2}}{f}\sin^{2}\theta_{\nu}-a^{2}\sin^{2}\theta}\sqrt{\frac{g_{+}^{2}}{f}\sin^{2}\theta_{\bar{\nu}}-a^{2}\sin^{2}\theta}\right)\:.
\end{eqnarray*}
In the following we restrict our study to the neutrino-anti neutrino pairs that are scattering
at the rotational axis. For $\theta=0$ the scalar product reduces to the expression
\begin{eqnarray*}
\boldsymbol{p}_{\nu}\cdot\boldsymbol{p}_{\bar{\nu}} & = & \frac{\omega_{0\nu}\omega_{0\bar{\nu}}}{c^{2}}\left[g^{tt}+g^{\theta\theta}\left(\frac{g_{+}^{2}}{f}\cos\theta_{\nu}\cos\theta_{\bar{\nu}}+\frac{g_{+}^{2}}{f}\sin\theta_{\nu}\sin\theta_{\bar{\nu}}\right)\right]\\
 & = & -\frac{\omega_{0\nu}\omega_{0\bar{\nu}}}{c^{2}}g^{\theta\theta}\frac{g_{+}^{2}}{f}\left[1-\cos\theta_{\nu}\cos\theta_{\bar{\nu}}-\sin\theta_{\nu}\sin\theta_{\bar{\nu}}\right]\:,
\end{eqnarray*}
where we have used the relation $g^{tt}=-g_{\phi\phi}/(g_{t\phi}^{2}-g_{tt}g_{\phi\phi})$
and Eq.~(\ref{gppgththg2}), evaluated at $\theta=0$. By introducing the proper
energy of the neutrinos defined as
\[
\varepsilon_{\nu}=\hbar\omega_{0\nu}\left.\sqrt{\frac{g_{\phi\phi}}{g_{t\phi}^{2}-g_{tt}g_{\phi\phi}}}\right|_{\theta=0}=\hbar\omega_{0\nu}\sqrt{g^{\theta\theta}(r,0)\frac{g_{+}^2(r,0)}{f(r)}},
\]
 we obtain
 \begin{equation}
\boldsymbol{p}_{\nu}\cdot\boldsymbol{p}_{\bar{\nu}}=-\frac{\varepsilon_{\nu}\varepsilon_{\bar{\nu}}}{c^{2}}(1-\cos\theta_{\nu}\cos\theta_{\bar{\nu}}-\sin\theta_{\nu}\sin\theta_{\bar{\nu}}),
\end{equation}
or, in the general case,
\begin{equation}
\boldsymbol{p}_{\nu}\cdot\boldsymbol{p}_{\bar{\nu}}=-\frac{\varepsilon_{\nu}\varepsilon_{\bar{\nu}}}{c^{2}}[1-\cos\theta_{\nu}\cos\theta_{\bar{\nu}}-\sin\theta_{\nu}\sin\theta_{\bar{\nu}}\cos(\varphi_{\nu}-\varphi_{\bar{\nu}})],
\label{eq:pnupolnu}
\end{equation}
where $\varphi_{\nu}$ and $\varphi_{\bar{\nu}}$ are the azimuthal collision angles for the $\nu\bar{\nu}$ pair, as measured at the rotational axis.

This equation is the same as the one derived for the Kerr spacetime \cite{AsFu01}. Provided that neglecting the angular dependence of the function $g_{+}$, and the substitution of the separated geodesic equations (\ref{eq:p2r}) and (\ref{eq:p2th}) into the scalar product does not increase the error propagation significantly, the expression (\ref{eq:pnupolnu}) is a good approximation for any stationary and axially symmetric spacetime. We can therefore use the same procedure to calculate the neutrino-antineutrino annihilation energy deposition rate at the rotational axis of neutron and quarks stars as the one used for the black holes.

\section{Energy deposition rate from $\nu\bar{\nu}$ pair annihilation along
the rotational axis}\label{3}

We consider a geometrically thin accretion disk around a rotating
compact object, with an inner edge at the curvature radius $R_{in}$, and a cutoff radius
at $R_{out}$. We assume that the mass-energy of the disk
produces only a negligible effect on the spacetime geometry of the
central object. Thus any massless particle can propagate along null geodesics
in the vicinity of the disk-compact object system, with the global geometry determined by the central massive object. We also suppose
that the disk is the primary source of neutrinos in the system, and
we neglect any interaction between the disk and the neutrino/anti neutrino
radiation after their generation in the rotating plasma. Furthermore we
assume that these particles can freely propagate along null geodesics
even in the interior of the central object, and have only negligible energy losses in the
interactions with the stellar matter. In the following we determine the electron-positron energy deposition rate (EDR) produced due to the neutrino-anti neutrino pair annihilations into $e^{-}e^{+}$ pairs
along the rotation axis of the compact object, where the baryon contamination
related to the matter content of the disk corona is minimal.

For any point ${\boldsymbol{r}=(r,\theta,\phi)}$ of the $\nu-\bar{\nu}$ collision, the EDR per unit volume via electron-positron pair creation due to neutrino-anti neutrino annihilation is given by
\begin{equation}\label{eq:dotq}
\dot{q}(\boldsymbol{r})=\frac{dE_{0}}{dtdV}=\int\int f_{\nu}(\mathbf{p}_{\nu},{\boldsymbol{r}})f_{\bar{\nu}}({\mathbf{p}}_{_{\bar{\nu}}},{\boldsymbol{r}})\{\sigma_{\nu\bar{\nu}}|{\boldsymbol{v}}_{\nu}-
{\boldsymbol{v}}_{\bar{\nu}}|\varepsilon_{\nu}\varepsilon_{\bar{\nu}}\}\frac{\varepsilon_{\nu}+
\varepsilon_{\bar{\nu}}}{\varepsilon_{\nu}\varepsilon_{\bar{\nu}}}d^{3}{\mathbf{p}}_{\nu}d^{3}
{\mathbf{p}}_{\bar{\nu}},
\end{equation}
where ${\mathbf{p}}_{\nu}$ and ${\mathbf{p}}_{\bar{\nu}}$ are the 3-momenta in the momentum
space, $f_{\nu}$
and $f_{\bar{\nu}}$ are the neutrino and anti-neutrino number densities, and
${\boldsymbol{v}}_{\nu}$, $\varepsilon_{\nu}$, ${\boldsymbol{v}}_{\bar{\nu}}$
and $\varepsilon_{\bar{\nu}}$ are the 3-velocities and the energy
of the colliding neutrino-antineutrino pairs, respectively \cite{Go87}.
The cross-section of the collision is denoted by $\sigma_{\nu\bar{\nu}}$, and can be calculated by using the formula
\[\sigma_{\nu\bar{\nu}}=KG^2_F(\varepsilon_{\nu}\varepsilon_{\bar{\nu}}-c^2{\mathbf{p}}_{\nu}\cdot{\mathbf{p}}_{{\bar{\nu}}})=-KG^2_F {\boldsymbol{p}}_{\nu}\cdot{\boldsymbol{p}}_{{\bar{\nu}}}\]
(special relativity is considered), where $K=(1\pm4\sin^{2}\theta_{W}+8\sin^{4}\theta_{W})/6\pi$, with $\sin^{2}\theta_{W}=0.23$, and $G_{F}^{2}=5.29\times10^{-44}\;\mathrm{cm}^{2}\;\mathrm{MeV}^{-2}$, respectively.
 By applying the decompositions
${\mathbf{p}}_{\nu} =\varepsilon_{\nu }\boldsymbol{\Omega }_{\nu }$
and $d^{3}\mathbf{p}_{\nu}=\varepsilon_{\nu }^{2}d\varepsilon_{\nu}d\boldsymbol{\Omega}_{\nu}$,
with the solid angle vector
$\boldsymbol{\Omega}_{\nu }$
pointing in the direction of ${\mathbf{p}}_{\nu}$, and with the assumption that the neutrino source in the disk emits particles isotropically, the integral in Eq.~(\ref{eq:dotq}) can be separated into an energy
integral and an angular part.

After evaluating the
energy integral for fermions and inserting the general relativistic version  of the scalar product of the 4-momenta given by Eq.~(\ref{eq:pnupolnu}) into the cross-section equation for the colliding $\nu\bar{\nu}$ pair, for a stationary and axially symmetric geometry we obtain the EDR at the rotational axis ($\theta=0$) per unit volume per unit time ($dtdV=\sqrt{-g}d^{4}x$) in the form \cite{AsFu01}
\begin{eqnarray}
\dot{q}(r)&=&\frac{dE_{0}}{dtdV}=2cKG_{F}^{2}\Theta(r)\int\int f_{\nu}f_{\bar{\nu}}(\varepsilon_{\nu}+\varepsilon_{\bar{\nu}})\varepsilon_{\nu}^{3}\varepsilon_{\bar{\nu}}^{3}d\varepsilon_{\nu}d\varepsilon_{\bar{\nu}}\nonumber\\
& = & \frac{21\pi^{4}}{4}\zeta(5)\frac{KG_{F}^{2}}{h^{6}c^{5}}k^{9}T_{eff}^{9}(3r_{g})\Theta(r).\label{EDR0}
\end{eqnarray}
Here  $T_{eff}(3r_g)$ is the effective neutrino temperature evaluated at 3 $\times r_g=6GM/c^2$, and $\zeta $ is the Riemann zeta function. The angular part $\Theta(r)$ of the EDR is given by
\begin{eqnarray}
\Theta(r) & = & \frac{1}{T_{eff}^{9}(3r_{g})}\left(\left.\frac{g_{\phi\phi}}{g_{t\phi}^{2}-g_{tt}g_{\phi\phi}}\right|_{\theta=0}\right)^{4}\int_{\theta_{m}}^{\theta_{M}}d\theta_{\nu}\sin\theta_{\nu}\nonumber\\
 &  & \times\int_{\theta_{m}}^{\theta_{M}}d\theta_{\bar{\nu}}\sin\theta_{\bar{\nu}}\int_{0}^{2\pi}d\varphi_{\nu}\int_{0}^{2\pi}d\varphi_{\bar{\nu}}T_{0}^{5}(R_{\nu})\nonumber\\
 &  & \times T_{0}^{4}(R_{\bar{\nu}})[1-\sin\theta_{\nu}\sin\theta_{\bar{\nu}}\cos(\varphi_{\nu}-\varphi_{\bar{\nu}})\nonumber\\
 &  & -\cos\theta_{\nu}\cos\theta_{\bar{\nu}}]^{2},
 \label{Fr}
 \end{eqnarray}
with the neutrino temperature $T_0$ observed at infinity, and the angular boundaries $\theta_m$ and $\theta_M$ of the collision angle for the (anti)neutrinos arriving form the disk edges at $R_{in}$ and $R_{out}$, respectively. The expression in parenthesis is the redshift factor for massless particles.

Integrating this expression over the azimuthal collision angles $\varphi_{\nu}$ and $\varphi_{\bar{\nu}}$ of the neutrino-antineutrino pairs yields the formula
\begin{eqnarray}
\Theta(r) & = & \frac{2\pi^{2}}{T_{eff}^{9}(3r_{g})}\left(\left.\frac{g_{\phi\phi}}{g_{t\phi}^{2}-g_{tt}g_{\phi\phi}}\right|_{\theta=0}\right)^{4}\nonumber\left[2\int_{\mu_{m}}^{\mu_{M}}d\mu_{\nu}T_{0}^{5}(\mu_{\nu})\int_{\mu_{m}}^{\mu_{M}}d\mu_{\bar{\nu}}T_{0}^{4}(\mu_{\bar{\nu}})\right.\nonumber\\
 &  & +\int_{\mu_{m}}^{\mu_{M}}d\mu_{\nu}(1-\mu_{\nu}^{2})T_{0}^{5}(\mu_{\nu})\int_{\mu_{m}}^{\mu_{M}}d\mu_{\bar{\nu}}(1-\mu_{\bar{\nu}}^{2})T_{0}^{4}(\mu_{\bar{\nu}})\nonumber\\
 &  & +2\int_{\mu_{m}}^{\mu_{M}}d\mu_{\nu}\mu_{\nu}^{2}T_{0}^{5}(\mu_{\nu})\int_{\mu_{m}}^{\mu_{M}}d\mu_{\bar{\nu}}\mu_{\bar{\nu}}^{2}T_{0}^{4}(\mu_{\bar{\nu}})\nonumber\\
 &  & \left.-4\int_{\mu_{m}}^{\mu_{M}}d\mu_{\nu}\mu_{\nu}T_{0}^{5}(\mu_{\nu})\int_{\mu_{m}}^{\mu_{M}}d\mu_{\bar{\nu}}\mu_{\bar{\nu}}T_{0}^{4}(\mu_{\bar{\nu}})\right]\:,\label{Th}\end{eqnarray}
where $\mu_{\nu}=\cos\theta_{\nu}$, $\mu_{\bar{\nu}}=\cos\theta_{\bar{\nu}}$,
$\mu_{m}=\cos\theta_{m}$ and $\mu_{M}=\cos\theta_{M}$.

In order to determine the boundary angles $\theta_m$ and $\theta_M$ (or, in turn, $\mu_m$ and $\mu_M$), we traced the null geodesics back from a given point on the rotational axis to its intersection with the equatorial plane.
In the ray tracing code we supplemented the coordinates of $r$ and $\theta$ with their momenta $p_r$ and $p_{\theta}$, and used the dynamical equations for the extended set of variables.  We present some details of this method in Appendix B. For a given set of canonical data of the final point at the rotational axis the code derived the canonical data of the initial point in the equatorial plane. That is, for any radial coordinate $r$ along the axis, the code can provide the values of $p_{\theta}$ and $p_r$ for which null geodesics starting from the equatorial plane at a given radius $R$ will hit the axis at $r$. The collision angle $\theta_{\nu}$ is determined by the scalar product of the normal vector pointing into the direction of the 3-momentum ${\mathbf p}_{\nu}=((p_{\nu})^r,(p_{\nu})^{\theta},(p_{\nu})^{\phi})$ of the neutrino passing through the axis,  and the normal vector ${\mathbf e}_z=(1/\sqrt{g_{rr}},0,0)$ of the rotational axis,
\begin{equation}
\cos\theta_{\nu}=\frac{{\mathbf p}_{\nu}\cdot{\mathbf e}_z}{|{\mathbf p}_{\nu}|}
=\frac{(p_{\nu})_r}{\sqrt{g_{rr}[(p_{\nu})^2_r/g_{rr}+(p_{\nu})^2_{\theta}/g_{\theta\theta}]}}\;, \label{costh0}
\end{equation}
where we have used the expression ${\mathbf p}_{\nu}\cdot{\mathbf e}_z=g_{rr}(p_{\nu})^r(e_z)^r=\sqrt{g_{rr}}(p_{\nu})^r$, and the fact that the norm of the 3-momentum is reduced to  $|{\mathbf p}_{\nu}|=\sqrt{g^{rr}(p_{\nu})^2_r+g^{\theta\theta}(p_{\nu})^2_{\theta}}$,
since $(p_{\nu})_{\phi}=L_{\nu}$ is zero for the particles reaching the rotational axis.

We can express the square of $p_r$ from the geodesic equation (\ref{eq:geodeq1}) as
\begin{equation}
\frac{(p_{\nu})^2_r}{g_{rr}} = \frac{g_{\phi\phi}(\omega_0/c)^2}{g^2_{t\phi}-g_{tt}g_{\phi\phi}} - \frac{(p_{\nu})^2_{\theta}}{g_{\theta\theta}}=\frac{g^2_{+}(\omega_0/c)^2f^{-1}-(p_{\nu})^2_{\theta}}{g_{\theta\theta}}\;.\label{p2rgrr}
\end{equation}
Here the constant of motion $\omega_0$ for the particle energy is arbitrary ($<c^2$), provided the mass shell condition ${\boldsymbol p}_{\nu}\cdot{\boldsymbol p}_{\nu}=0$ for null particles holds. Combining Eqs.~ (\ref{costh0}) and (\ref{p2rgrr}) together we obtain
a relation between the collision angle and  $p_{\theta}$:
\begin{equation}
\cos\theta_{\nu}=\pm\sqrt{1-\left[\frac{(p_{\nu})_{\theta}}{\omega_0/c}\right]^2\frac{f(r)}{g^2_{+}(r,0)}}\label{costh}
\end{equation}
or
\begin{equation}
\sin^2\theta_{\nu}=\left[\frac{(p_{\nu})_{\theta}}{\omega_0/c}\right]^2\frac{f(r)}{g^2_{+}(r,0)}\:.\label{sin2th}
\end{equation}

From the latter  formula in the Kerr spacetime we have $\sin\theta_{\nu}=\rho_{\nu} \sqrt{\Delta}/{(r^2+a^2)}$,  with the horizon function $\Delta=r^2+a^2-2Mr$, and $\rho_{\nu}=c(p_{\nu})_{\theta}/\omega_0=cJ/\omega_0$ at the rotational axis \cite{AsFu01}.
We used the relation (\ref{costh}) to calculate $\theta_{\nu}$ for any particle traced back from a given point at $r$ on the rotational axis, with a given value of $p_{\theta}$, to its starting point at $R$ in the equatorial plane. For $R=R_{in}$ and $R=R_{out}$, Eq.~(\ref{costh}) provides the values of $\theta_m$,  $\theta_M$ (o, equivalently, $\mu_m$ and $\mu_M$), for any value of $r$.

For neutrinos propagating from the accretion
disk with zero angular momentum  in the direction of the disk rotation, the energy is redshifted
by the gravitational potential, and Lorenz boosted due to the disk
motion measured in the locally non-rotating frame (LNRF) of the spacetime.
Then the observed neutrino temperature can be written as
\begin{equation}
T_{0}(R)=\frac{\sqrt{g_{t\phi}^{2}-g_{tt}g_{\phi\phi}}}{\Gamma g_{\phi\phi}}T_{eff}(R)\:,\label{eq:T0R}
\end{equation}
where $\Gamma=(1-v^{2}/c^{2})^{-1/2}$, and the square of the disk
velocity measured in the LNRF is
\[
v^{2}=\frac{g_{\phi\phi}^{2}}{g_{t\phi}^{2}-g_{tt}g_{\phi\phi}}\left(\Omega_K-\omega\right)^{2}
\]
with the angular frequency $\Omega_K$ of the Keplerian disk, and the frame dragging frequency $\omega$ of the spacetime given by
\[\Omega_K  =\frac{d\phi }{dt}=c\frac{-g_{t\phi ,r}+\sqrt{(g_{t\phi
,r})^{2}-g_{tt,r}g_{\phi \phi ,r}}}{g_{\phi \phi ,r}}\:,\qquad\omega=-\frac{cg_{t\phi}}{g_{\phi\phi}}\;,
\]
respectively. If we solve Eq.~(\ref{costh}) for $p_{\theta}$ at a given value of $\theta_{\nu}$, and use the result to prescribe the canonical data of any final point on the axis, then we can trace null particles back to their initial point in the equatorial plane. Hence we can determine the radial coordinate $R$ measured along the disk at which $T_0$ has to be evaluated in the integral given by Eq.~(\ref{Th}) for any $\mu_{\nu}=\arccos \theta_{\nu}$.

The energy deposition rate per unit volume for a distant observer is
given by Eq.~(\ref{EDR0}), with an integral over the volume $dV=\sqrt{-g}drd\theta d\phi$. Since we estimate the EDR only along the rotational axis over
$r_{min}$ and $r_{max}$ with $\theta\rightarrow0$, it is proportional
to the integral
\begin{eqnarray}
\left.\frac{dE_{0}}{dt}\right|_{\theta=0} & \simeq & 2\frac{d}{d\theta}\left.\int_{0}^{2\pi}\int_{0}^{\pi/2}\int_{r_{min}}^{r_{max}}\frac{dE_{0}}{dtdV}\sqrt{-g}drd\theta d\phi\right|_{\theta=0}\nonumber\\
 & = & 21\pi^{5}\zeta(5)\frac{KG_{F}^{2}}{h^{6}c^{5}}k^{9}T_{eff}^{9}(3r_{g})r_{g}^{2}\int_{r_{min}}^{r_{max}}G(r)dr\:,
 \label{dE0dt}
 \end{eqnarray}
where the dimensionless quantity $G(r)$ is defined by
\begin{equation}
G(r)\equiv r_{g}^{-2}\left.\frac{\sqrt{-g(r,\theta)}}{\sin\theta}\right|_{\theta=0}\Theta(r)\:.\label{G}
\end{equation}

We note that since $g\sim\sin^{2}\theta$
holds for the metric determinant of stationary and axially symmetric spacetimes in both the curvature and isotropic coordinate systems, $d\sqrt{-g}/d\theta|_{\theta=0}$
can be written in the form $(\sqrt{-g}/\sin\theta)(d\sin\theta/d\theta)_{\theta=0}=\sqrt{-g}/\sin\theta$,
which has a finite value at $\theta=0$. For Kerr the metric the cofactor
of $\Theta(r)$ in the Eq.~(\ref{G}) is just $(r^{2}+a^{2})/r_{g}^{2}$, and $G$ reduces to its definition introduced in \cite{AsFu01}.

\section{Equations of state and stellar models}\label{eos}

In order to obtain a consistent and realistic physical description of the rotating general relativistic neutron and quark stars, as a first step we have to adopt the equations of state for the dense neutron and quark matter, respectively. In the present study we consider the following equations of state for neutron and quark matter:

1) Akmal-Pandharipande-Ravenhall  (APR) EOS \citep{Ak98}. EOS APR has been obtained by using the variational chain summation methods and the  Argonne $v_{18}$ two-nucleon interaction. Boost corrections to the two-nucleon interaction, which give the leading relativistic effect of order $(v/c)^2$, as well as three-nucleon interactions, are also included in the nuclear Hamiltonian.  The density range is from $2\times 10^{14}$ g/cm$^3$ to $2.6\times 10^{15}$ g/cm$^3$. The maximum mass limit in the static case for this EOS is $2.20 M_{\odot}$.  We join this equation of state to the composite BBP ($\epsilon /c^2>4.3\times10^{11}$g/cm$^3$) \citep{Ba71a} - BPS ($10^4$ g/cm$^3$ $<4.3\times 10^{11}$g/cm$^3$) \citep{Ba71b} - FMT ($\epsilon/c^2<10^4$ g/cm$^3$) \citep{Fe49} equations of state, respectively.

2) Douchin-Haensel (DH) EOS \citep{DoHa01}. EOS DH is an equation of state of the neutron star matter, describing both the neutron star crust and the liquid core. It is based on the effective nuclear interaction SLy of the Skyrme type, which is particularly suitable for the application to the calculation of the properties of very neutron rich matter. The structure of the crust, and its EOS, is calculated in the zero temperature approximation, and under the assumption of the ground state composition. The EOS of the liquid core is calculated assuming (minimal) $npe\mu $ composition. The density range is from $3.49\times 10^{11}$ g/cm$^3$ to $4.04\times 10^{15}$ g/cm$^3$. The minimum and maximum masses of the static neutron stars for this EOS are $0.094M_{\odot}$ and $2.05 M_{\odot}$, respectively.

3) Shen-Toki-Oyamatsu-Sumiyoshi (STOS) EOS \citep{Shen}. The STOS equation of state of nuclear matter is obtained by using the relativistic mean field theory with nonlinear $\sigma $ and $\omega $ terms in a wide density and temperature range, with various proton fractions. The EOS was specifically designed for the use of supernova simulation and for the neutron star calculations. The  Thomas-Fermi approximation is used to describe inhomogeneous matter, where heavy nuclei are formed together with free nucleon gas. The temperature is mentioned for each STOS equation of state, so that, for example, STOS0 represents the STOS EOS for $T=0$ MeV. For the proton fraction we chose  the value $Y_p=10^{-2}$ in order to avoid the negative pressure regime for low baryon mass densities.

4) Relativistic Mean Field (RMF) equations of state with isovector scalar mean field corresponding to the $\delta $-meson- RMF soft and RMF stiff EOS \citep{Kubis}.  While the $\delta $-meson mean field vanishes in symmetric nuclear matter, it can influence properties of asymmetric nuclear matter in neutron stars. The Relativistic mean field contribution due to the $\delta $-field to the nuclear symmetry energy is negative. The energy per particle of neutron matter is then larger at high densities than the one with no $\delta $-field included. Also, the proton fraction of $\beta $-stable matter increases. Splitting of proton and neutron effective masses due to the $\delta $-field can affect transport properties of neutron star matter. The equations of state can be parameterized by the coupling parameters $C_{\sigma }^2=g_{\sigma }^2/m_{\sigma }^2$, $C_{\omega }^2=g_{\omega }^2/m_{\omega }^2$, $\bar{b}=b/g_{\sigma}^3$ and $\bar{c}=c/g_{\sigma}^4$, where $m_{\sigma }$ and $m_{\omega }$ are the masses of the respective mesons, and $b$ and $c$ are the coefficients in the potential energy $U\left(\sigma \right)$ of the $\sigma $-field. The soft RMF EOS is parameterized by $C_{\sigma }^2=1.582$ fm$^2$, $C_{\omega }^2=1.019$ fm$^2$, $\bar{b}=-0.7188$ and $\bar{c}=6.563$, while the stiff RMF EOS is parameterized by $C_{\sigma }^2=11.25$ fm$^2$, $C_{\omega }^2=6.483$ fm$^2$, $\bar{b}=0.003825$ and $\bar{c}=3.5\times 10^{-6}$, respectively.

5) Baldo-Bombaci-Burgio (BBB) EOS \citep{baldo}. The BBB EOS is an EOS for asymmetric nuclear matter, derived from the Brueckner-Bethe-Goldstone many-body theory with explicit three-body forces. Two EOS's are obtained, one corresponding to the Argonne AV14 (BBBAV14), and the other to the Paris two-body nuclear force (BBBParis), implemented by the Urbana model for the three-body force. The maximum static mass configurations are $M_{max} = 1.8 M_{\odot}$ and  $M_{max} = 1.94 M_{\odot}$ when the AV14 and Paris interactions are used, respectively.  The onset of direct Urca processes occurs at densities $n\geq 0.65$ fm$^{-3}$ for the AV14 potential and $n\geq 0.54$ fm$^{-3}$ for the Paris potential. The comparison with other microscopic models for the EOS shows noticeable differences. The density range is from $1.35\times 10^{14}$ g/cm$^3$ to $3.507\times 10^{15}$ g/cm$^3$.

6) Bag model equation of state for quark matter (Q) EOS \citep{It70, Bo71, Wi84, Ch98}. For the description of the quark matter we adopt first a simple phenomenological description, based on the MIT bag model equation of state, in which the pressure $p$ is related to the energy density $\rho $ by
\begin{equation}
p=\frac{1}{3}\left(\rho-4B\right)c^2,
\end{equation}
where $B$ is the difference between the energy density of the perturbative and
non-perturbative QCD vacuum (the bag constant), with the value $4B=4.2\times 10^{14}$ g/cm$^3$.

7) It is generally agreed today that the color-flavor-locked (CFL) state is
likely to be the ground state of matter, at least for asymptotic densities,
and even if the quark masses are unequal \citep{cfl1,cfl2, HoLu04, cfl3}. Moreover,
the equal number of flavors is enforced by symmetry, and electrons are
absent, since the mixture is automatically neutral. By assuming that the
mass $m_{s}$ of the $s$ quark is not large as compared to the chemical
potential $\mu $, the thermodynamical potential of the quark matter in CFL
phase can be approximated as \citep{LuHo02}
\begin{equation}
\Omega _{CFL}=-\frac{3\mu ^{4}}{4\pi ^{2}}+\frac{3m_{s}^{2}}{4\pi ^{2}}-%
\frac{1-12\ln \left( m_{s}/2\mu \right) }{32\pi ^{2}}m_{s}^{4}-\frac{3}{\pi
^{2}}\Delta ^{2}\mu ^{2}+B,
\end{equation}%
where $\Delta $ is the gap energy. With the use of this expression the
pressure $P$ of the quark matter in the CFL phase can be obtained as an
explicit function of the energy density $\varepsilon $ in the form \citep%
{LuHo02}
\begin{equation}\label{pres}
P=\frac{1}{3}\left( \varepsilon -4B\right) +\frac{2\Delta ^{2}\delta ^{2}}{\pi
^{2}}-\frac{m_{s}^{2}\delta ^{2}}{2\pi ^{2}},
\end{equation}
where
\begin{equation}
\delta ^{2}=-\alpha +\sqrt{\alpha ^{2}+\frac{4}{9}\pi ^{2}\left( \varepsilon
-B\right) },
\end{equation}%
and $\alpha =-m_{s}^{2}/6+2\Delta ^{2}/3$. In the following the value of the gap energy $\Delta $ considered in each case will be also mentioned for the CFL equation of state, so that, for example, CFL200 represents the CFL EOS with $\Delta =200$. For the bag constant $B$ we adopt the value $4B=4.2\times 10^{14}$ g/cm$^3$, while for the mass of the strange quark we take the value $m_s=150$ MeV.


The pressure-density relation is presented for the considered equations of state in Fig.~{\ref{fig1}.

\begin{figure}[tbp]
\centering
\includegraphics[width=8.15cm]{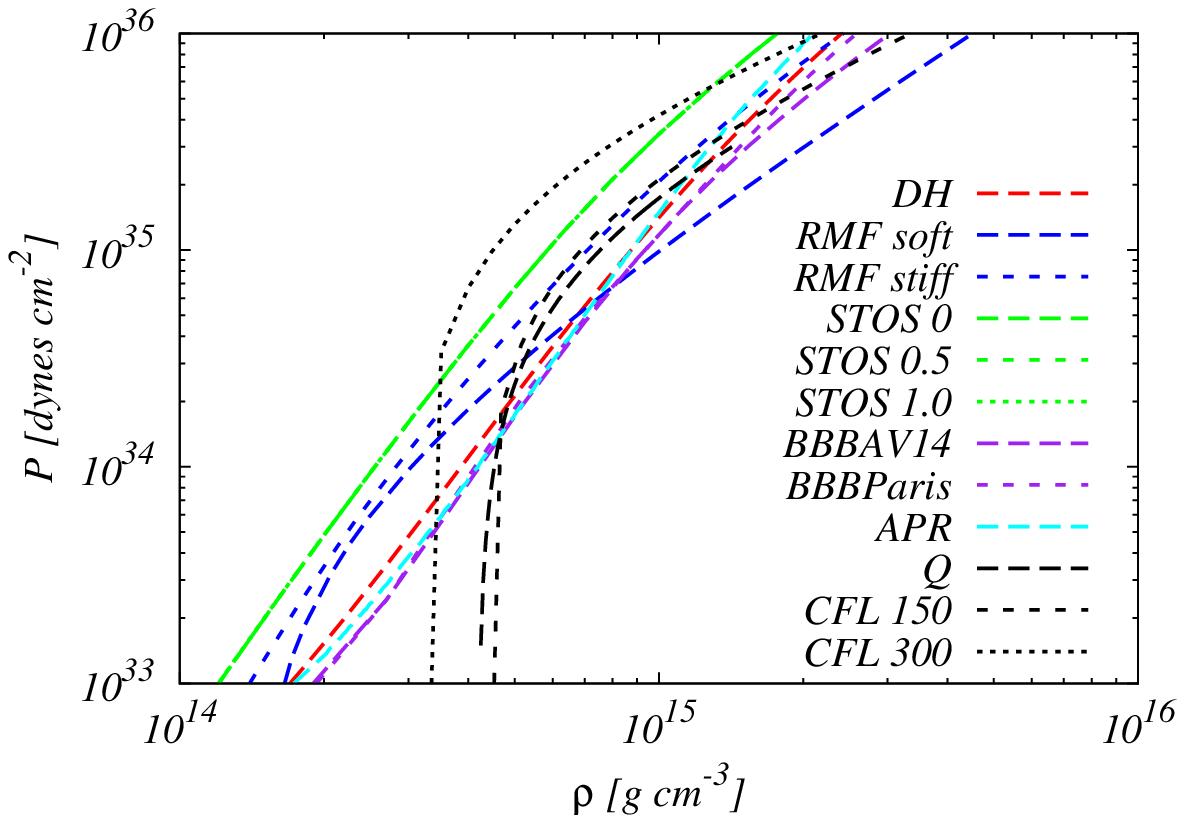}
\caption{Pressure as a function of density (in a logarithmic scale) for the equations of state  DH, RMF soft, RMF stiff, STOS 0. STOS 0.5, STOS 1, BBBAV14, BBBParis, APR, Q,  CFL150, and CFL300, respectively.}
\label{fig1}
\end{figure}

 In order to calculate the equilibrium configurations of the rotating neutron and quark stars with the EOS's presented here  we use the RNS code, as introduced in Stergioulas \& Friedman (1995), and discussed in details in Stergioulas (2003). This code was used for the study of different models of  rotating neutron stars \citep{No98}, and for the study of the rapidly rotating strange stars \citep{Ste99}.
The metric outside the rotating compact general relativistic stars can be described,
in quasi-isotropic coordinates, as
\begin{equation}
ds^2=-e^{{\gamma }+{\rho }}dt^2+e^{2{\alpha } }(d\bar{r}^2+\bar{r}^2d\theta ^2)+
e^{{\gamma }-{\rho }}\bar{r}^2\sin ^2\theta (d\phi -{\omega}dt)^2,\label{ds20}
\end{equation}
where the metric potentials ${\gamma }$, ${\rho }$, ${\alpha }$ and the angular velocity of the stellar fluid relative to the local inertial frame ${\omega}$ are all functions of the quasi-isotropic radial coordinate $\bar{r}$ and of the poloidal angle $\theta$. The RNS code computes the metric functions in a quasi-spheroidal coordinate system and write them as functions of the compactified dimensionless distance $s=\bar{r}/\left(\bar{r}+\bar{r}_e\right)$, where $\bar{r}_e$ is the equatorial radius of the star, and the cosine of the poloidal angle. The quasi-spheroidal radial coordinate is converted into curvature coordinate $r$ according to the equation $r=\bar{r}\exp\left\{\left[\gamma(\bar{r},\theta)-\rho(\bar{r},\theta)\right]/2\right\}$.

\section{The isothermal thin disk}\label{5}

With the assumption that the surface of the thin accretion disk is in an
isothermal state, (which is definitely a crude approximation), the effective
temperature $T_{eff}$ of the disk surface can be set to a constant over the
entire disk. Following \cite{AsFu01},  we use a constant effective temperature measured at $3r_{g}$, $T_{eff}=T_{eff}(3r_{g})$, with $T_{eff}$ in the range of 10 to 20 MeV.
In this case the radial dependence of the neutrino temperature $T_{0}$ is determined only by the
gravitational redshift, and the Lorentz boost, as can be seen in Eq.~(\ref{eq:T0R}).
Since the neutrinosphere either does not exist (if the central body is not compact enough), or its equatorial radius is smaller than the stellar radius, for the  configurations of the neutron and quark star models presented here, the entire surface of the accretion disk can be considered as a neutrino source \cite{Ko09}.

As it has been showed before, if black holes spin up the energy deposition via the neutrino pair annihilation is increasing at the rotational axis of the black hole  \cite{AsFu01,Mi03}. This phenomenon is a consequence of the decreasing radius of the marginally stable orbit, i.e., of the increasing surface area of the accretion disk with the increasing spin. However, this enhancement is moderated by the redshift of the neutrinos emerging from the innermost region of the disk, which causes a smaller and smaller increment in the EDR as the spin parameter is approaching one. The Doppler shift due to the disk rotation has also to be taken into account when we calculate the net temperature effect on the axial distribution of the EDR.

If the central body is a neutron or quark star, then the situation is even more complicated. In many cases there is no marginally stable orbit outside the star and the inner edge of the accretion disk is located at the stellar surface. Then the proportionality between the area of the disk surface and the spin is no longer guaranteed, and it indeed breaks down at higher rotational frequencies for some physical configurations of stars, and for some EOS types where inner disk edge touches the stellar surface.

\begin{figure}
\includegraphics[width=8.15cm]{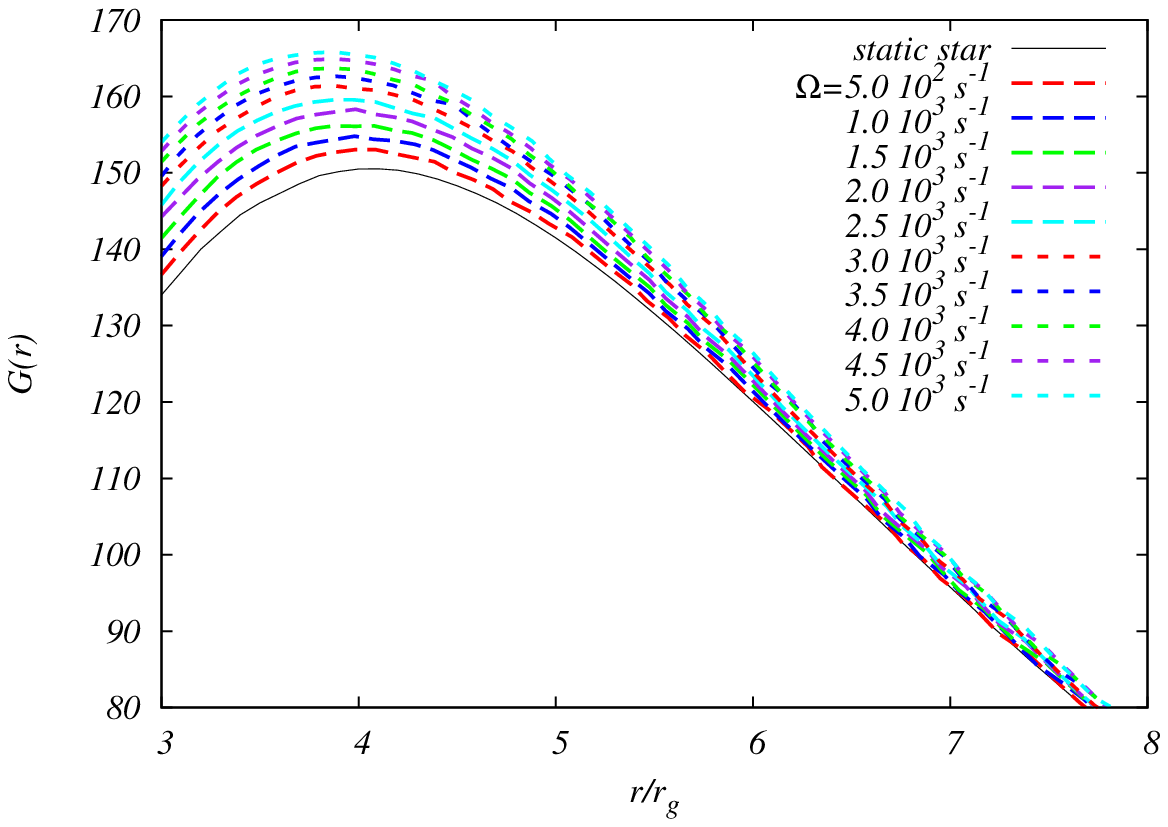}
\includegraphics[width=8.15cm]{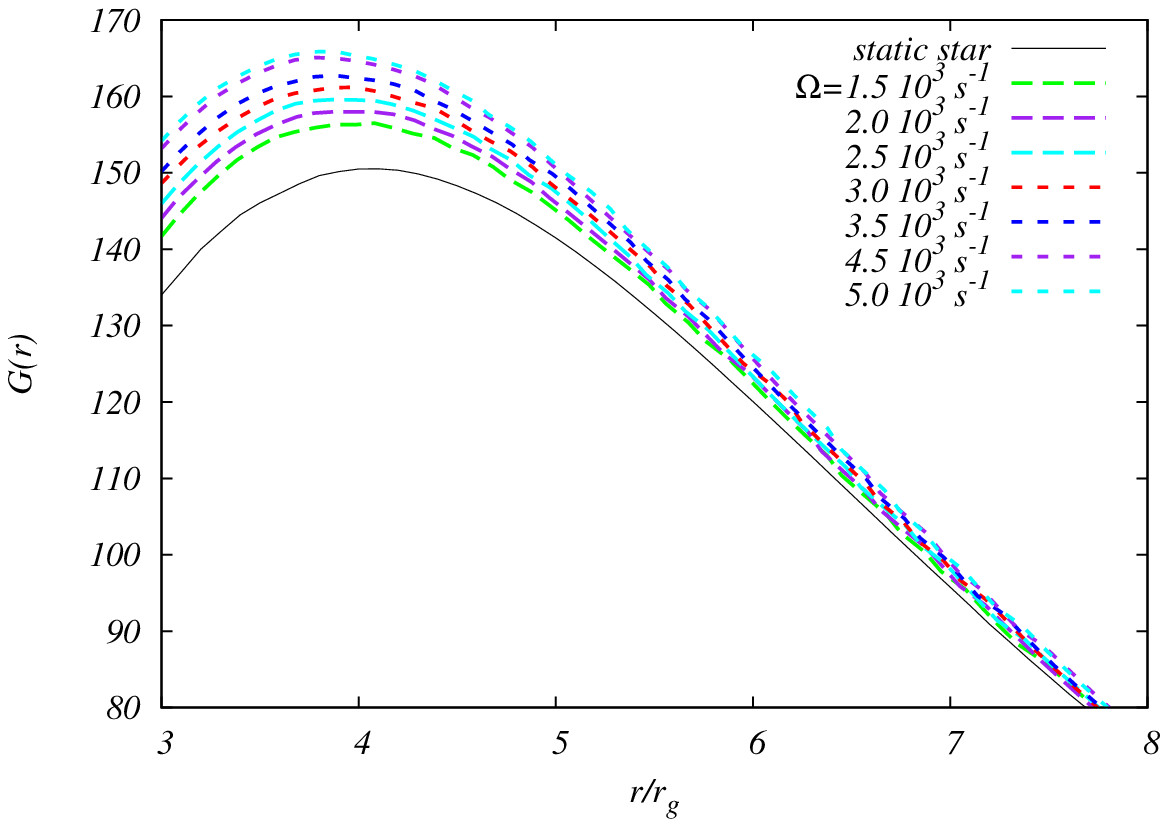}
\includegraphics[width=8.15cm]{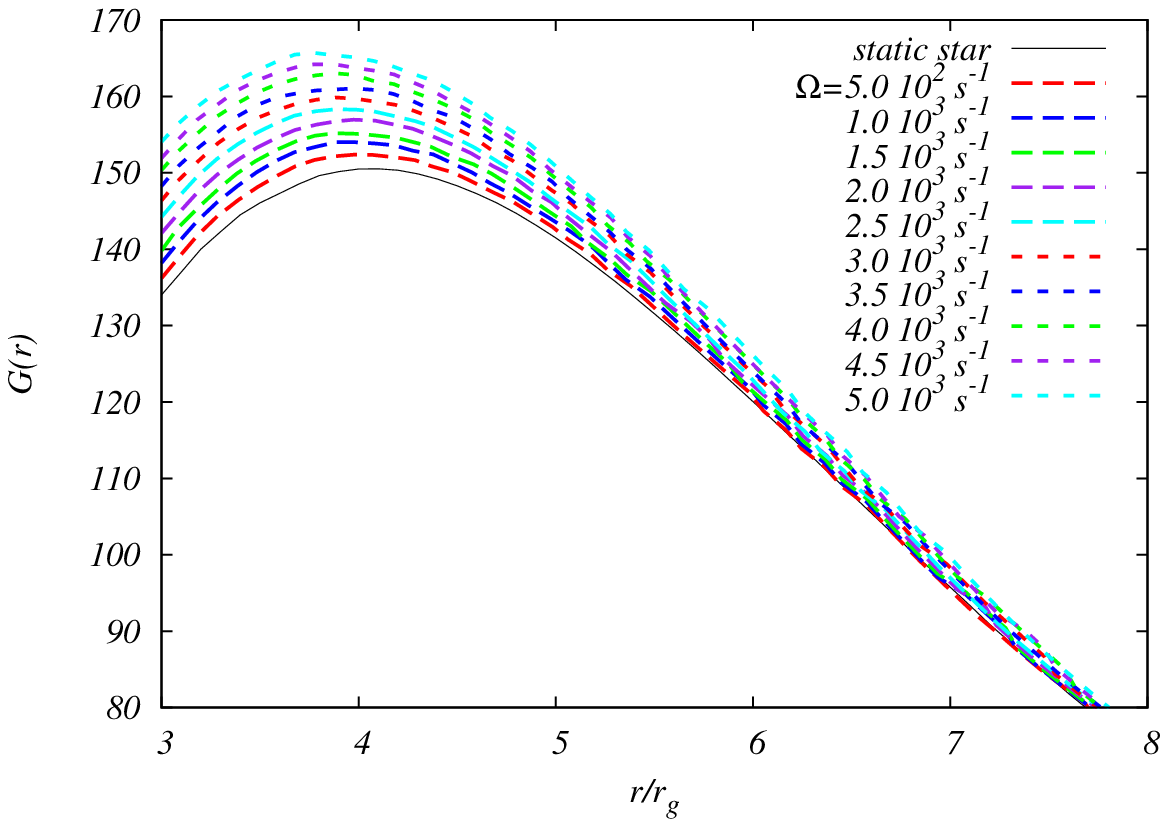}
\includegraphics[width=8.15cm]{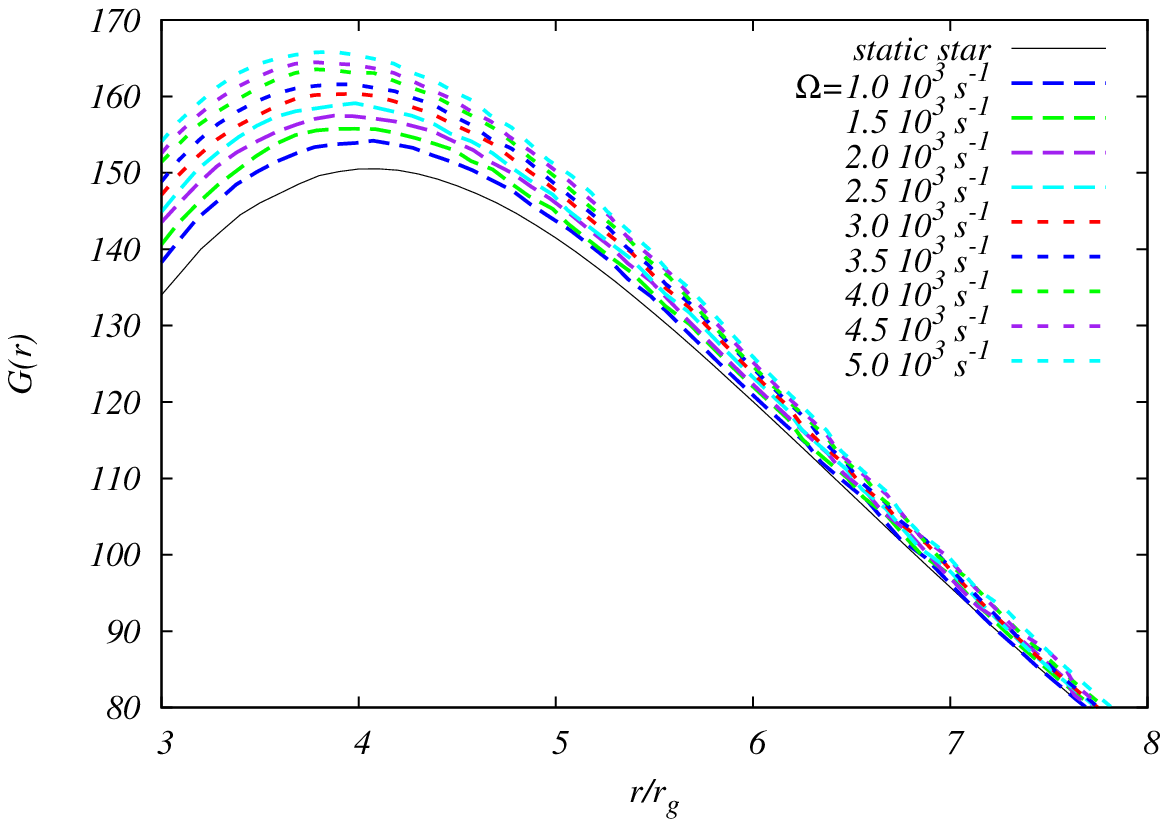}
\centering
\caption{The axial profile of the EDR
characterized by the dimensionless quantity $G(r)$ as a function of $\Omega$ for an isothermal accretion disk and neutron star models with DH (top left hand side), APR (top right hand side), BBBAV14 (bottom left hand side) and BBBParis (bottom right hand side) type EOS. The total stellar mass is $1.8 M_{\odot}$. For comparison we also plotted $G(r)$ for a static spacetime geometry (black solid curve).}
\label{fig2}
\end{figure}

In Table~\ref{table0} we present the relation between the spin parameter, the radius of the inner edge of the disk,  and the enhancement in the EDR. Here we have integrated the EDR given by Eq.~(\ref{dE0dt}) over $r=3-10r_g$ for each configuration, ${\rm EDR}=\int_{3r_g}^{10r_g}(dE_0/dt)_{\theta=0}dr$,
and we have computed $\Delta$EDR=(EDR-${\rm EDR}_s$)/${\rm EDR}_s$, where ${\rm EDR}_s$ is derived from the Schwarzschild geometry. The cutoff radius $R_{out}$ of the accretion disk was set to $10r_g$. In the calculations we used the metric provided by the RNS code for the total mass $M=1.8 M_{\odot}$, and different angular velocities between $5\times10^2 \;{\rm s}^{-1}$ and $5\times10^3\; {\rm s}^{-1}$. The tables containing the physical parameters of these stellar configurations can be found in Appendix C.

For the first four stellar models presented in Table~\ref{table0}, namely those with DH, APR, BBBAV14 and BBBParis type EOS's, the EDR is increasing as the disk surface is increasing ($R_{in}$ is decreasing) with the increasing spin of the star. Since at the rotation frequency of $5\times10^3\;{\rm s}^{-1}$ the value of the spin parameter  is only about $0.3-0.35$, the increase in the EDR as compared to the EDR obtained for a static star or black hole is only  $\sim7$\%. For a black hole spinning up to $a_*=0.99$, this enhancement is already 30\% \cite{AsFu01}

\begin{table*}
\centering
\begin{tabular}{|l|l|c|c|c|c|c|c|c|c|c|c|}
\hline
EOS & $\Omega\;[10^3{\rm s}^{-1}]$	& 0.5 & 1.0 & 1.5 & 2.0 & 2.5 & 3.0 & 3.5 & 4.0 & 4.5 & 5.0\\
\hline
DH & $a_*$  				& 0.03 &0.07&0.10 &0.13 & 0.17 & 0.21 & 0.24 & 0.27 & 0.31  & 0.35 \\
& $R_{in}/r_g$ 			& 2.94 & 2.89 & 2.85 & 2.79 & 2.75 & 2.70 & 2.67 & 2.63 & 2.60 & 2.56 \\
& $\Delta {\rm EDR}$ (isothermal) [\%]& 1.11 & 1.95 & 2.59 & 3.38 & 4.17 & 5.20 & 5.98 & 6.19 & 6.96 & 7.56 \\
& $\Delta {\rm EDR}$ ($T_{eff}(R))\quad$ [\%]& 12.8 & 25.0 & 37.0 & 51.9 & 66.9 & 85.9 & 100.6 & 115.8 & 133.3 & 149.6\\
\hline
APR & $a_*$ 				&  -  &  -  & 0.10 &0.14&0.17& 0.20 & 0.24 &  0.27& 0.31 & 0.35 \\
& $R_{in}/r_g$ 			&   - &  -  & 2.84 & 2.79 & 2.75 & 2.71 & 2.66 & 2.63 & 2.59 & 2.56 \\
& $\Delta {\rm EDR}$ (isothermal) [\%]&-&-&1.51&3.35&4.16&5.00&5.82&6.44&6.70&7.59\\
& $\Delta {\rm EDR}$ ($T_{eff}(R))\quad$ [\%]& - & - & 36.5 & 53.1 & 66.6 & 83.3 & 102.0 & 115.7 & 134.6 & 149.1\\
\hline
BBBAV14& $a_*$ 				& 0.03&0.05 & 0.08 &0.11 & 0.13 &0.16 & 0.19 & 0.23  & 0.26  & 0.29  \\
& $R_{in}/r_g$ 			& 2.95 & 2.91 & 2.88 & 2.83 & 2.79 & 2.75 & 2.71 & 2.66 & 2.62 & 2.58  \\
& $\Delta {\rm EDR}$ (isothermal) [\%]& 0.19 & 1.12 & 1.81 & 2.73 & 3.54 & 4.37 & 4.91 & 5.62 & 6.30 & 7.14 \\
& $\Delta {\rm EDR}$ ($T_{eff}(R))\quad$ [\%]& 9.8 & 19.4 & 28.5 & 41.2 & 53.5 & 68.8 & 83.7 & 102.2 & 120.2 & 143.2\\
\hline
BBBParis&$a_*$ 				&  -  & 0.06 &0.09  &0.12 &0.15 & 0.18  & 0.21 & 0.24 & 0.27 &  0.31 \\
&$R_{in}/r_g$ 			&  -  &  2.90 & 2.86 & 2.81 & 2.77 & 2.73 & 2.69 & 2.64 & 2.61 & 2.57 \\
& $\Delta {\rm EDR}$ (isothermal) [\%] & - & 1.71 & 2.49 & 3.38 & 4.11& 4.84 & 5.57& 6.39 & 6.47 & 6.94 \\
& $\Delta {\rm EDR}$ ($T_{eff}(R))\quad$ [\%]& - & 21.0 & 32.8 & 47.3 & 60.2 & 74.4 & 90.0 & 111.5 & 124.5 & 144.7\\
\hline
RMF stiff &$a_*$ 				&0.05 &0.09 & 0.14&0.20 & 0.25 & 0.30 & 0.36 & 0.43&  0.51  & 0.63  \\
&$R_{in}/r_g$ 			& 2.92 & 2.87 & 2.81 & 2.77 & 2.75 & 2.63 & 2.67 & 2.75 & 2.83 & 2.96\\
& $\Delta {\rm EDR}$ (isothermal) [\%] & 1.65 & 2.62 & 3.65 & 4.37 & 4.88 & 6.03 & 5.72 & 4.50 & 3.29 & 2.09 \\
& $\Delta {\rm EDR}$ ($T_{eff}(R))\quad$ [\%] &16.8 & 32.0 & 48.0 & 62.4 & 70.0 & 114.6 & 96.8 & 68.7 & 40.8 & 6.5\\
\hline
STOS0 &$a_*$ 				& 0.06&0.12 &  0.17   &0.24 &0.30  & 0.38 & 0.46  &  0.56 & 0.69 & 0.86  \\
&$R_{in}/r_g$ 			& 2.92 & 2.87 & 2.88 & 2.90 & 2.96 & 3.00 & 3.10 & 3.21 & 3.43 & 3.84\\
& $\Delta {\rm EDR}$ (isothermal) [\%]& 1.88 & 2.69 & 3.05 & 2.96 & 1.96 & 1.65 & -0.41  & -1.49 & -6.08 & -12.36\\
& $\Delta {\rm EDR}$ ($T_{eff}(R))\quad$ [\%]& 21.7 & 36.2 & 34.7 & 27.3 & 14.8 & 4.5 & -15.0 & -31.2 & - 56.6 & -80.0 \\
\hline
Q&$a_*$ 				&  -  & -   & 0.11&0.15 &  0.19& 0.23 & 0.28& 0.32& 0.37 & 0.41 \\
&$R_{in}/r_g$ 			&  -  & -   & 2.83 & 2.78 & 2.74 & 2.71 & 2.67 & 2.67 & 2.67 & 2.67 \\
& $\Delta {\rm EDR}$ (isothermal) [\%]&-&-&2.99&4.01&4.54&5.11&5.48&5.82&5.92&6.37\\
& $\Delta {\rm EDR}$ ($T_{eff}(R))\quad$ [\%] & - & - & 40.2 & 55.8 & 70.5 & 84.1 & 96.3 & 101.6 & 102.5 & 100.2\\
\hline
CFL150&$a_*$ 				&  -  & 0.08 & 0.13&0.17 & 0.21 & 0.26 & 0.31& 0.36 & 0.41  & 0.47 \\
&$R_{in}/r_g$ 			&  -  & 2.88 & 2.82 & 2.77 & 2.73 & 2.71 & 2.69 & 2.70 & 2.73 & 2.79 \\
& $\Delta {\rm EDR}$ (isothermal) [\%]&-&2.15&3.07&4.10&4.59&5.06&5.30&5.47&5.20&5.06\\
& $\Delta {\rm EDR}$ ($T_{eff}(R))\quad$ [\%] & - & 28.7 & 44.1 & 60.3 & 72.5 & 82.3 & 88.6 & 88.3 & 78.0 & 59.0\\
\hline
\end{tabular}
\caption{The spin parameter $a_*$ of the star, the inner edge $R_{in}$ of the accretion disk, and the variation in the EDR integrated over the rotational axis for neutron and quark star models, with a total mass of $1.8M_{\odot}$ and angular velocities from $5\times10^2 \;{\rm s}^{-1}$ to $5\times10^3 \;{\rm s}^{-1}$. Here $\Delta$EDR = (EDR - ${\rm EDR}_s$)/${\rm EDR}_s$ and ${\rm EDR}_s$ is derived for the static case ($\Omega=0$). The variation in EDR is calculated for both isothermal disks ($T_{eff}$=const) and disks with the temperature profile $T_{eff}(R)=T_{eff}\times3r_g/R$.}
\label{table0}
\end{table*}

\begin{figure}
\includegraphics[width=8.15cm]{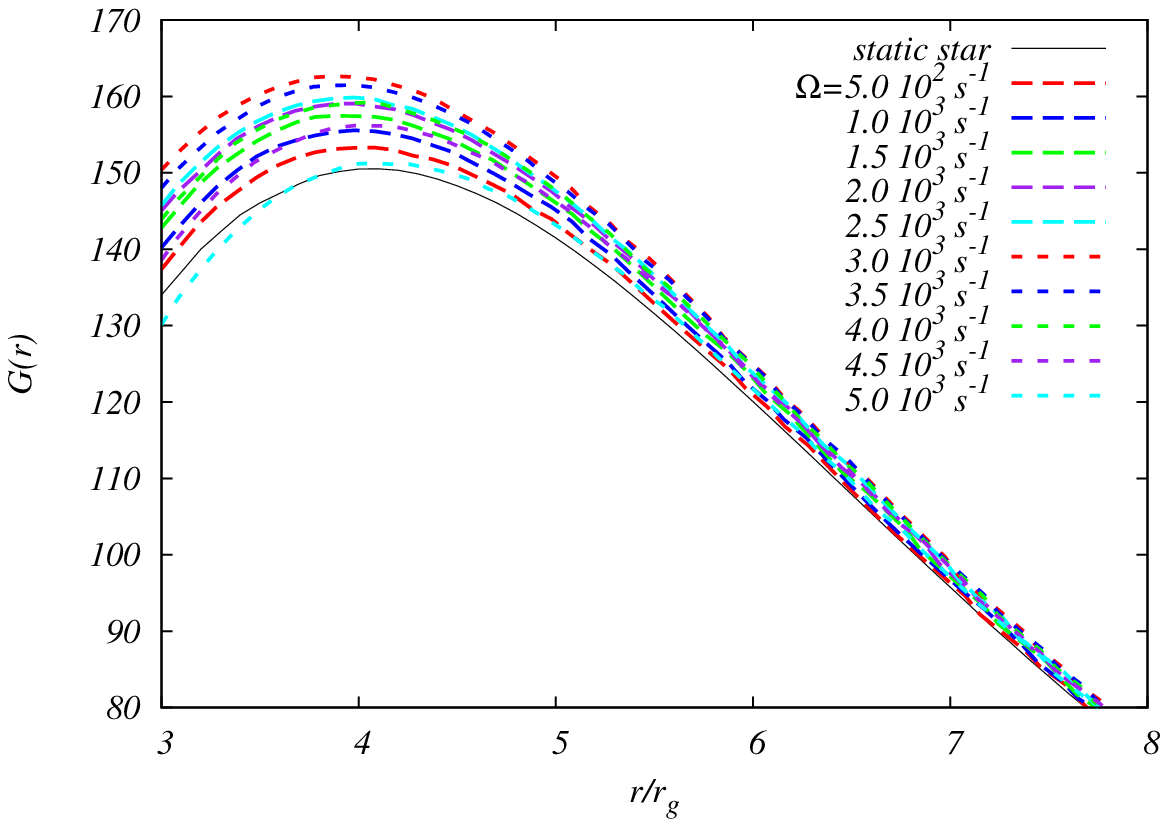}
\includegraphics[width=8.15cm]{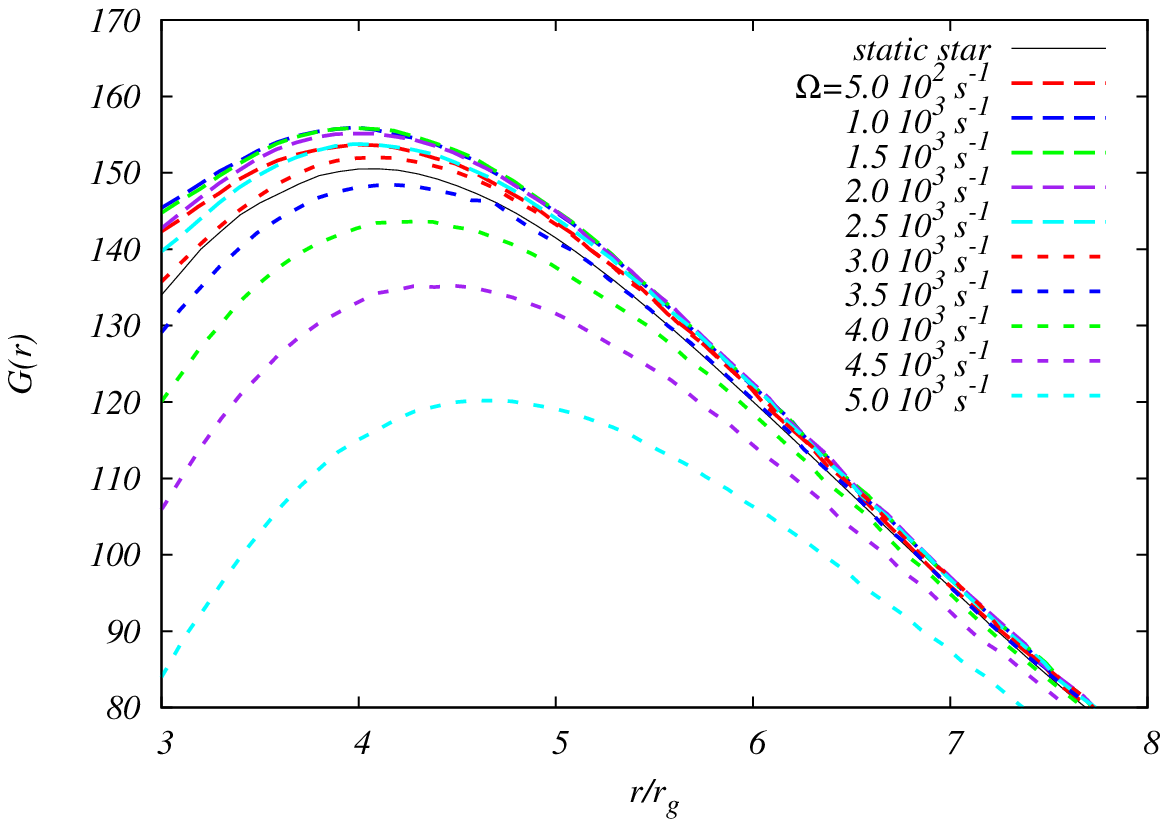}
\includegraphics[width=8.15cm]{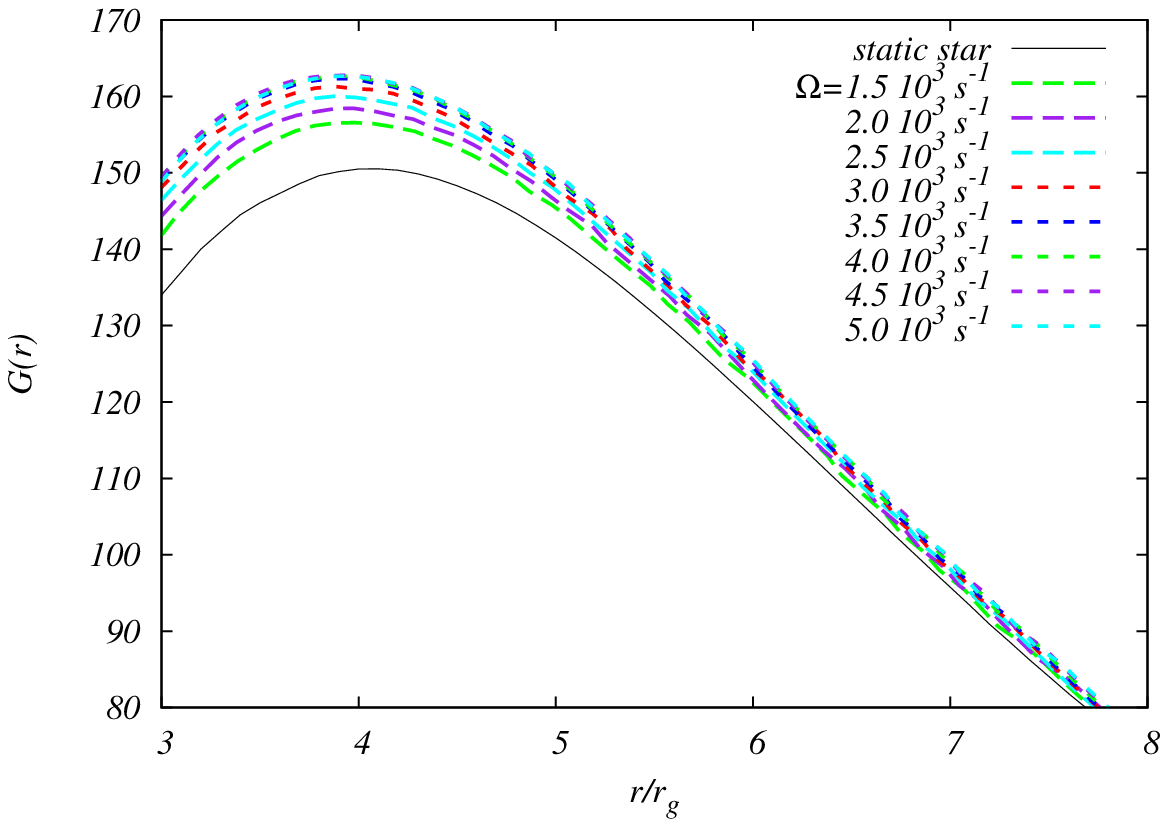}
\includegraphics[width=8.15cm]{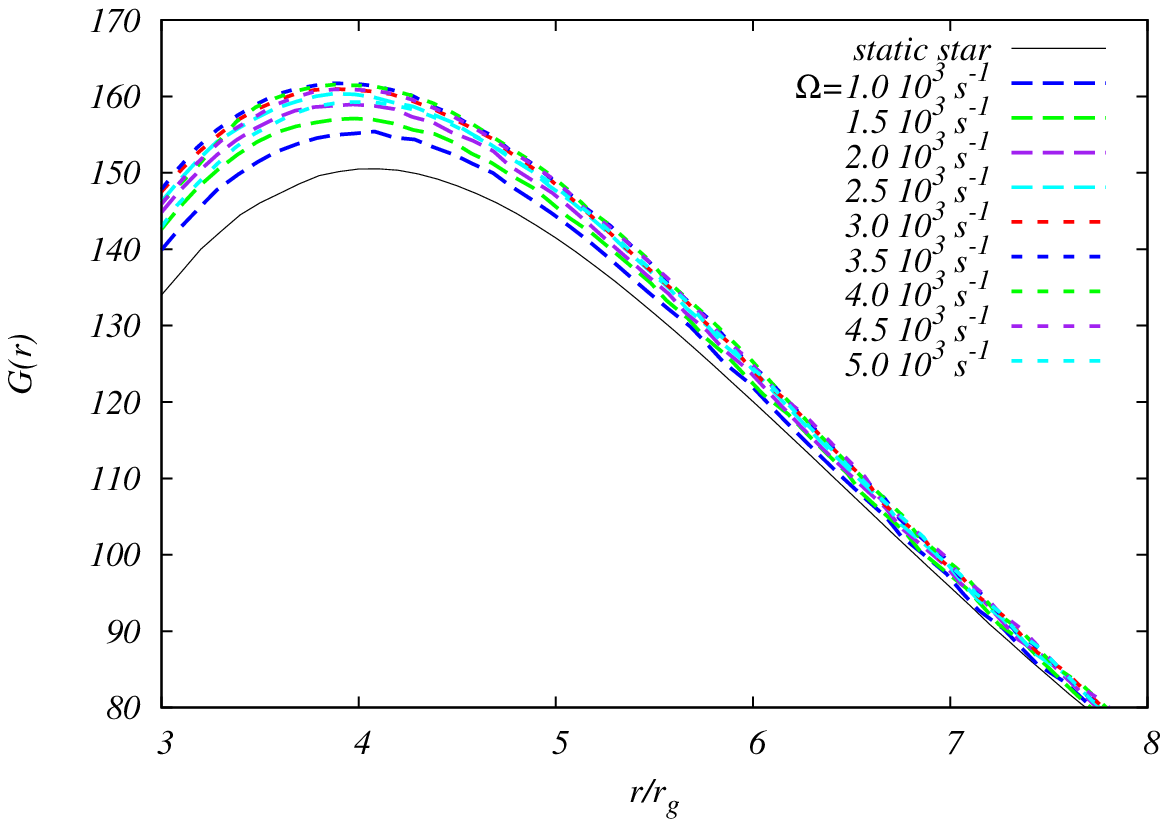}
\centering
\caption{The axial profile of EDR
characterized by the dimensionless quantity $G(r)$ as a function of $\Omega$ for an isothermal disk and neutron star models with RMF stiff (top left hand side) and STOS0 (top right hand side) type EOSs,  and for quark stars with Q (bottom left hand side) and CFL 150 (bottom right hand side) type EOSs.}
\label{fig3}
\end{figure}

We also present the axial distribution of the EDR for these stellar models in Fig.~\ref{fig2},  where we have plotted the quantity $G(r)$, defined in Eq.~(\ref{G}), as a function of the curvature coordinate $r$. The graphs show that the proportionality between the maximum of the deposition rate and the spin parameter of the star is also preserved: the faster the neutron star rotates, the higher the maximal value of the EDR along the axis. Nevertheless, these maxima shift closer to the stellar surface for higher angular velocities, which involves the baryon contamination problem.
Besides the uncertainties arisen from baryon contamination, one must be careful with the interpretation of these results, since the application of a formalism developed for black holes may produce considerable errors at higher values of $\Omega$. As the error due to the separation of the geodesic equation is increasing with the higher rotational speed of the central object, the calculations produce less reliable results for high $\Omega $.
 Hence it is not clear that the moderate increase in the EDR at higher angular frequencies is the result of the effect of the redshift of the neutrino/anti neutrino energy alone, or it is caused by the underestimation of the real values in the approximation. The latter must at least not dominate over the real physical effects, as far as we consider configurations with lower values of the spin parameter (see Appendix A).

From the off-axis calculations of the spatial distribution of the EDR in the vicinity of Kerr black holes we already know that the increase in the rotational speed of the black hole enhances the energy deposition along both the equatorial plane, and the axis of rotation \cite{Mi03}. This may indicate that the behavior presented in Fig. ~\ref{fig2} is in accordance with the previous results on the EDR distribution along the equatorial plane for the same family of stellar EOSs \cite{Ko09}. It was shown that for the APR type EOS the maximum of the energy deposition from $\nu\bar{\nu}$ pair annihilation is increasing along the surface of the thin accretion disk as the star spins up.  The other types of EOS shown in Fig.~\ref{fig2} exhibit similar behaviors. Therefore we can conclude that they follow the pattern found in the case of Kerr black holes.

\begin{figure}
\includegraphics[width=8.15cm]{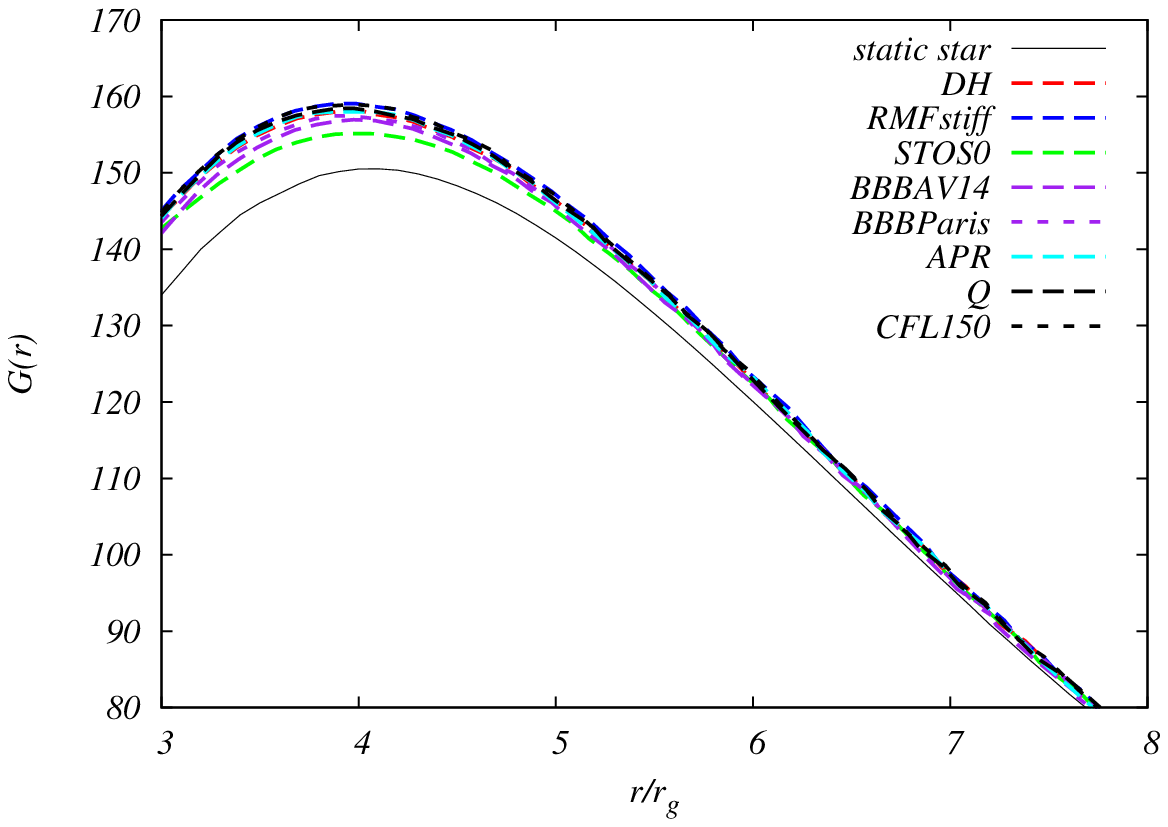}
\includegraphics[width=8.15cm]{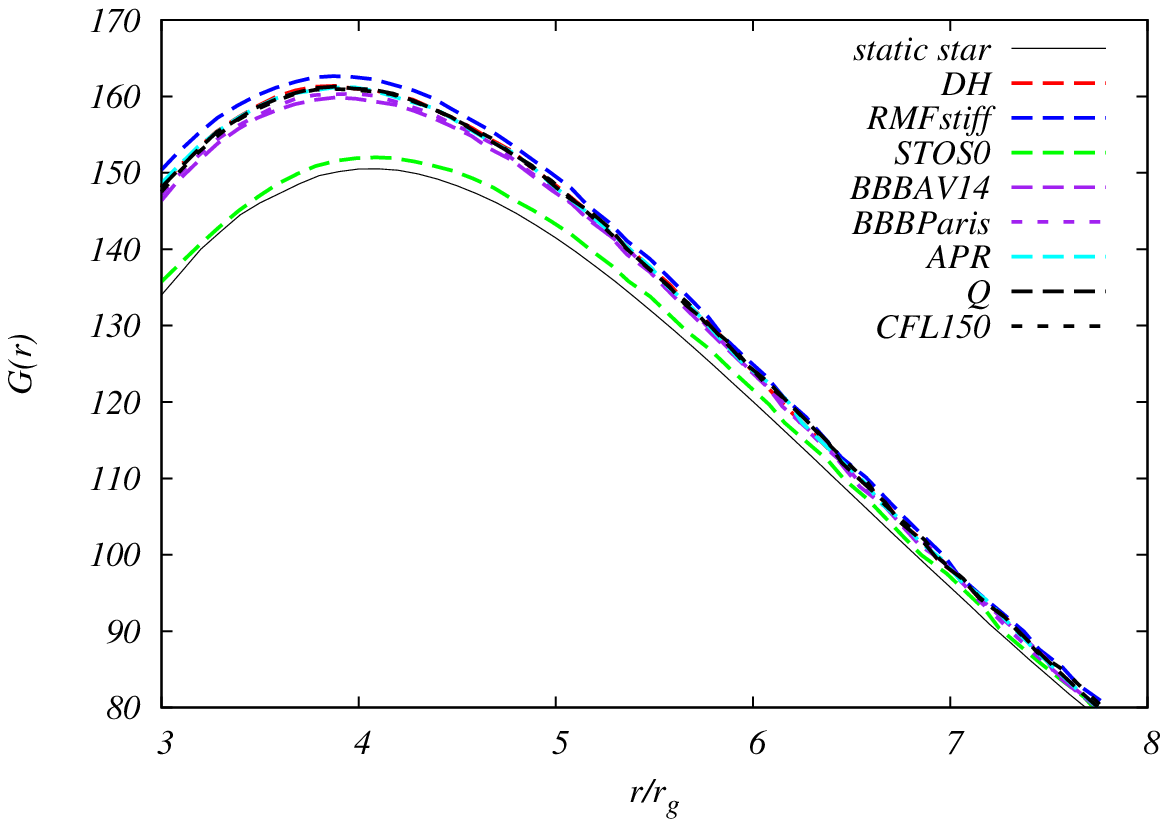}
\includegraphics[width=8.15cm]{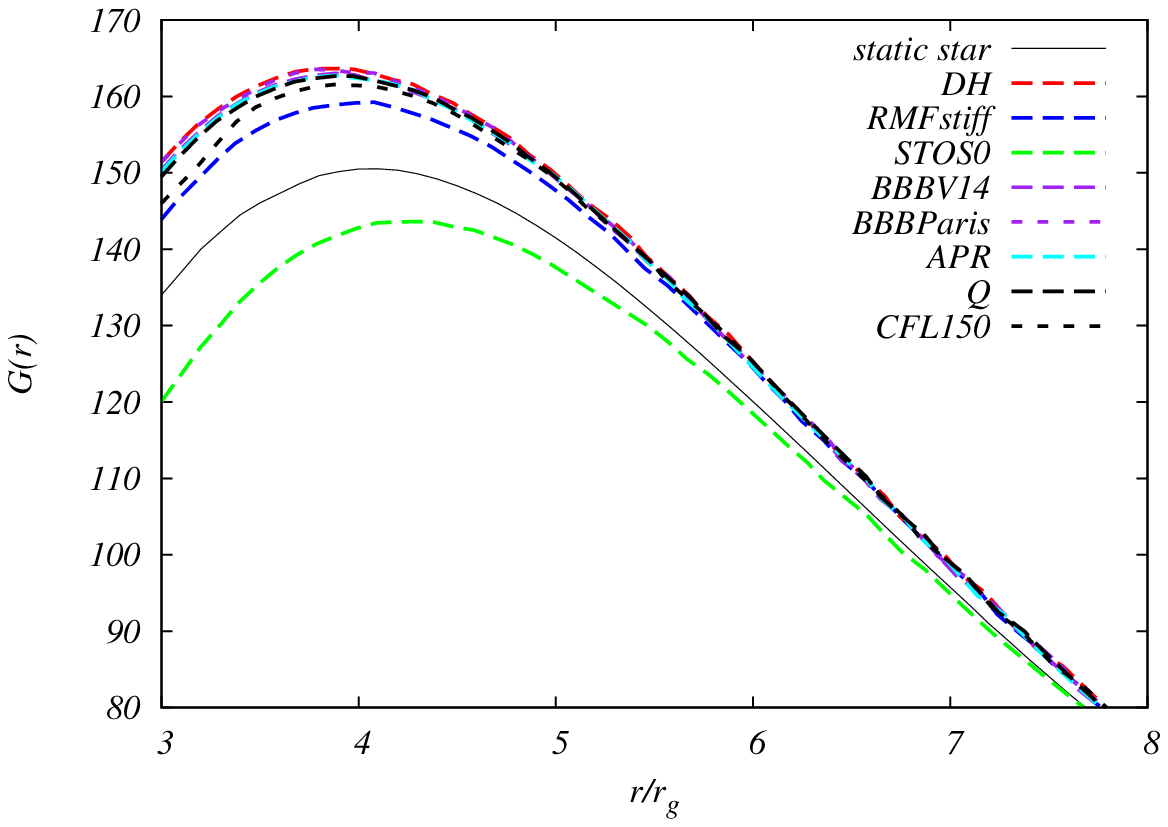}
\includegraphics[width=8.15cm]{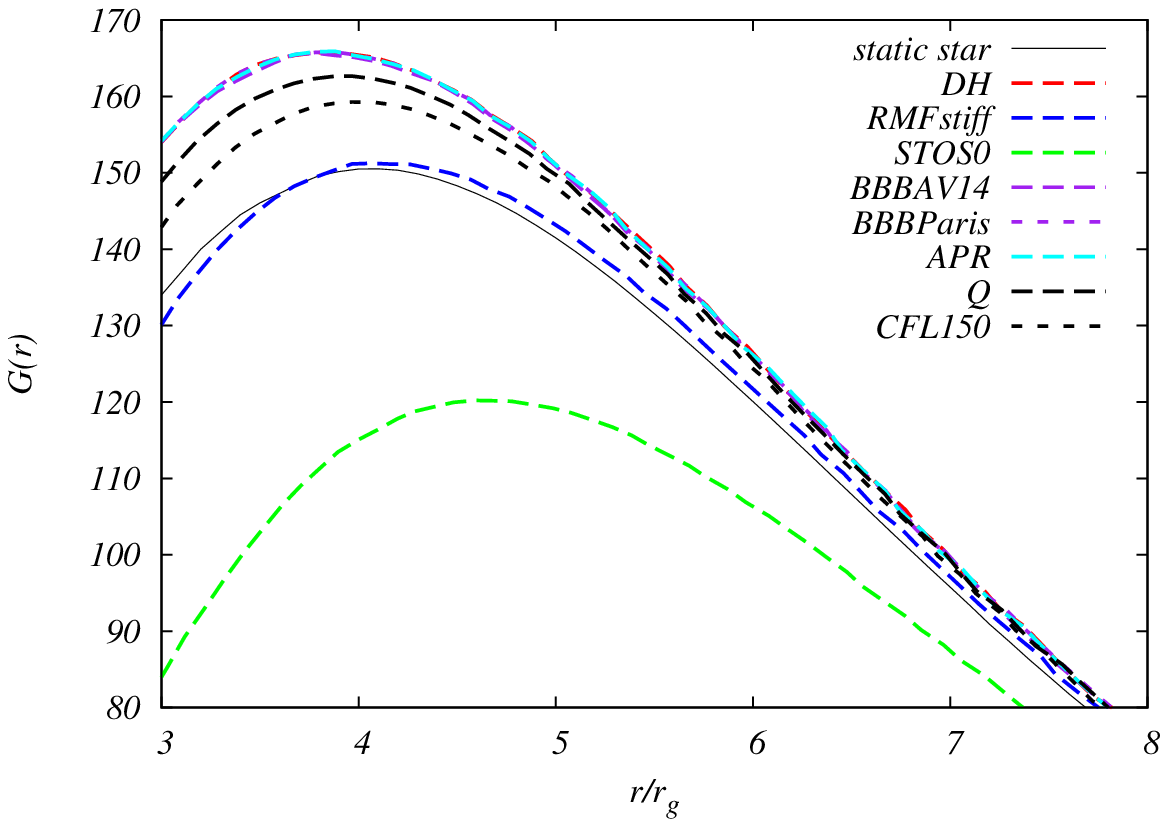}
\centering
\caption{The axial profile of the EDR,
characterized by the dimensionless quantities $G(r)$ for an isothermal disk, and different neutron and quark star models with the total mass $M=1.8 M_{\odot}$ and rotational velocities $\Omega=2\times 10^3\;{\rm s}^{-1}$ (top left hand side panel), $3\times 10^3\;{\rm s}^{-1}$ (top right hand side panel), $4\times 10^3\;{\rm s}^{-1}$ (bottom left hand side panel) and $5\times 10^3\;{\rm s}^{-1}$ (bottom right hand side panel).}
\label{fig4}
\end{figure}

The neutron star models with RMF stiff and STOS0 type EOS, together with the Q and CFL 150 type quark stars do not produce a simple relation between the spin and the EDR, as shown in the bottom four lines of Table~{\ref{table0}}. In Fig.~\ref{fig3} we also present the dependence of the EDR on $\Omega$ for these models.
In the case of the neutron stars with RMF stiff and STOS0 type EOSs there is some enhancement in the EDR for rotational velocities increasing up to $\sim 3\times10^3 \;{\rm s}^{-1}$ (RMF stiff) and $\sim 1.5\times10^3 \;{\rm s}^{-1}$  (STOS0), but at higher values of $\Omega$ the maxima of the ERD starts to decrease. As the star rotates faster its surface reaches higher equatorial orbits, and reduces the surface area of the accretion disk by pushing the inner edge of the disk to higher radii.
At rotational frequencies approaching $ 5\times10^3 \;{\rm s}^{-1}$ the amount of the energy deposition for the stellar model with RMF stiff type EOS reduces to the one derived in the static case.  Since the stellar radius is already much greater than $r_{ms}=3r_{g}$ of the static case, we obtain even smaller values of the EDR for the STOS0 model then those derived for a static star (see the negative values in Table~\ref{table0}).

The quark star models with Q and CFL150 type EOS do not deviate so drastically from the spin-EDR relation characterizing the first family of EOSs and the black holes. In Fig.~\ref{fig3} we can see that the maximum of the EDR has an increase, turns over, and then decreases  with increasing spin parameter. However, its reduction after the turn over is minimal,  and it might be a result of the combination of several physical effects, such as the variation of the disk surface area, the redshift and the Lorenz boost, together with the artifacts caused by the break-down of our approximation for higher values of the spin parameter. The total EDR integrated along the rotational axis remains increasing for the Q type EOS, but the differences are small, and may be dominated by the uncertainties for fast stellar rotation. These results are still in good agreement with the fact that the amount of energy liberated in the $\nu\bar{\nu}$ pair annihilation along the equatorial plane is proportional to the angular velocity of the quark star with Q type EOS, as far as we compare the configurations for $\Omega=3\times10^{3}\; {\rm s}^{-1}$ and $5\times10^{3} \;{\rm s}^{-1}$ with each other \cite{Ko09}.

In Fig.~\ref{fig4} we also compare the EDR distributions calculated along the rotational axis for different EOS types at constant angular frequencies ranging from $\Omega=2\times10^{3}\;{\rm s}^{-1}$ to $5\times10^{3}\;{\rm s}^{-1}$, respectively. The plots show that the quark stars and the neutron stars with RMF stiff type EOS produce the highest values of the EDR at low rotational frequencies, although the differences are not considerable. This relation is similar to  the one derived in the equatorial plane, where the EDR is higher for quark stars than for neutron stars, apart from those with RMF stiff and STOS0 EOS type \cite{Ko09}.
The only difference is the behavior of the STOS0 model, which has somewhat lower values in the EDR  along the rotational axis than other neutron star models do. However, the EDR values in the equatorial plane were derived at $\Omega=5\times10^{3}\;{\rm s}^{-1}$ in \cite{Ko09}, and we should compare them with the plots displaying configurations with high rotational velocities.

The comparison shows that quark stars and  neutron stars with RMF stiff and STOS0 type EOS generate less energy along the rotational axis from the neutrino-anti neutrino annihilation than the neutron stars with other EOS types do at  $\Omega=5\times10^{3}{\rm s}^{-1}$, as opposed to the relations between the values of EDR calculated in the equatorial plane. In fact, there is no discrepancy between the two results, as the EDR at a given point $R$ in the equatorial plane is calculated by integrating the angular part of Eq.~(\ref{eq:dotq}) between $R$ and $R_{out}$ \cite{SaWi99}, whereas the calculation of the EDR at a point $r$ on the rotational axis always involves the integral of Eq.~(\ref{eq:dotq}) over the entire accretion disk \cite{AsFu01}. As a result, the first method provides $\dot{q}$ for any value $R>R_{in}$, which does not depend on $R_{in}$ (the latter one is only a lower boundary of the domain of $\dot{q}$), whereas the EDR derived by the second method always depends on $R_{in}$, i.e, the area of the disk surface.

Nevertheless, the results presented here are already uncertain in the regime of high angular frequencies, and the EDR along the rotational axis should already be calculated for rapidly rotating stars by applying the fully 3D techniques of off-axis calculations.



\section{Thin disk with radiation in thermodynamical equilibrium}\label{6}

In this Section we use for the calculation of the EDR along the rotational axis for various rotating
compact objects  a more realistic model of the thermal distribution
of the neutrino flux emerging from the thin accretion disk. This model is based on the thermodynamical equilibrium
approximation of the state of the accretion disk, where the temperature gradient
is given by a profile inversely proportional to the radius. Following Asano \& Fukuyama (2001), we replace the constant temperature profile $T_{eff}=T_{eff}(3r_g)$  with $T_{eff}(R)=T_{eff}(3r_g)\times3r_g/R$, and we recalculate the EDR along the rotational axis for all the constant disk temperature configurations considered before.

\begin{figure}
\includegraphics[width=8.15cm]{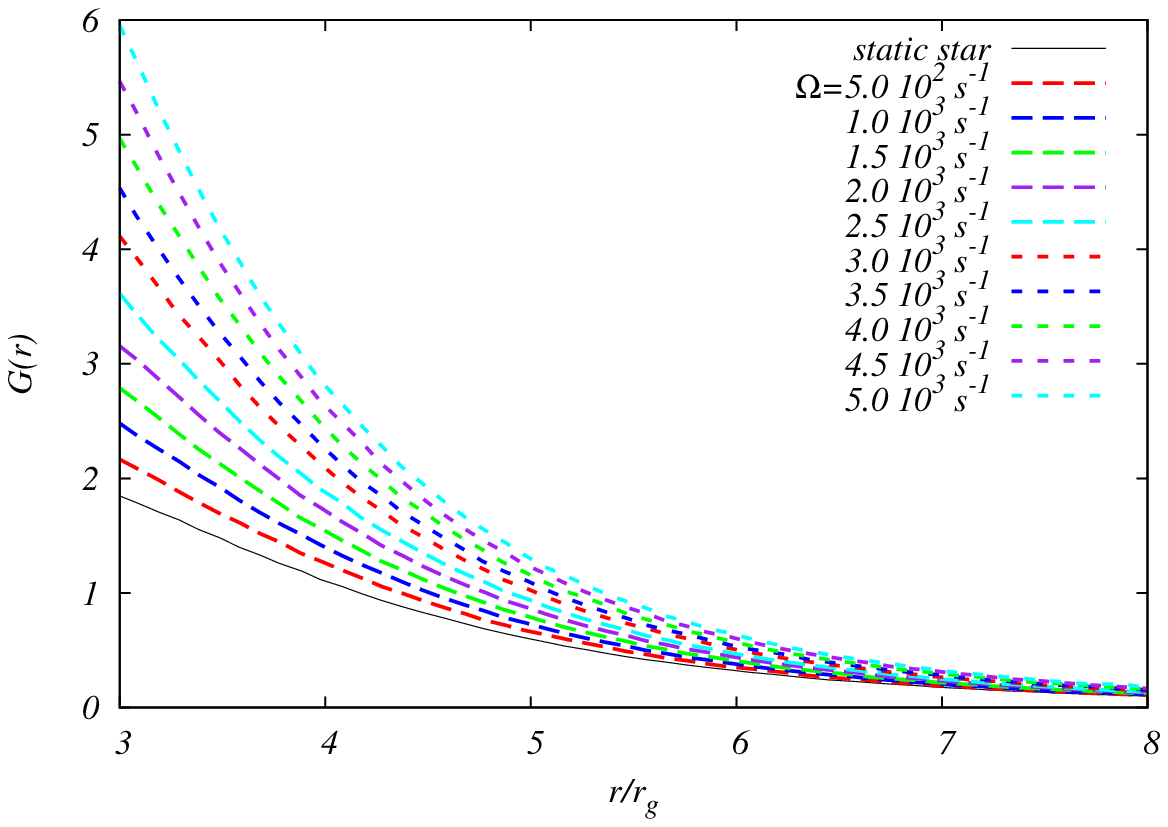}
\includegraphics[width=8.15cm]{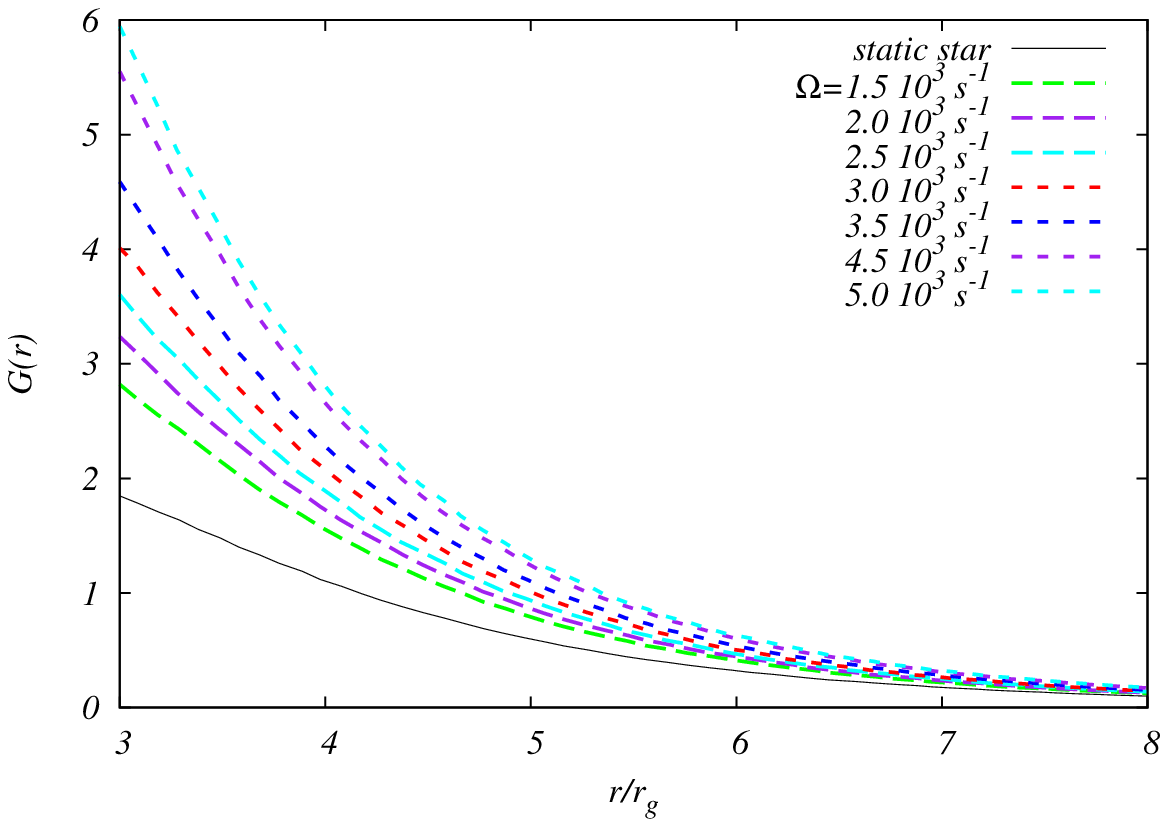}
\includegraphics[width=8.15cm]{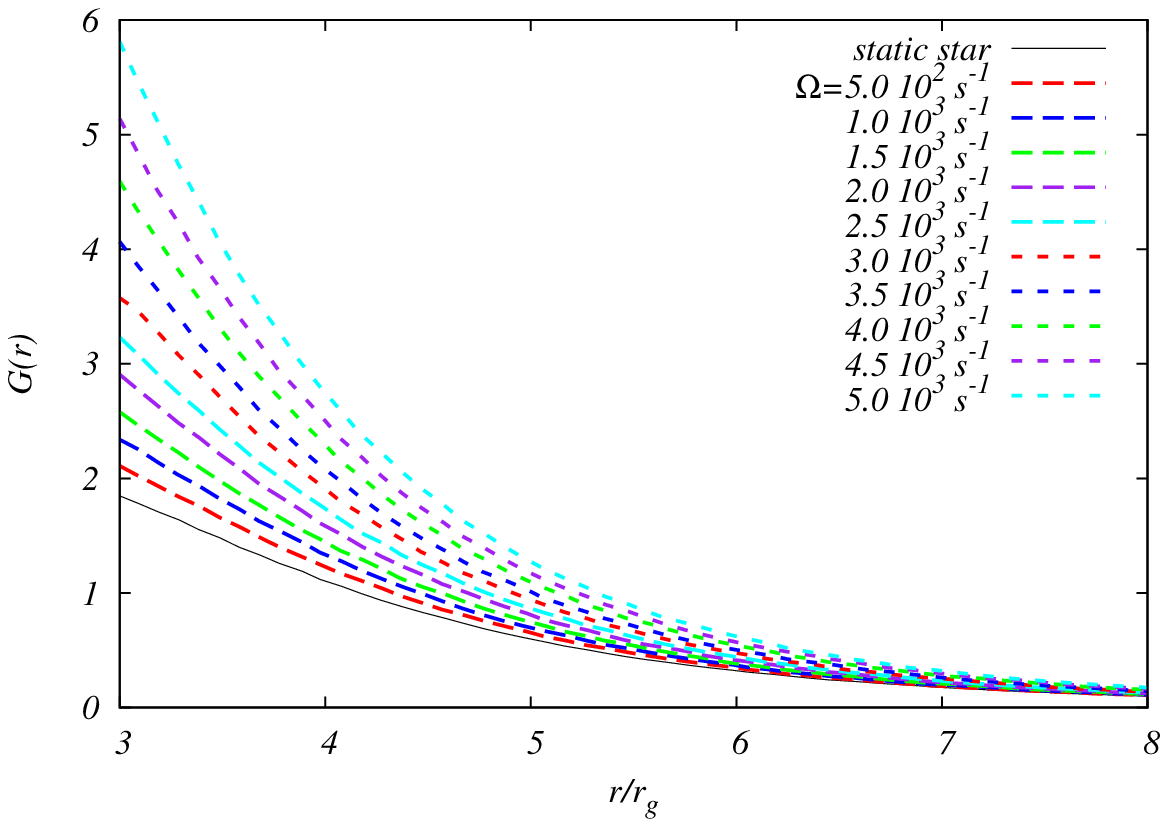}
\includegraphics[width=8.15cm]{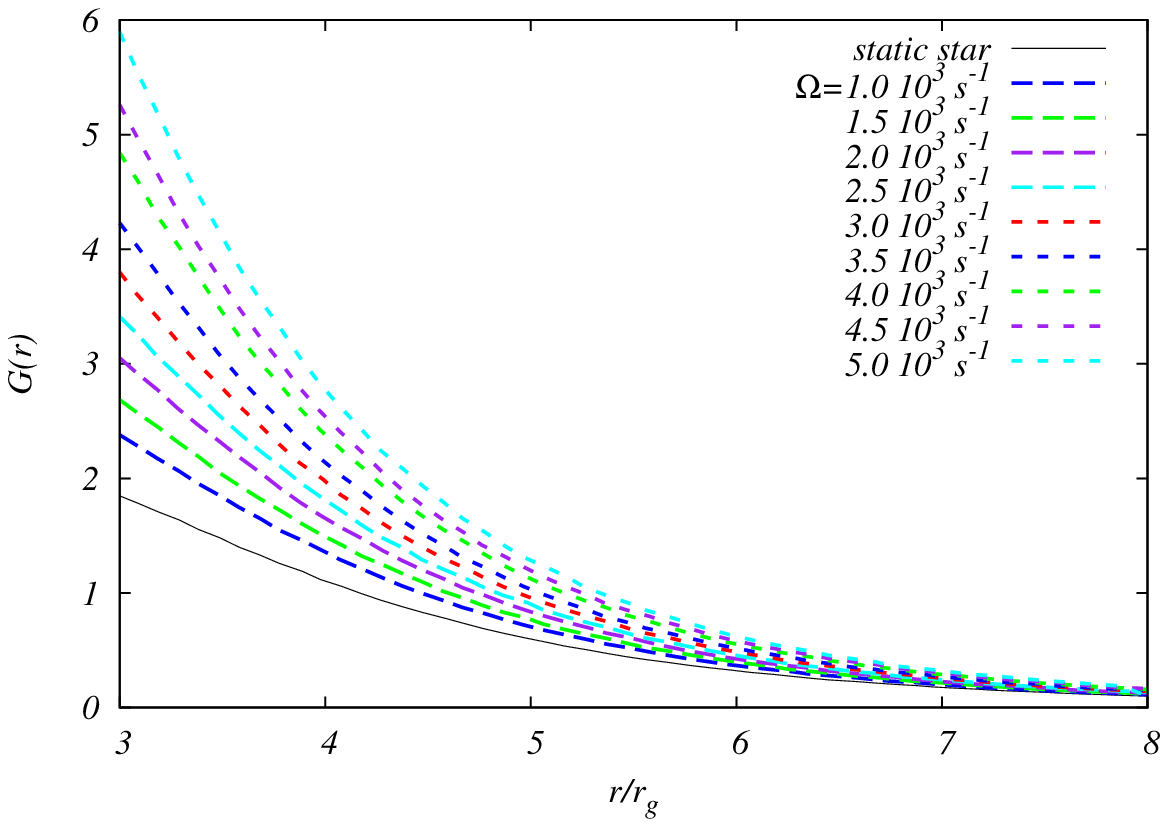}
\centering
\caption{The axial profile of the EDR
characterized by the dimensionless quantity $G(r)$ for $T_{eff}\propto R^{-1}$ and different angular velocities of the neutron star models with DH (top left hand side), APR (top right hand side), BBBAV14 (bottom left hand side) and BBBParis (bottom right hand side) type EOSs. (The total mass of the central object is $1.8 M_{\odot}$.) For comparison we have also plotted $G(r)$ for a static spacetime geometry (black solid curve).}
\label{fig5}
\end{figure}
\begin{figure}
\includegraphics[width=8.15cm]{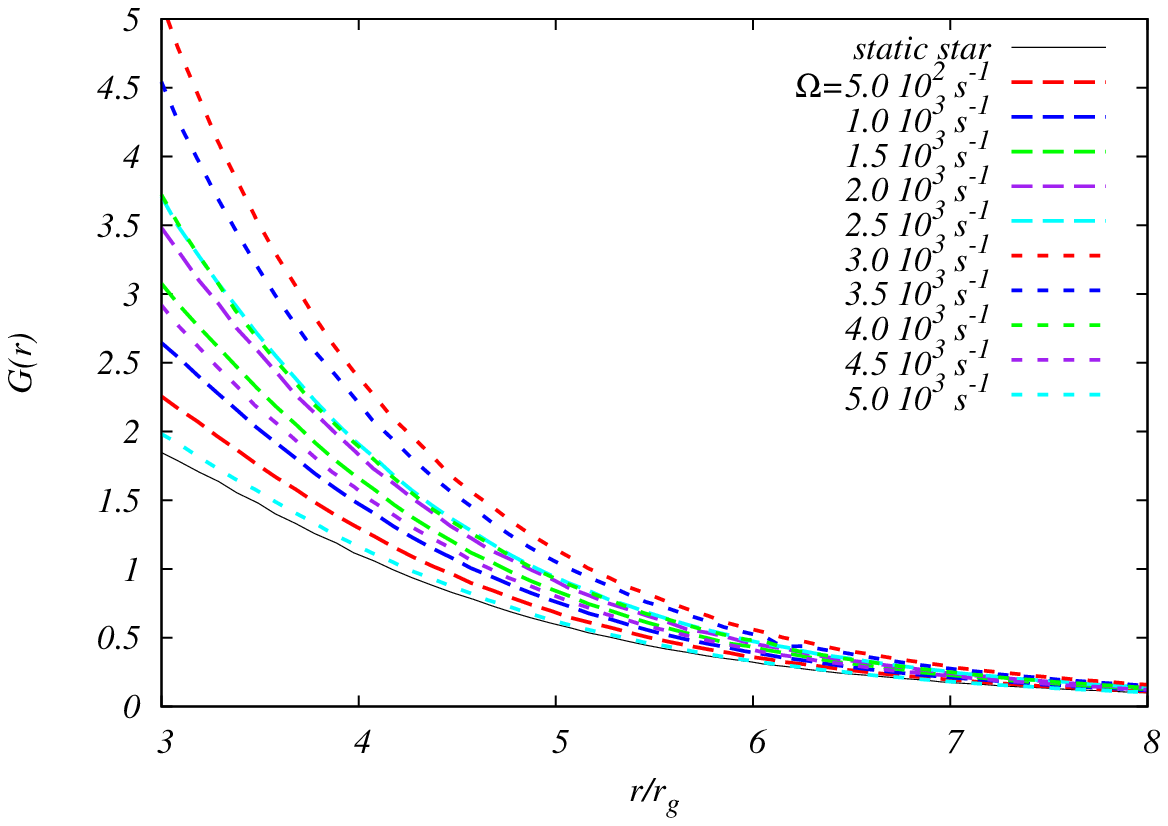}
\includegraphics[width=8.15cm]{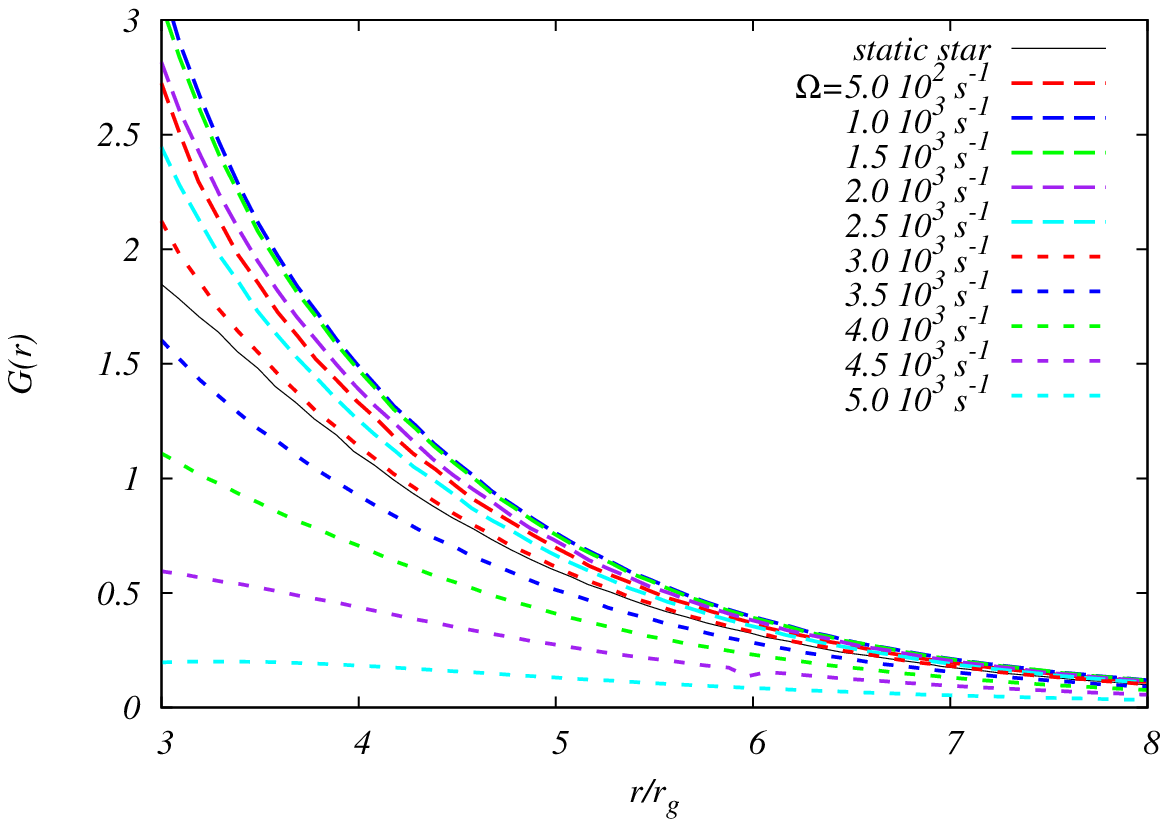}
\includegraphics[width=8.15cm]{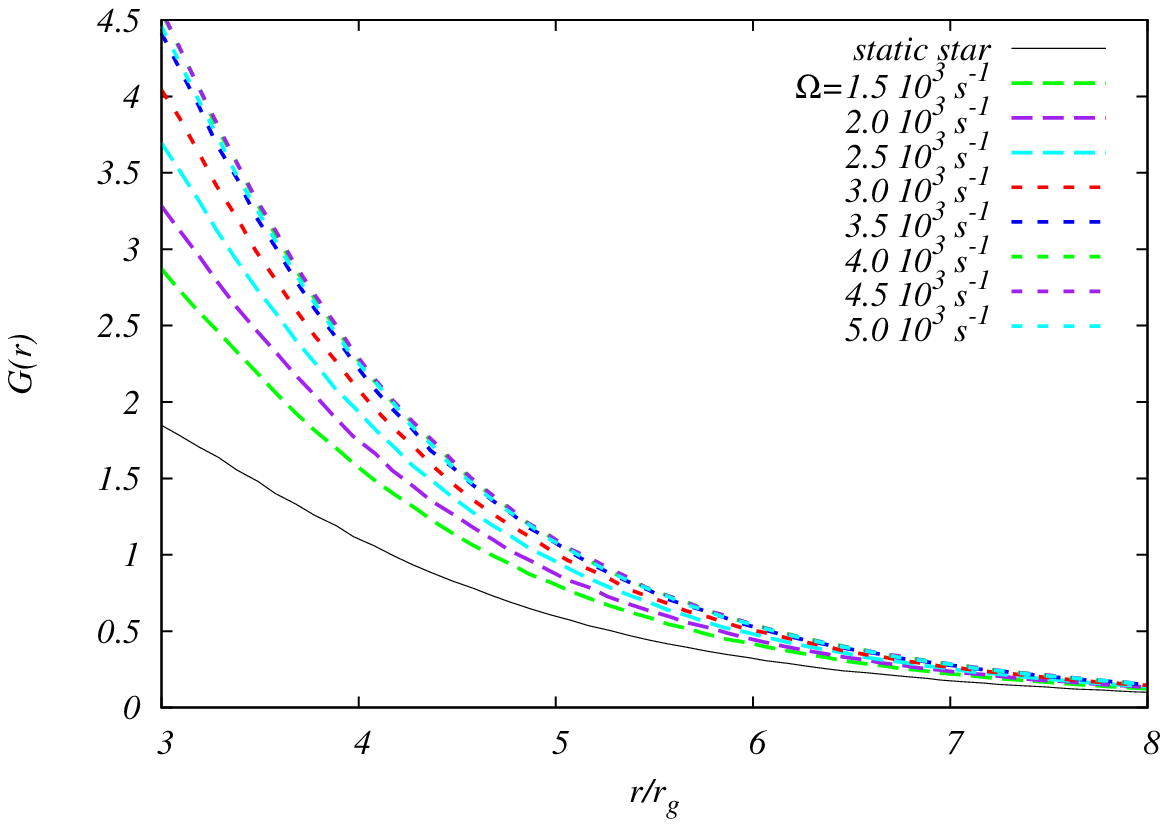}
\includegraphics[width=8.15cm]{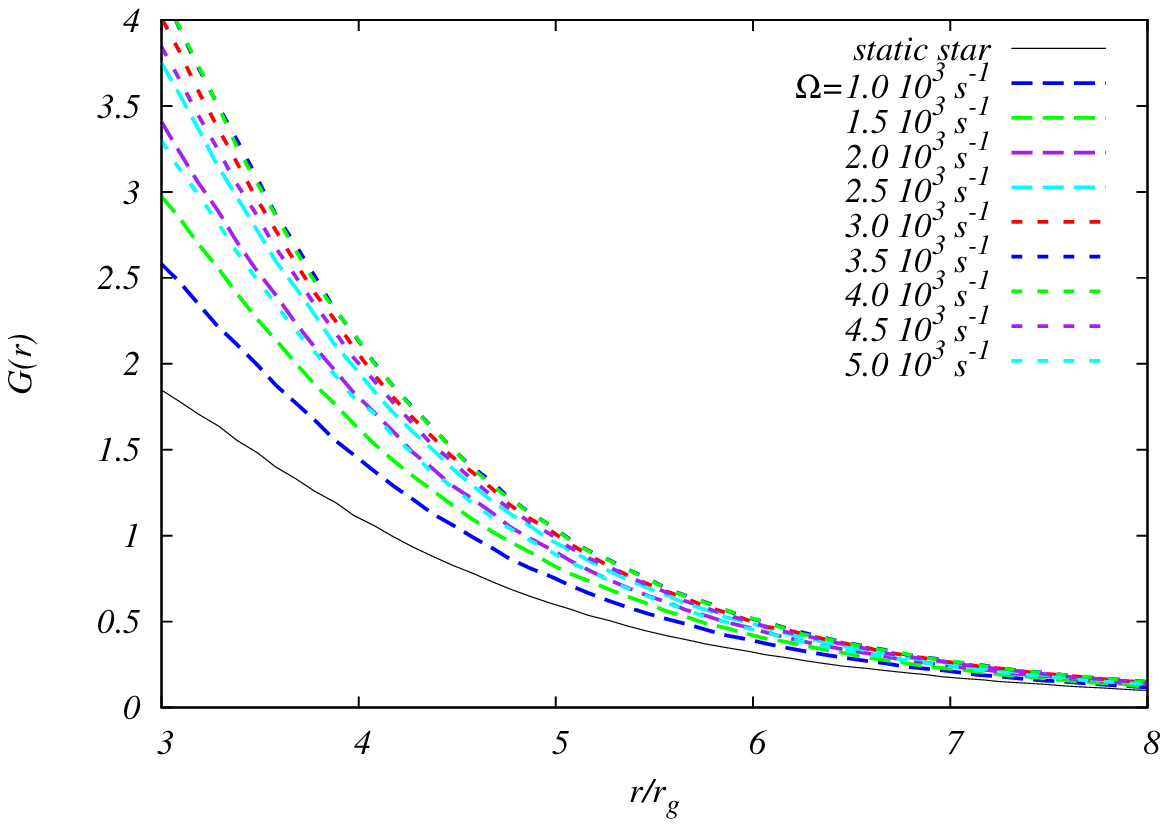}
\centering
\caption{The axial profile of the EDR
characterized by the dimensionless quantity $G(r)$ for $T_{eff}\propto R^{-1}$, and different angular velocities of the neutron star models with RMF stiff (top left hand side) and STOS0 (top right hand side) type EOSs, and for quark stars with Q (bottom left hand side) and CFL 150 (bottom right hand side) type EOSs.}
\label{fig6}
\end{figure}

In Figs.~\ref{fig5} and \ref{fig6} we present the axial distribution of the EDR derived for the accretion disk with temperature gradient. If we compare the results with those obtained in the isothermal case for the same $T_{eff}(3r_g)$, we see that the EDR is reduced by a factor of 100 at low rotational frequencies of the central body.
Even for the highest values of $\Omega$, the EDR in the isothermal case is about 25 times higher than the one calculated with temperature gradient. Only the dynamical range of the EDR became wider for the first group of EOS in Fig.~\ref{fig5}: in the isothermal case its values at $r\sim3r_g$ have a $\sim10$\% increase for $\Omega=5\times10^{3}\; {\rm s}^{-1}$, as compared with those for $\Omega=5\times10^{2} \;{\rm s}^{-1}$. The difference increases to a factor of $3$ for the disk with the temperature gradient. Table~\ref{table0} also contains the differences of the total EDR integrated over $r=3r_g-10r_g$ for these configurations for a disk with the $R$-dependent temperature profile. The differences in the total EDR compared with the static case are from $\sim10$\% to $\sim150$\%, which span a much wider range than those for the isothermal disk do, between $\sim1$ and $\sim7$\%.
Although in the models with a disk temperature gradient the efficiency of the energy deposition compared with the static case increases in a much higher rate with increasing $\Omega$, the comparison of Figs.~\ref{fig2} and \ref{fig5} shows that even at high rotational velocities the EDR is much smaller for $T_{eff}\propto R^{-1}$ than the one calculated for isothermal disks. The plots also demonstrate that the EDR decreases more rapidly with $r$ for the latter model than it does for isothermal disks, especially at higher angular velocities. This means that the axial distribution of the EDR is more concentrated to the region close to the star, and the energy is produced here mainly by the neutrinos and anti - neutrinos emitted from the innermost region of the disk. For the temperature profile  $T_{eff}\propto R^{-1}$,  this region is hot, but the neutrinos
leaving the rapidly rotating disk surface suffer a large Doppler shift, which, together with the large gravitational redshift, prevents them from depositing much energy at higher radii.

For the second group of the EOS types shown in Fig.~\ref{fig6}, the EDR still preserves its non-linear dependence on the rotational velocity of the central body.
Over $\Omega=3\times10^{3}\; {\rm s}^{-1}$ for the RMF stiff model and over $10^{3} \;{\rm s}^{-1}$ for the STOS0 EOS type their relation is inverted, and the EDR is approaching to, or even takes smaller values, than the one obtained in the static configuration. This trend is also present in the differences of the total EDR integrated along the axis, as can be seen in Table~\ref{table0}, where $\Delta$EDR has a considerable reduction at higher values of $\Omega$, as compared with its maxima.

The quark stars follow the similar trend, but with moderate deviations, as in the isothermal case. For higher rotational speeds in Table~\ref{table0} the proportionality between the spin and the EDR is inverted,  and we obtain a slight decrease in the total EDR as compared with the cases of slower rotation at $3.5-4\times10^3\;{\rm s}^{-1}$.

\begin{figure}
\includegraphics[width=8.15cm]{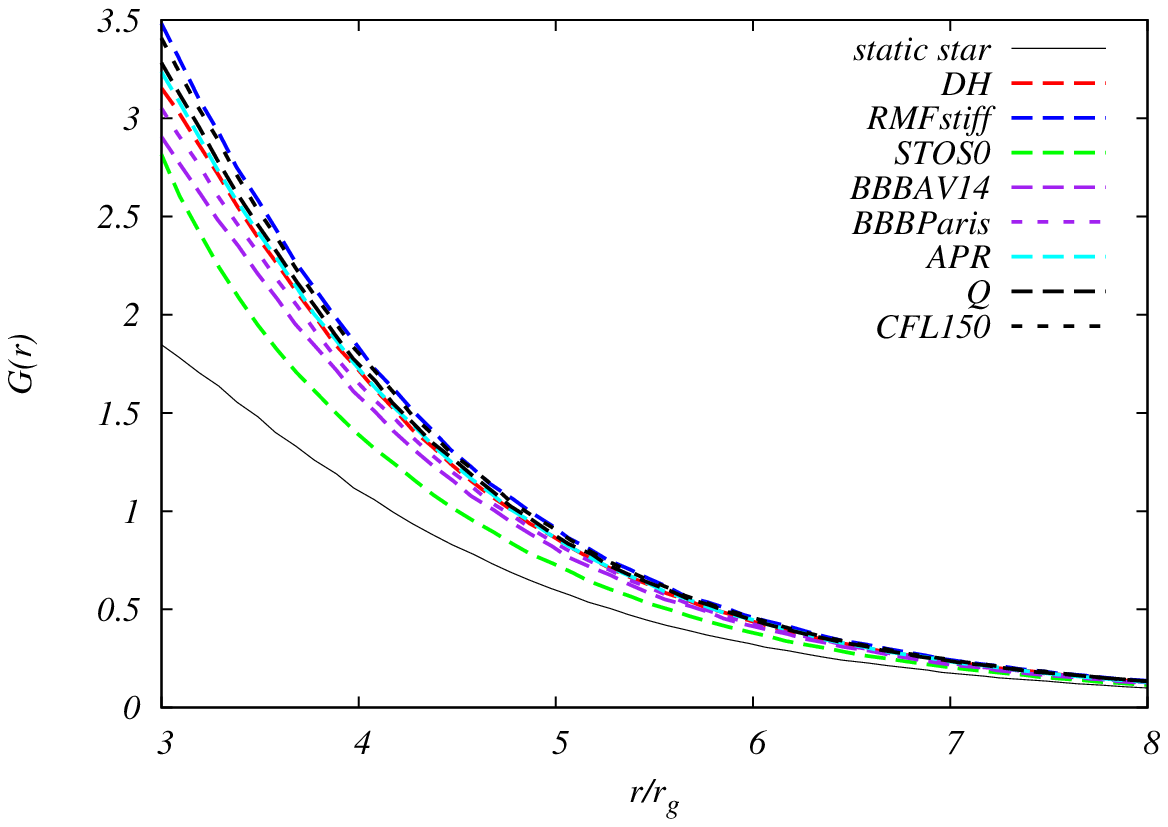}
\includegraphics[width=8.15cm]{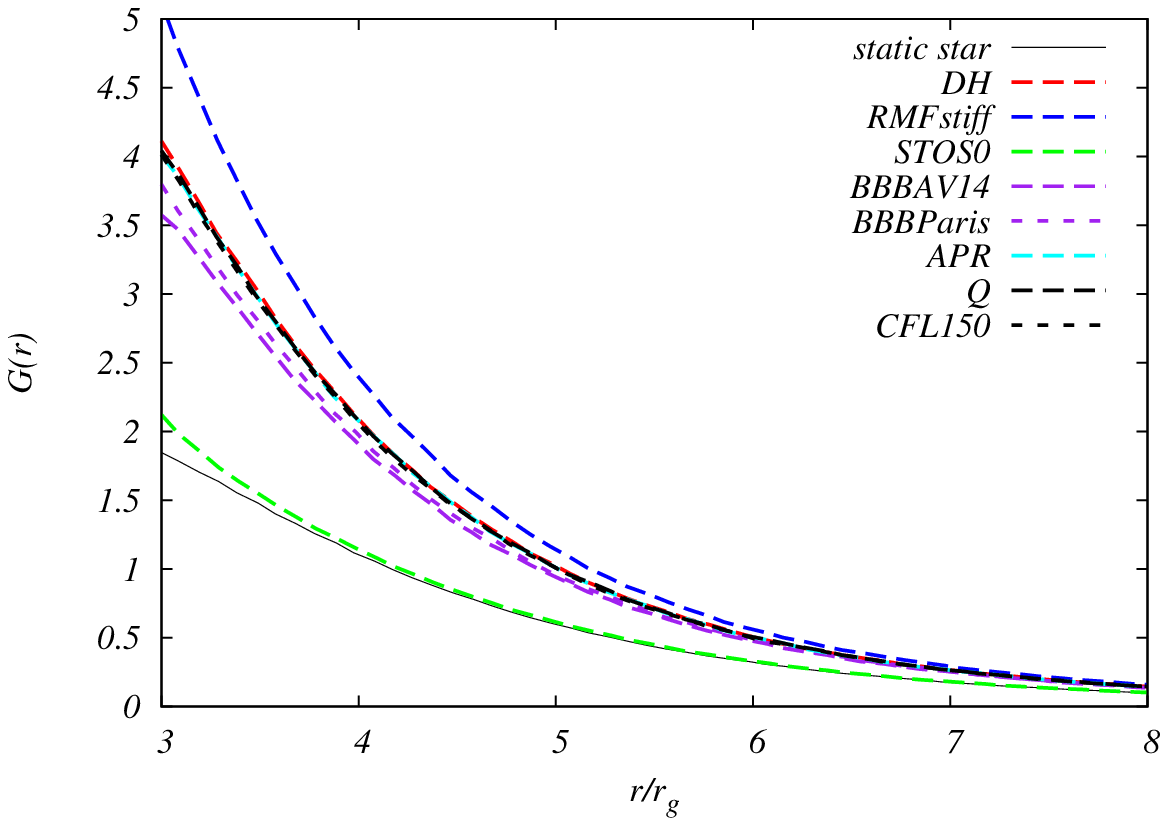}
\includegraphics[width=8.15cm]{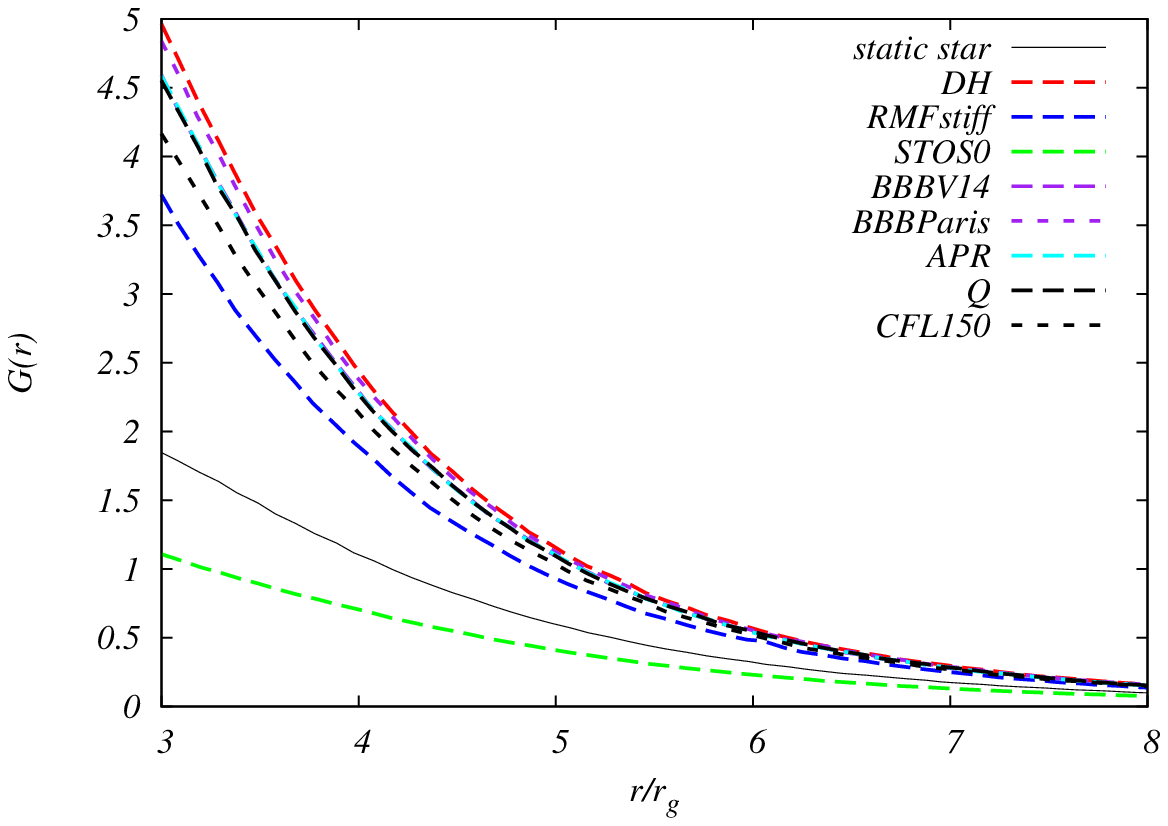}
\includegraphics[width=8.15cm]{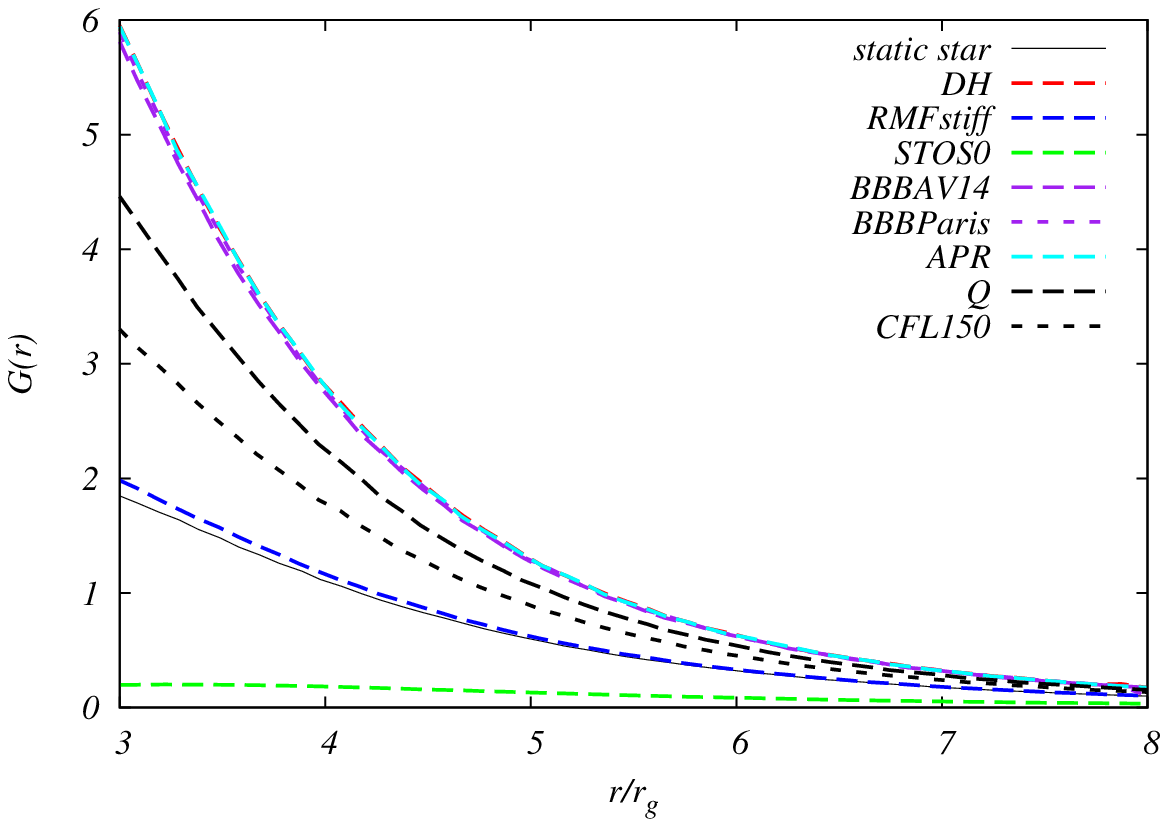}
\centering
\caption{The axial profiles of the EDR
characterized by the dimensionless quantities $G(r)$ for an isothermal disk, and different neutron and quark star models, with the total mass of the central object $M=1.8 M_{\odot}$, and rotational velocities $\Omega=2\times 10^3\;{\rm s}^{-1}$ (top left hand side panel), $3\times 10^3\;{\rm s}^{-1}$ (top right hand side panel), $4\times 10^3\;{\rm s}^{-1}$ (bottom left hand side panel) and $5\times 10^3\;{\rm s}^{-1}$ (bottom right hand side panel).}
\label{fig7}
\end{figure}

In Fig.~\ref{fig7} we present $G(r)$ for accretion disks with a $T_{eff}\propto R^{-1}$ temperature profiles, rotating around stars with different EOSs types, but with fixed values $\Omega$. These plots also show the same trends as those found in the case of the isothermal disk:  at lower rotational frequencies the EDR is somewhat higher for the quark stars than the one derived for neutron stars - apart from the stellar model with RMF stiff type EOS - but with increasing $\Omega$ its value for the Q and the CFL150 type EOS drops below those of the stellar models with DH, APR, BBBAV14, and BBBParis EOSs, respectively. The only neutron stars producing less EDR at higher frequencies than the one obtained for the quark stars are the models with RMF stiff and STOS0 type EOS.

\section{Discussions and final remarks}\label{7}

In the present paper we have considered the energy deposition rate from the neutrino-antineutrino annihilation process on the rotation axis of  rotating neutron and quark stars. The energy deposition rate  has been obtained numerically for several equations of state of the neutron matter, and for two types of quark stars, respectively.  All the general relativistic correction factors, related to this process, can be obtained from the metric of the central compact object. In the present paper we have neglected the possible effects of the stellar magnetic field on the external geometry of the spacetime around the compact general relativistic effects, as well as on the motion of the neutrinos. Generally, we assume that the magnetic energy of the disk, $B^2/8\pi $ is much smaller than the $\rho
V_K^2/2$, the kinetic energy density of the disk, $B^2/8\pi << \rho V_K^2/2$, where $V_K$ is the Keplerian velocity. Hence we neglect the possible influence of the magnetic field on the temperature profile of the disk.  Due to the differences in the space-time structure, the quark stars  present some important differences with respect to the energy deposition rate, as compared to the neutron stars. As a general result we have found that there is a general correlation between  the energy deposition rate and the spin parameter of the star. There is also a strong dependence between the temperature profile of the disk and the EDR,  and at high rotational velocities the EDR is much smaller for the model with $T_{eff}\propto R^{-1}$ than the one calculated for the constant temperature (isothermal) disks. Our study is based on the main assumption of the separability of the geodesic equations of motion for the neutrinos. Even that from a purely theoretical point of view the geodesic and the Hamilton-Jacobi equations cannot be separated, from a practical-computational point of view we have found that such a procedure can work, at least for low angular speeds, with a good precision, with the error introduced by this approximation not exceeding 1\%.

In order to compare the properties of the quark and neutron stars we have chosen a common mass value $M=1.8M_{\odot}$.  Generally, quark stars may have masses greater than $1.8M_{\odot}$. However, for some equations of state of the neutron matter the maximum allowable mass could be smaller than the maximum mass of quark stars. Therefore we have chosen this particular value of the stellar mass in order to be able to compare sequences of quark stars and neutron stars for a large number of equations of state.

A possible astrophysical application of our results could be in the explanation of the physical processes that lead to the formation of the Gamma-Ray Bursts (GRBs). The so-called fireball model
can basically explain the observational facts well, and thus it is strongly favored, and widely accepted today (for recent reviews on GRBs see M\'esz\'aros (2006) and Zhang (2007), respectively). Newborn neutron stars, surrounded by hyper-accreting and neutrino-cooled disks may exist in some gamma-ray bursts and/or supernovae.  As compared with the black hole disk, the neutrino annihilation luminosity above the neutron star disk is higher \cite{ZhDa09}.  Although a heavily mass-loaded outflow from the neutron star surface at early times of neutron star formation prevents the outflow material from being accelerated to a high bulk Lorentz factor, if the disk accretion rate and the neutrino emission luminosity from the surface boundary layer are sufficiently high, an energetic ultra-relativistic jet via neutrino annihilation can be produced above the stellar polar region at late times. The energy deposition along the rotation axis will form a narrow cone jet, which encounters much less material, and hence it can maintain a  large Lorentz factor.

The neutrino emission rate is also strongly dependent on the temperature. In the standard models of GRBs it is assumed that the central object is surrounded by a degenerate accretion disk, which allows super-Eddington accretion rates, of the order of one solar mass per second \cite{Zh07}. If the central compact object is a neutron or a quark star, such super-Eddington accretion rates can maintain the compact object at very high temperatures, and thus allowing very high neutrino luminosities, as well as a high rate of electron-positron pair production. The neutrino temperature can be estimated by assuming that the accretion power $\eta \dot{M}c^2$, where $\eta $ is the efficiency of the energy conversion and $\dot{M}$ is the accretion rate, is equal to the radiation power $4\pi R^2\sigma T_{\nu }^4$, which gives the temperature as $T_{\nu }=\left(\eta \dot{M}c^2/4\pi R^2\sigma \right)^{1/4}$. By taking $\eta =0.1$, $R=10^6$ cm, and an accretion rate of $\dot{M}=1M_{\odot}/10\;{\rm s}$, we obtain $T_{\nu }=7.14\times 10^{10}$K, a temperature which is of the order of MeV. Therefore, if the accretion disk is fed at a high rate, like, for example, by the fallback material after a supernova explosion, a high neutrino-antineutrino emission rate can be maintained, and this could explain some of the basic properties of the GRB phenomenon.

The neutrino annihilation and the electron-positron pair production also plays an essential role in the astrophysical processes associated with the phase transitions that could take place inside neutron stars.  For example, the sudden phase transition from normal nuclear matter to a mixed phase of quark and nuclear matter induces temperature and density oscillations at the neutrinosphere. Consequently, pulsating neutrino/antineutrino and leptonic pair fluxes will be emitted \cite{Ch09}. During this stage several mass ejecta can be ejected from the stellar surface by the neutrinos and antineutrinos. These ejecta
can be further accelerated to relativistic speeds by the electron/positron pairs, created by the neutrino and antineutrino annihilation outside the stellar surface. In order to produce the Gamma-Ray Bursts, a high neutrino emission rate is necessary. On the other hand, it is important to note that electron-positron pairs can deposit energy much more efficiently than the neutrinos, and the dominant energy deposition process is the neutrino-antineutrino annihilation process \cite{Ch09}. In fact most pairs are created outside the star. A large fraction of the neutrino energy, will be absorbed by the matter very near the stellar surface. When this amount of energy exceeds the gravitational binding energy, some mass near the stellar surface will be ejected, and this mass will be further accelerated by absorbing pairs created from the neutrino and antineutrino annihilation processes outside the star. This process may be a possible mechanism for short Gamma-Ray Bursts \cite{Ch09}.

\section*{Acknowledgments}

K. S. C. is supported by the GRF grant number HKU 7011/09P of the government of the Hong Kong SAR. The work of T. H. is supported by the GRF grant number 702507P of the Government of the Hong Kong SAR. Z.K. is indebted to Katsukai Asano and Gabe Perez-Giz for valuable discussions.

\appendix
\section{The separability of the geodesic equation}
\begin{figure}
\centering
\includegraphics[width=8.15cm]{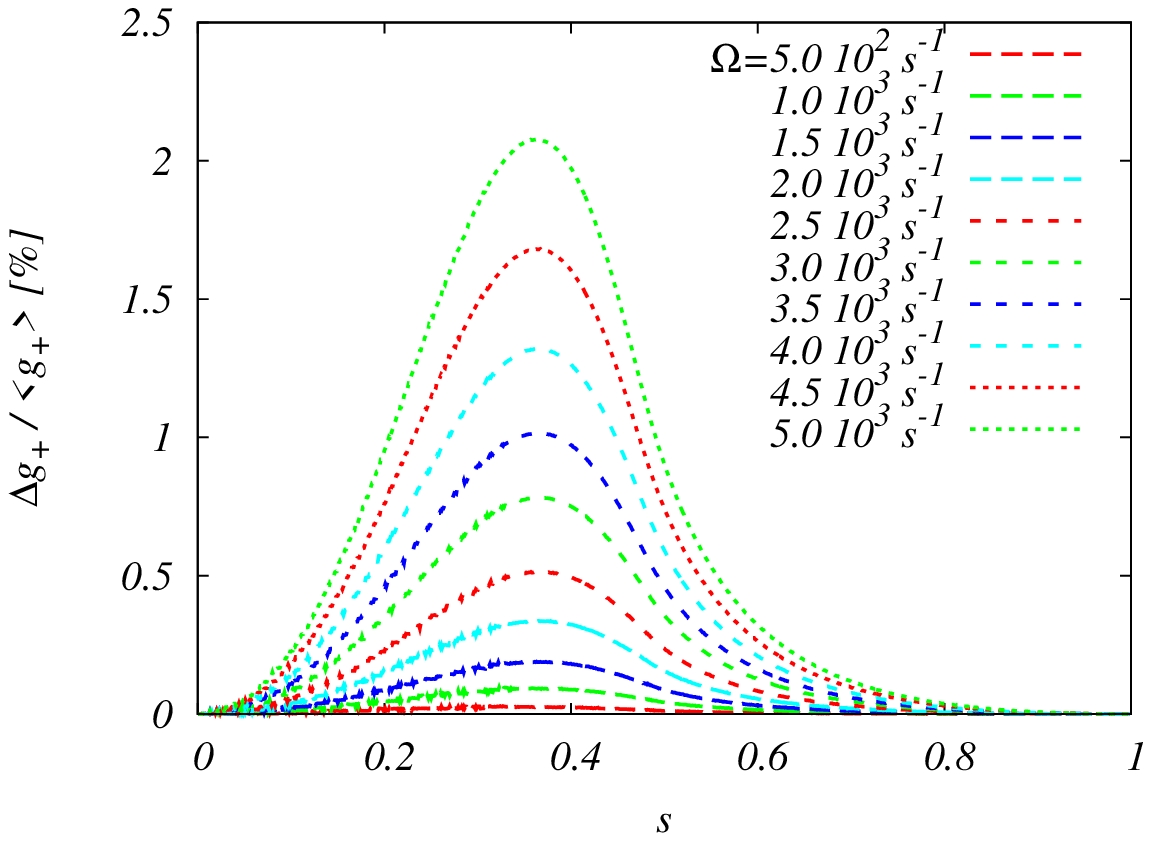}
\includegraphics[width=8.15cm]{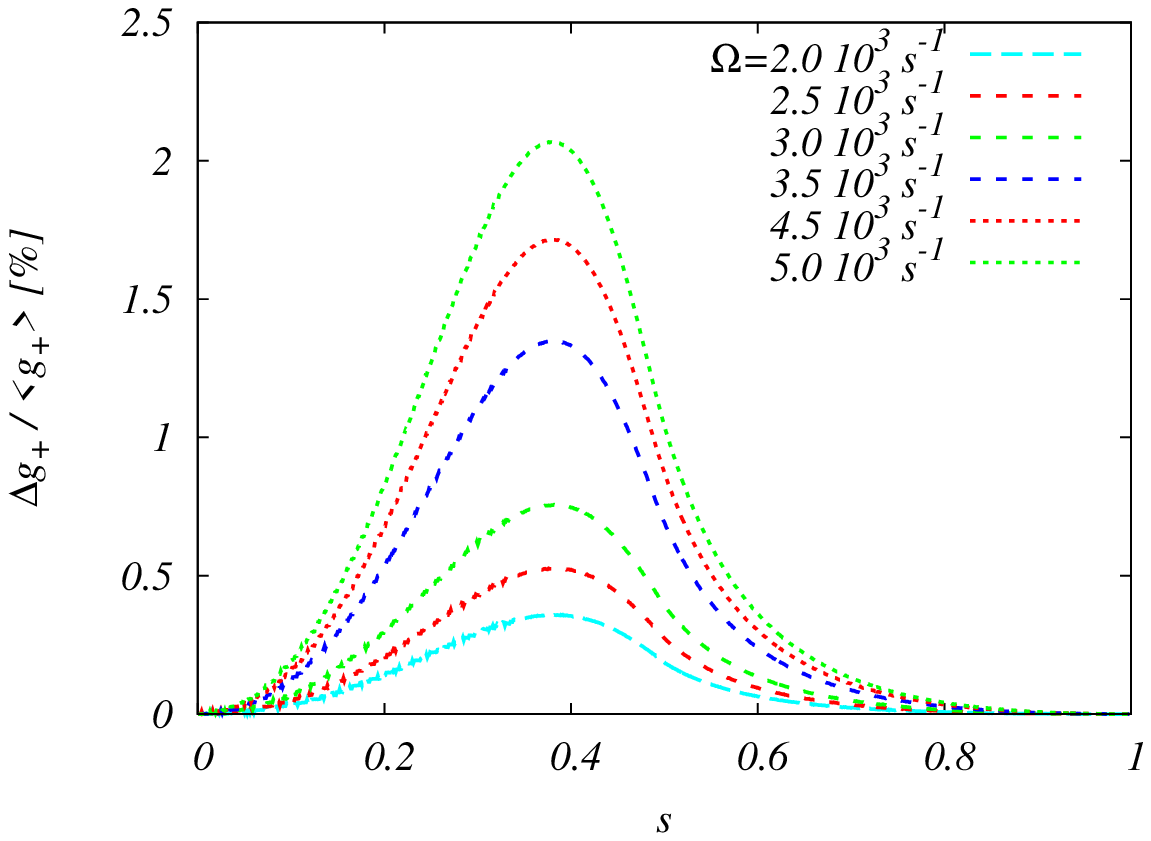}
\includegraphics[width=8.15cm]{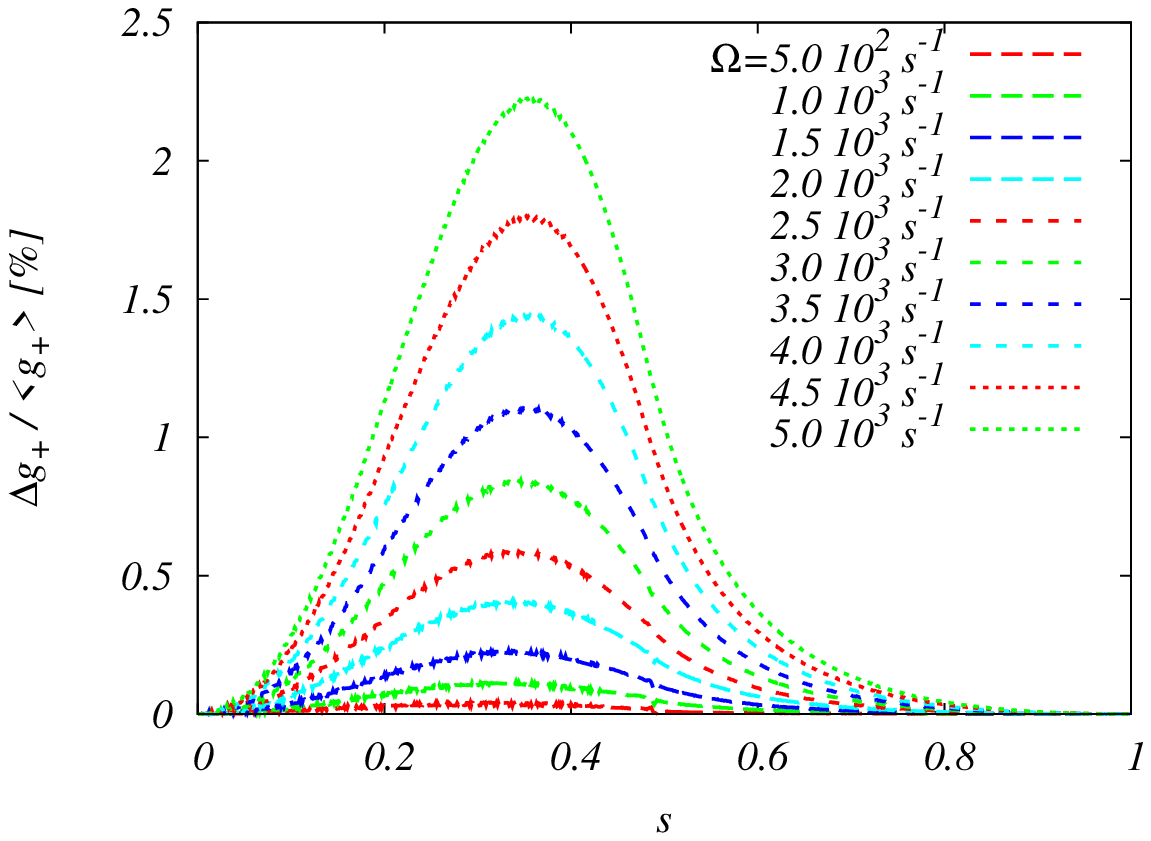}
\includegraphics[width=8.15cm]{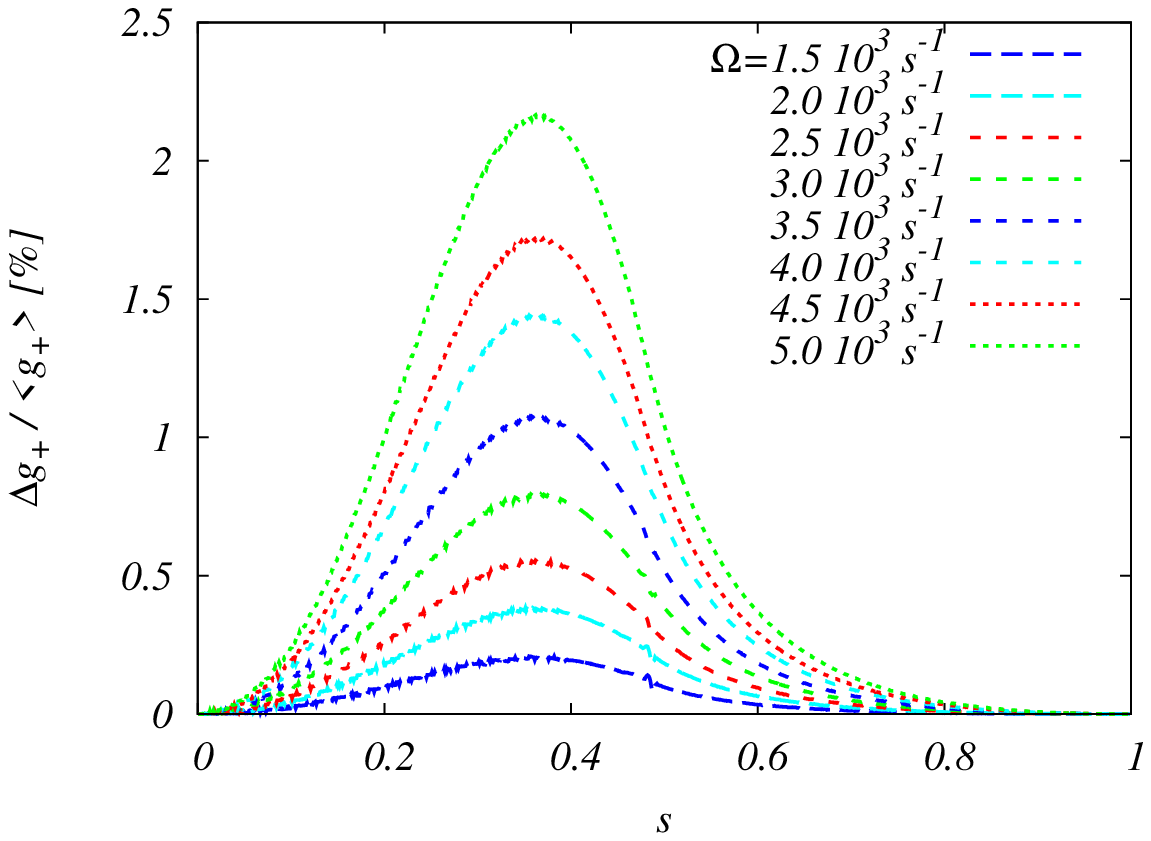}
\caption{
The $\Delta g_{+}/\langle g_{+} \rangle$ vs. $s$ relation for neutron star models with various EOS types, such as DH, shown in Table~{\ref{table1}} (top left hand side plot), APR ,shown in Table~{\ref{table2}} (top right hand side plot), BBBAV14, shown in Table~{\ref{table3}} (bottom left hand side plot) and BBBParis shown in Table~ {\ref{table4}} (bottom right hand side plot).
}
\label{figA1}
\end{figure}
\begin{figure}
\centering
\includegraphics[width=8.15cm]{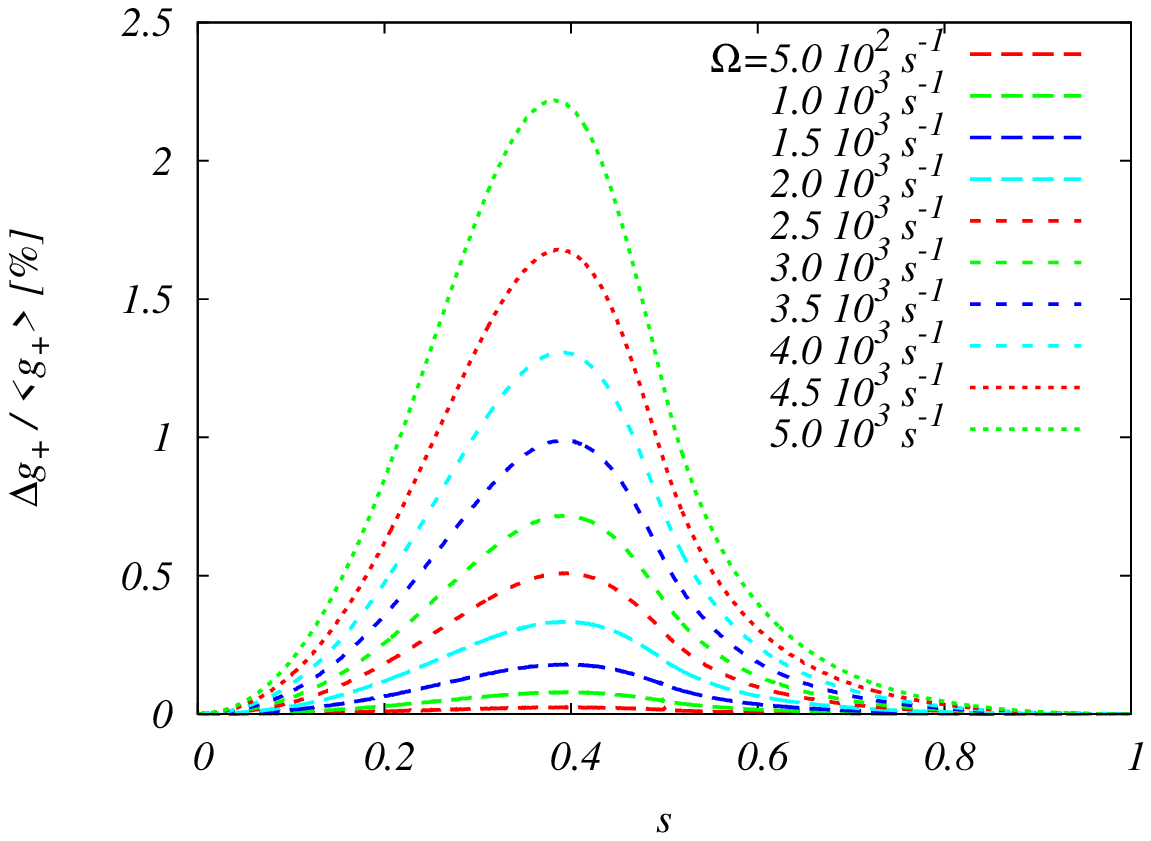}
\includegraphics[width=8.15cm]{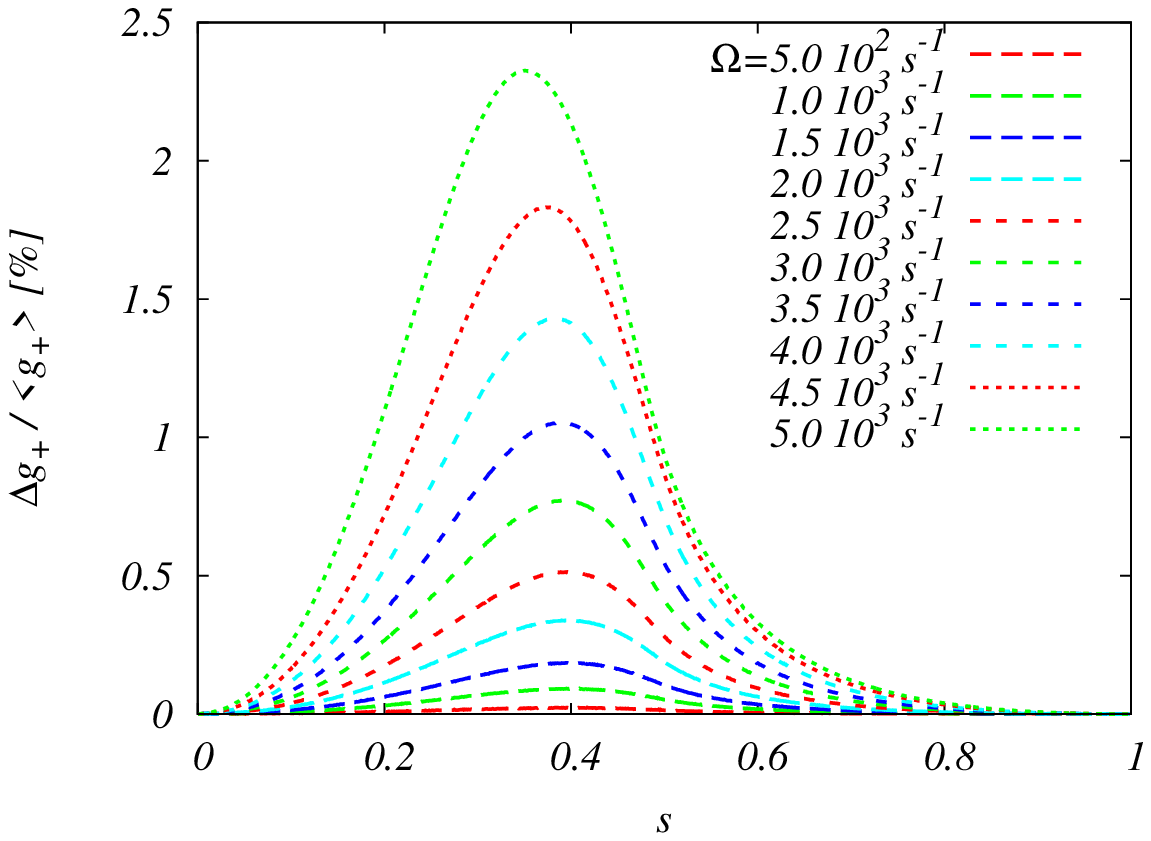}
\includegraphics[width=8.15cm]{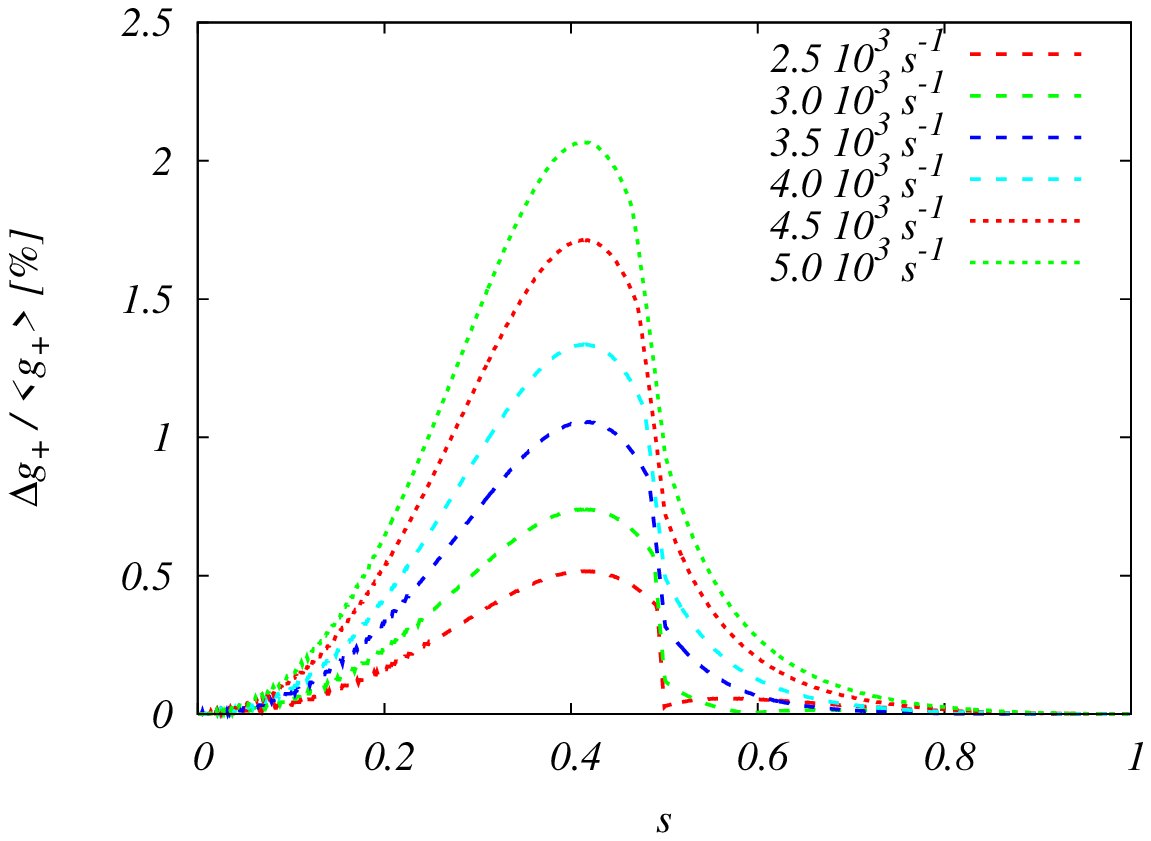}
\includegraphics[width=8.15cm]{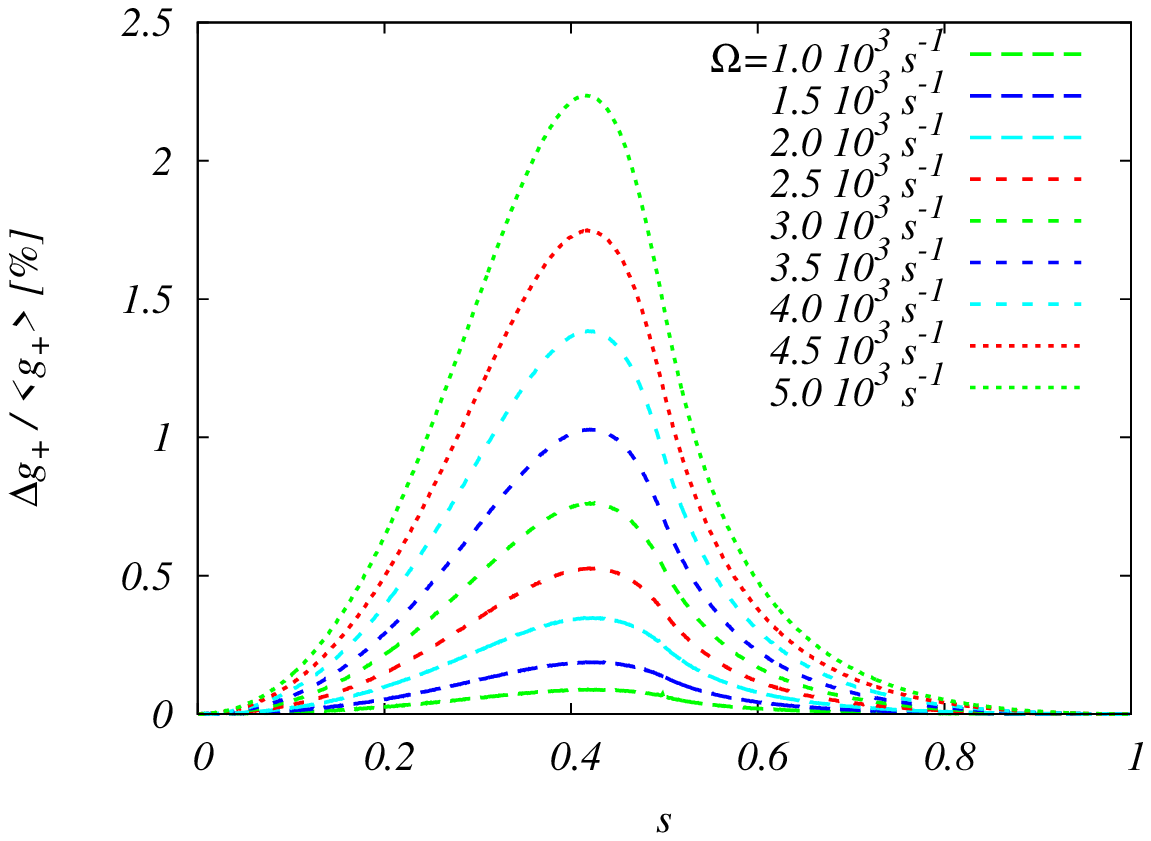}
\caption{
The $\Delta g_{+}/\langle g_{+} \rangle$ vs. $s$ relation for neutron and quark star models with various EOS types, such as RMF stiff, shown in Table~{\ref{table5}} (top left hand side plot), STOS0 ,shown in Table~{\ref{table6}} (top right hand side plot), Q, shown in Table~{\ref{table7}} (bottom left hand side plot) and CFL150 shown in Table~ {\ref{table8}} (bottom right hand side plot).}
\label{figA2}
\end{figure}

The condition for the separability of the geodesic equation Eq.~(\ref{eq:geodeq1}) for $L=0$ is
that the functions $f$ and $g_+$, given by Eqs.~(\ref{grrgthth}) and (\ref{eq:gpm}), respectively, must depend only on the radial coordinate $r$. This condition does not hold for general stationary and axially symmetric spacetimes, since the metric components are also functions of the polar angle. Now we examine if we can use the approximation of neglecting the angular dependence of the functions $f$ and $g_+$ for the spacetime geometry of the neutron and quark stars which are considered in the present paper. If we insert the components $g_{\alpha\beta}$ of the metric (\ref{ds20}), derived by the RNS code for these types of stars into Eqs.~(\ref{grrgthth}) and (\ref{eq:gpm}), we obtain $f(\bar{r})=\bar{r}^2$, and an expression for $g_+$ which still depends on both $\bar{r}$ and $\theta$. Since $f$ is  the square of the radial coordinate only, the question is the angular dependence of $g_{+}$. We consider the quantity
$\Delta g_{+}/\langle g_{+} \rangle$ as a function of the dimensionless radial coordinate $s=\bar{r}/(\bar{r}+{\bar{r}_e})$ (compactified in the domain [0,1]) and angle variable $\mu=\cos\theta$. Here $\Delta g_{+}$ is the maximal difference in the values of $g_{+}$, and $\langle g_{+} \rangle$ is the average of $g_{+}$ for a given $s$ over the entire domain [0,1] of $\mu$. If $g_+$ has no angular dependence, as in the case of the Kerr spacetime (in both Boyer-Lyndquist and quasi-isotropic coordinates), this quantity is just zero. Its deviation from zero characterizes the limits of the validity of our approximation. In Figs.~\ref{figA1} and \ref{figA2}  we plot this function versus the coordinate $s$ for all the stellar models we have studied in this work.

By definition of the coordinate $s$, the stellar surface in the equatorial plane is located at $s=0.5$ and still remains  close to this value at higher latitude. Therefore, the half of the right hand side  of each plot represents the region of the spacetime outside the star. The accretion disk is rotating in this region, and the null geodesics emanating from the disk and intersecting the rotational axis at radii higher than $3r_g$ mostly remain located in this region as well. The trajectories originating at the inner disk edge might run inside the star, but we consider only those particles which hit the axis at $3r_g$ or higher radii. However, this minimal value is high enough to keep these trajectories close to the stellar surface ($s\la0.5$).
In the case of the low mass stellar models with $M=1.8 M_{\odot}$, $\Delta g_{+}/\langle g_{+} \rangle$  is under 0.5\% for $\Omega=2\times10^3 {\rm s}^{-1}$, while it is more the 1\% for angular velocities higher than $4\times10^3 {\rm s}^{-1}$. Then  on the one hand we do not expect that the separation of Eq.~(\ref{eq:geodeq1})
causes large errors in the computation of scalar product ${\boldsymbol p}\cdot {\boldsymbol p}$ for slowly rotating objects, with $\Omega\sim2\times10^3 {\rm s}^{-1}$. On the other hand, the results derived for $\Omega\ga 4\times10^3 {\rm s}^{-1}$ have an increasing uncertainty as
the error propagation grows due to the non-separability of Eq.~(\ref{eq:geodeq1}),
containing $g_{+}(r,\theta)$, which does depend on $\theta$.

\section{The ray tracing algorithm}

The Hamilton-Jacobi equation (\ref{eq:geodeq1})
can be written as
\begin{equation}
\frac{g_{\theta\theta}}{g_{rr}}\dot{r}^{2}+\dot{\theta}^{2}=g_{\theta\theta}V(\omega_{0},L)\:,
\label{rdotthetadot}
\end{equation}
where the effective potential is given by
\[
V(\omega_{0},L)=\frac{g_{\phi\phi}(\omega_0/c)^{2}+2g_{t\phi}(\omega_0/c)L+g_{tt}\widetilde{L}^{2}}{g_{t\phi}^{2}-g_{tt}g_{\phi\phi}}\:.
\]
This is generally not a separable differential equation for the coordinates
$r$ and $\theta$.  For given initial values of the coordinates at
some time $\tau=\tau_{0}$, and for fixed constants of motion $\omega_{0}$ and $L$,
we can, at least numerically, integrate this equation, to obtain the world-line
projected onto the $r-\theta$ plane of a particle moving in the effective potential $V$. The only technical
difficulty in the integration may stem from the non-separability of
Eq.~(\ref{rdotthetadot}). However, this can be easily circumvented
by introducing the momenta $p_{r}$ and $p_{\theta}$  and by replacing Eq.~(\ref{rdotthetadot})
with their canonical equations of motion \cite{PGL09}.
Then the complete set of the evolution equations is
\begin{eqnarray}
\dot{r} & = & \frac{p_{r}}{g_{rr}},\label{rdot}\\
\dot{\theta} & = & \frac{p_{\theta}}{g_{\theta\theta}},\label{thetadot}\\
\dot{p}_{r} & = &\frac{1}{2}\left(\frac{g_{rr,r}}{g_{rr}^{2}}p_{r}^{2}+\frac{g_{\theta\theta,r}}{g_{\theta\theta}^{2}}p_{\theta}^{2}-V_{,r}\right),\label{p_rdot}\\
\dot{p}_{\theta} & = &\frac{1}{2}\left(\frac{g_{rr,\theta}}{g_{rr}^{2}}p_{r}^{2}+\frac{g_{\theta\theta,\theta}}{g_{\theta\theta}^{2}}p_{\theta}^{2}-V_{,\theta}\right),\label{p_thetadot}
\end{eqnarray}
together with the decoupled geodesic equation for the coordinates $t$ and $\phi$.
 This is a system of ordinary
differential equations, with the unknown functions $r(\tau)$, $\theta(\tau)$
$p_{r}(\tau)$ and $p_{\theta}(\tau)$, respectively. The parameters appearing in
the system are the constants of motion $p_{t}=-\omega_0/c$ and $p_{\phi}=L$.
If the initial values of the unknown functions and of the parameters
are given, so that the mass-shell condition $\boldsymbol{p}\cdot\boldsymbol{p}=0$
holds (a constraint imposed on the initial data), then the system (\ref{rdot})-(\ref{p_thetadot}) can be integrated numerically. We have applied the fourth
order Runge-Kutta method for this system of differential equations to implement a ray
tracing algorithm by reversing the proper time evolution.

In order to input the metric components $g_{\alpha\beta}$ in these equations, we have used numerical tables of the metric functions $\alpha$, $\gamma$, $\rho$ and $\omega$, generated by the code RNS with a high resolution mesh ($1001\times501$) of the compactified quasi-isotropic radial coordinate $s=\bar{r}/(\bar{r}+\bar{r}_e)$, and the angular coordinate $\mu=\cos\theta$. Since the numerical scheme of the ray tracing code needs to evaluate the functions
$g_{\alpha\beta}=g_{\alpha\beta}(\alpha(s,\mu),\gamma(s,\mu),\rho(s,\mu),\omega(s,\mu))$,
defined over the 2-dimensional grid $(s,\mu)$ at any arbitrary point $(\bar{r},\theta)$, we
applied bi-cubic interpolation between the discrete points of the numerical tables. The derivatives with respect to $r$ and $\theta$ (or $\bar{r}$ and $\theta$) in the evolution equations were replaced by finite differences, with a fine mesh, which we could calculate in any point from the quasi continuous interpolated function values.
For a given set of canonical data $(\bar{r},\theta=0,p_{\bar{r}},p_{\theta})$,  we traced the (anti)neutrinos back from the rotational axis to the equatorial plane where they are supposed to be produced. The parameter $\omega_0$ was set to an arbitrary value, e.g. $0.9c$, and the condition of the vanishing angular momentum, $L=0$, gives the trajectories that intersect the rotational axis.

With these values for the constants of motion we can determine the magnitude of $p_{\bar{r}}$ from the norm $\boldsymbol{p}\cdot\boldsymbol{p}=0$, and we have a consistent set of canonical data at the axis for a given $p_{\theta}$. Since Eq.~(\ref{costh}) determines the relation between $p_{\theta}$ and the collision angle $\theta_{\nu}$, the ray tracing code can provide the radius $\bar{R}$ at which a null particle hitting the axis at a given radius $\bar{r}$ and with a given collision angle left the surface of the disk in the equatorial plane.

\section{Physical parameters of neutron and quark stars}

Here we present all the data for the physical parameters of the compact stars with total mass $M\approx 1.8 M_{\odot}$, and different rotational frequencies, which were obtained by the RNS code. In the Tables $\rho _c$ is the central density, $M$ is the gravitational mass, $M_0$ is the rest mass, $R_e$ is the circumferential radius at the equator, $\Omega $ is the angular velocity, $\Omega _p$ is the angular velocity of a particle in circular orbit at the equator, $T/W$ is the rotational-gravitational energy ratio, $J$ is the angular momentum, $I$ is the moment of inertia, $\Phi_2$ gives the mass quadrupole moment, $h_+$ is the height from the surface of the last stable co-rotating circular orbit in the equatorial plane, $h_{-}$ is the height from surface of the last stable counter-rotating circular orbit in the equatorial plane, $Z_p$ is the polar redshift, $Z_b$ is the backward equatorial resdhift, $Z_f$ is the forward equatorial redshift, $\omega _c/\Omega$ is the ratio of the central value of the potential $\omega $ to $\Omega $, $r_e$ is the coordinate equatorial radius, and $r_p/r_e$ is the axes ratio (polar to equatorial), respectively.

We have obtained $R_{in}$ from the formula $R_{in}=R_e+h_+$, i.e., as a the sum of the stellar radius in the equator and the height from the surface of the last stable co-rotating circular orbit in the equatorial plane. Thus the inner edge of the accretion disk is located at the radius of the innermost stable circular orbit $r_{ms}$, or at the equatorial radius $R_e$ of the stellar surface, if $h_+$ is zero. For the RMF stiff, STOS0, Q and CFL150 type EOS the code could not provide a physically reasonable values for $h_+$, and we have used a truncated form of the analytical approximation given by \cite{ShSa98},
\begin{eqnarray}
r_{ms}&\approx &3r_g\left(1-0.54433a_*-0.22619a^2_*+0.17989Q_2-0.23002a^3_*+\right.\nonumber\\
&& \left. 0.26296a_*Q_2-0.29693a^4_*+0.44546a^2_*Q_2\right),\label{rms}
\end{eqnarray}
where $a_*=cJ/GM^2$ and $Q_2=c^4\Phi_2/G^2M^3$ are the dimensionless spin parameter and the dimensionless mass quadrupole moment, defined in terms of the spin angular momentum $J$ and the mass quadrupole moment $\Phi_2$.
The results were checked by calculating the zeros of the second order derivative of the effective potential $V$ with respect to the radial coordinate. For the calculation we applied a polynomial fit of the  metric components in the equatorial plane, and computed the first and the second order derivatives of the polynomials, both analytically and numerically.  Since the analytic forms of the derivatives of the polynomial fit were proved to be a good fit for the numerical derivatives too, we could use them to calculate $V_{,rr}$. The zeros of $V_{,rr}$ were in an agreement with the results of Eq.~(\ref{rms}) within 10\%.

\begin{table*}
\centering
\begin{tabular}{|l|l|l|l|l|l|l|l|l|l|l|}
\hline
$\rho_c\;[10^{15}{\rm g}/{\rm cm}^{3}]$
 & 1.4250 & 1.4200 & 1.4100 & 1.4000 & 1.3900 & 1.3800 & 1.3700 & 1.3400 & 1.3200 & 1.2940\\
\hline
$M\;[M_{\odot}]$ & 1.8019 & 1.8015 & 1.7992 & 1.7989 & 1.7998 & 1.8043 & 1.8076 & 1.8022 & 1.8054 & 1.8071\\
\hline
$M_0\; [M_{\odot}]$ & 2.0701 & 2.0689 & 2.0649 & 2.0632 & 2.0627 & 2.0664 & 2.0687 & 2.0582 & 2.0589 & 2.0570\\
\hline
$R_e [{\rm km}]$ & 11.2738 & 11.2942 & 11.3274 & 11.3734 & 11.4277 & 11.5055 & 11.5771 & 11.7052 & 11.8398 & 12.0091\\
\hline
$\Omega [10^3{\rm s}^{-1}]$ & 0.5246 & 1.0481 & 1.5119 & 2.0268 & 2.5064 & 3.0834 & 3.5030 & 4.0075 & 4.5071 & 4.9978\\
\hline
$\Omega_p [10^3{\rm s}^{-1}]$ & 12.8662 & 12.7900 & 12.6968 & 12.5925 & 12.4873 & 12.3606 & 12.2508 & 12.0334 & 11.8445 & 11.6131\\
\hline
$T/W [10^{-2}]$ & 0.0329 & 0.1317 & 0.2764 & 0.5016 & 0.7753 & 1.1888 & 1.5535 & 2.0921 & 2.7103 & 3.4355\\
\hline
$cJ/GM_{\odot}^2$ & 0.1109 & 0.2219 & 0.3208 & 0.4324 & 0.5385 & 0.6710 & 0.7707 & 0.8909 & 1.0198 & 1.1535\\
\hline
$I [10^{45}{\rm g}\;{\rm cm}^2]$ & 1.8570 & 1.8611 & 1.8651 & 1.8749 & 1.8883 & 1.9126 & 1.9336 & 1.9538 & 1.9885 & 2.0283\\
\hline
$\Phi_2 [10^{42}{\rm g}\;{\rm cm}^2]$ & 0.7995 & 3.0606 & 6.3866 & 11.5847 & 17.9592 & 27.6936 & 36.3965 & 49.5047 & 64.8507 & 83.3478\\
\hline
$h_+ [{\rm km}]$ & 4.3861 & 4.0805 & 3.7865 & 3.4791 & 3.2005 & 2.8939 & 2.6645 & 2.2884 & 1.9918 & 1.6754\\
\hline
$h_- [{\rm km}]$ & 4.9786 & 5.2653 & 5.4976 & 5.7775 & 6.0495 & 6.4120 & 6.6746 & 6.8938 & 7.1985 & 7.4826\\
\hline
$Z_p$ & 0.3761 & 0.3761 & 0.3753 & 0.3755 & 0.3762 & 0.3787 & 0.3806 & 0.3788 & 0.3807 & 0.3820\\
\hline
$Z_f$ & 0.3390 & 0.3021 & 0.2690 & 0.2334 & 0.2006 & 0.1623 & 0.1342 & 0.0969 & 0.0620 & 0.0263\\
\hline
$Z_b$ & 0.4137 & 0.4516 & 0.4847 & 0.5234 & 0.5607 & 0.6087 & 0.6446 & 0.6837 & 0.7287 & 0.7737\\
\hline
$\omega _c/\Omega$  & 0.5987  & 0.5981  & 0.5965  & 0.5954  & 0.5946  & 0.5948  & 0.5947  & 0.5901  & 0.5888  & 0.5865\\
\hline
$r_e  [{\rm km}]$ & 8.4046 & 8.4247 & 8.4604 & 8.5045 & 8.5543 & 8.6196 & 8.6815 & 8.8138 & 8.9366 & 9.0959\\
\hline
$r_p/r_e$ & 0.9989 & 0.9955 & 0.9906 & 0.9830 & 0.9738 & 0.9600 & 0.9479 & 0.9300 & 0.9097 & 0.8860\\
\hline
\end{tabular}
\caption{The physical parameters for a neutron star model with a DH type EOS, a total mass of $1.8M_{\odot}$ and angular velocities from $5\times10^2 {\rm s}^{-1}$ to $5\times10^3 {\rm s}^{-1}$.}
\label{table1}
\end{table*}
\begin{table*}
\centering
\begin{tabular}{|l|l|l|l|l|l|l|l|l|}
\hline
$\rho_c\;[10^{15}{\rm g}/{\rm cm}^{3}]$ & 1.2640 & 1.2745 & 1.2710 & 1.2650 & 1.2500 & 1.2400 & 1.2200 & 1.2130\\
\hline
$M\;[M_{\odot}]$ & 1.7788 & 1.7972 & 1.7996 & 1.8019 & 1.8038 & 1.8060 & 1.8000 & 1.8083\\
\hline
$M_0\; [M_{\odot}]$ & 2.0746 & 2.0988 & 2.1006 & 2.1016 & 2.1013 & 2.1016 & 2.0895 & 2.0976\\
\hline
$R_e [{\rm km}]$ & 10.9336 & 10.9627 & 11.0009 & 11.0542 & 11.1299 & 11.2026 & 11.3140 & 11.4073\\
\hline
$\Omega [10^3{\rm s}^{-1}]$ & 1.5271 & 2.0791 & 2.5175 & 3.0120 & 3.5665 & 4.0060 & 4.5288 & 4.9377\\
\hline
$\Omega_p [10^3{\rm s}^{-1}]$ & 13.3079 & 13.2923 & 13.2144 & 13.1141 & 12.9808 & 12.8637 & 12.6656 & 12.5504\\
\hline
$T/W [10^{-2}]$ & 0.2913 & 0.5373 & 0.7936 & 1.1478 & 1.6342 & 2.0917 & 2.7466 & 3.3127\\
\hline
$cJ/GM_{\odot}^2$ & 0.3215 & 0.4459 & 0.5438 & 0.6564 & 0.7860 & 0.8928 & 1.0189 & 1.1312\\
\hline
$I [10^{45}{\rm g}\;{\rm cm}^2]$ & 1.8505 & 1.8847 & 1.8985 & 1.9152 & 1.9369 & 1.9586 & 1.9772 & 2.0134\\
\hline
$\Phi_2 [10^{42}{\rm g}\;{\rm cm}^2]$ & 6.7680 & 12.4757 & 18.4095 & 26.6997 & 38.1889 & 49.1323 & 64.9632 & 79.0479\\
\hline
$h_+ [{\rm km}]$ & 3.9923 & 3.8447 & 3.6152 & 3.3529 & 3.0510 & 2.8164 & 2.4695 & 2.2982\\
\hline
$h_- [{\rm km}]$ & 5.7258 & 6.2159 & 6.4911 & 6.8003 & 7.1428 & 7.4236 & 7.6828 & 8.0085\\
\hline
$Z_p$ & 0.3901 & 0.3980 & 0.3996 & 0.4014 & 0.4033 & 0.4052 & 0.4042 & 0.4086\\
\hline
$Z_f$ & 0.2844 & 0.2532 & 0.2244 & 0.1918 & 0.1550 & 0.1258 & 0.0886 & 0.0623\\
\hline
$Z_b$ & 0.4993 & 0.5495 & 0.5847 & 0.6253 & 0.6717 & 0.7102 & 0.7524 & 0.7944\\
\hline
$\omega _c/\Omega$ & 0.5814  & 0.5876  & 0.5878  & 0.5878  & 0.5874  & 0.5870  & 0.5834  & 0.5848\\
\hline
$r_e  [{\rm km}]$ & 8.0918 & 8.0847 & 8.1153 & 8.1599 & 8.2259 & 8.2888 & 8.4026 & 8.4738\\
\hline
$r_p/r_e$ & 0.9907 & 0.9830 & 0.9750 & 0.9640 & 0.9490 & 0.9350 & 0.9150 & 0.8980\\
\hline
\end{tabular}
\caption{The physical parameters for a neutron star model with APR EOS, a total mass of $1.8M_{\odot}$ and angular velocities from $1.5\times10^3 {\rm s}^{-1}$ to $5\times10^3 {\rm s}^{-1}$.}
\label{table2}
\end{table*}
\begin{table*}
\centering
\begin{tabular}{|l|l|l|l|l|l|l|l|l|l|l|}
\hline
$\rho_c\;[10^{15}{\rm g}/{\rm cm}^{3}]$ & 3.4000 & 3.2000 & 3.1000 & 3.0000 & 2.9000 & 2.7000 & 2.5000 & 2.3500 & 2.1800 & 2.1500\\
\hline
$M\;[M_{\odot}]$ & 1.7891 & 1.7925 & 1.7946 & 1.7976 & 1.8002 & 1.8018 & 1.8006 & 1.8003 & 1.8067 & 1.8049\\
\hline
$M_0\; [M_{\odot}]$ & 2.1026 & 2.1073 & 2.1098 & 2.1130 & 2.1155 & 2.1159 & 2.1118 & 2.1083 & 2.1107 & 2.1077\\
\hline
$R_e [{\rm km}]$ & 9.3362 & 9.4389 & 9.4986 & 9.5670 & 9.6384 & 9.7808 & 9.9380 & 10.0843 & 10.3195 & 10.3498\\
\hline
$\Omega [10^3{\rm s}^{-1}]$ & 0.5525 & 1.0536 & 1.5232 & 2.0716 & 2.5101 & 3.0357 & 3.5133 & 4.0414 & 4.9989 & 5.0398\\
\hline
$\Omega_p [10^3{\rm s}^{-1}]$ & 17.0064 & 16.6910 & 16.4982 & 16.2884 & 16.0871 & 15.7139 & 15.3184 & 14.9718 & 14.4799 & 14.4116\\
\hline
$T/W [10^{-2}]$ & 0.0202 & 0.0764 & 0.1633 & 0.3095 & 0.4659 & 0.7178 & 1.0182 & 1.4153 & 2.3168 & 2.3808\\
\hline
$cJ/GM_{\odot}^2$ & 0.0873 & 0.1702 & 0.2490 & 0.3435 & 0.4221 & 0.5237 & 0.6215 & 0.7317 & 0.9427 & 0.9536\\
\hline
$I [10^{45}{\rm g}\;{\rm cm}^2]$ & 1.3887 & 1.4195 & 1.4368 & 1.4572 & 1.4781 & 1.5160 & 1.5547 & 1.5912 & 1.6573 & 1.6629\\
\hline
$\Phi_2 [10^{42}{\rm g}\;{\rm cm}^2]$ & 0.3292 & 1.1571 & 2.4929 & 4.7963 & 7.3192 & 11.6367 & 17.0855 & 24.4570 & 41.7102 & 43.1197\\
\hline
$h_+ [{\rm km}]$ & 6.2663 & 5.9751 & 5.7298 & 5.4476 & 5.2036 & 4.8317 & 4.4388 & 4.0492 & 3.4490 & 3.3853\\
\hline
$h_- [{\rm km}]$ & 6.7367 & 6.8894 & 7.0655 & 7.2848 & 7.4555 & 7.6162 & 7.7352 & 7.9136 & 8.3595 & 8.3522\\
\hline
$Z_p$ & 0.5173 & 0.5094 & 0.5060 & 0.5032 & 0.5002 & 0.4916 & 0.4815 & 0.4748 & 0.4718 & 0.4695\\
\hline
$Z_f$ & 0.4779 & 0.4348 & 0.3986 & 0.3578 & 0.3245 & 0.2801 & 0.2379 & 0.1954 & 0.1257 & 0.1208\\
\hline
$Z_b$ & 0.5571 & 0.5859 & 0.6173 & 0.6559 & 0.6866 & 0.7186 & 0.7456 & 0.7812 & 0.8597 & 0.8604\\
\hline
$\omega _c/\Omega$  & 0.7428  & 0.7328  & 0.7276  & 0.7226  & 0.7172  & 0.7050  & 0.6912  & 0.6805  & 0.6694  & 0.6666\\
\hline
$r_e  [{\rm km}]$ & 6.4255 & 6.5254 & 6.5820 & 6.6451 & 6.7119 & 6.8525 & 7.0128 & 7.1580 & 7.3745 & 7.4081\\
\hline
$r_p/r_e$ & 0.9993 & 0.9975 & 0.9947 & 0.9900 & 0.9850 & 0.9770 & 0.9675 & 0.9550 & 0.9270 & 0.9250\\
\hline
\end{tabular}
\caption{The physical parameters for a neutron star model with BBBAV14 EOS, a total mass of $1.8M_{\odot}$ and angular velocities from $5\times10^2 {\rm s}^{-1}$ to $5\times10^3 {\rm s}^{-1}$.}
\label{table3}
\end{table*}
\begin{table*}
\centering
\begin{tabular}{|l|l|l|l|l|l|l|l|l|l|}
\hline
$\rho_c\;[10^{15}{\rm g}/{\rm cm}^{3}]$ & 1.9000 & 1.8900 & 1.8800 & 1.8700 & 1.8500 & 1.8300 & 1.8000 & 1.7500 & 1.7000\\
\hline
$M\;[M_{\odot}]$ & 1.8017 & 1.8016 & 1.8033 & 1.8049 & 1.8055 & 1.8072 & 1.8085 & 1.8014 & 1.7989\\
\hline
$M_0\; [M_{\odot}]$ & 2.1170 & 2.1161 & 2.1171 & 2.1180 & 2.1169 & 2.1170 & 2.1157 & 2.1027 & 2.0949\\
\hline
$R_e [{\rm km}]$ & 10.1829 & 10.2058 & 10.2378 & 10.2697 & 10.3218 & 10.3809 & 10.4664 & 10.5682 & 10.7068\\
\hline
$\Omega [10^3{\rm s}^{-1}]$ & 1.0141 & 1.5161 & 2.0785 & 2.5091 & 3.0185 & 3.5098 & 4.0696 & 4.4813 & 5.0407\\
\hline
$\Omega_p [10^3{\rm s}^{-1}]$ & 14.9378 & 14.8458 & 14.7434 & 14.6556 & 14.5239 & 14.3900 & 14.2075 & 13.9766 & 13.7052\\
\hline
$T/W [10^{-2}]$ & 0.0957 & 0.2150 & 0.4071 & 0.5977 & 0.8768 & 1.2028 & 1.6522 & 2.0679 & 2.7144\\
\hline
$cJ/GM_{\odot}^2$ & 0.1897 & 0.2842 & 0.3918 & 0.4757 & 0.5768 & 0.6771 & 0.7952 & 0.8832 & 1.0106\\
\hline
$I [10^{45}{\rm g}\;{\rm cm}^2]$ & 1.6439 & 1.6473 & 1.6567 & 1.6662 & 1.6792 & 1.6954 & 1.7173 & 1.7322 & 1.7620\\
\hline
$\Phi_2 [10^{42}{\rm g}\;{\rm cm}^2]$ & 1.8153 & 4.0693 & 7.6582 & 11.2359 & 16.5747 & 22.8810 & 31.7467 & 40.2724 & 53.8111\\
\hline
$h_+ [{\rm km}]$ & 5.2697 & 5.0071 & 4.7255 & 4.5082 & 4.2300 & 3.9658 & 3.6472 & 3.3156 & 2.9369\\
\hline
$h_- [{\rm km}]$ & 6.2820 & 6.5233 & 6.8101 & 7.0316 & 7.2789 & 7.5282 & 7.8059 & 7.9256 & 8.1701\\
\hline
$Z_p$ & 0.4478 & 0.4477 & 0.4486 & 0.4495 & 0.4497 & 0.4505 & 0.4509 & 0.4461 & 0.4441\\
\hline
$Z_f$ & 0.3764 & 0.3415 & 0.3036 & 0.2747 & 0.2399 & 0.2068 & 0.1686 & 0.1363 & 0.0956\\
\hline
$Z_b$ & 0.5209 & 0.5576 & 0.6007 & 0.6346 & 0.6744 & 0.7144 & 0.7605 & 0.7885 & 0.8337\\
\hline
$\omega _c/\Omega$
  & 0.6546  & 0.6536  & 0.6531  & 0.6526  & 0.6509  & 0.6494  & 0.6469  & 0.6400  & 0.6343\\
\hline
$r_e  [{\rm km}]$ & 7.2806 & 7.3023 & 7.3290 & 7.3559 & 7.4037 & 7.4561 & 7.5343 & 7.6451 & 7.7814\\
\hline
$r_p/r_e$ & 0.9969 & 0.9931 & 0.9870 & 0.9810 & 0.9722 & 0.9620 & 0.9480 & 0.9350 & 0.9150\\
\hline
\end{tabular}
\caption{The physical parameters for a neutron star model with BBBParis EOS with a total mass of $1.8M_{\odot}$ and angular velocities from $10^3 {\rm s}^{-1}$ to $5\times10^3 {\rm s}^{-1}$.}
\label{table4}
\end{table*}
\begin{table*}
\centering
\begin{tabular}{|l|l|l|l|l|l|l|l|l|l|l|}
\hline
$\rho_c\;[10^{15}{\rm g}/{\rm cm}^{3}]$
 & 0.7290 & 0.7250 & 0.7200 & 0.7100 & 0.7000 & 0.6850 & 0.6670 & 0.6400 & 0.6100 & 0.5700\\
\hline
$M\;[M_{\odot}]$ & 1.8065 & 1.8053 & 1.8069 & 1.8062 & 1.8078 & 1.8054 & 1.8062 & 1.8000 & 1.7993 & 1.8153\\
\hline
$M_0\; [M_{\odot}]$ & 2.0443 & 2.0421 & 2.0429 & 2.0399 & 2.0395 & 2.0336 & 2.0304 & 2.0172 & 2.0092 & 2.0173\\
\hline
$R_e [{\rm km}]$ & 13.5283 & 13.5644 & 13.6313 & 13.7402 & 13.8690 & 14.0370 & 14.2746 & 14.6091 & 15.0696 & 15.9090\\
\hline
$\Omega [10^3{\rm s}^{-1}]$ & 0.5335 & 0.9922 & 1.5077 & 2.0600 & 2.5368 & 3.0031 & 3.5078 & 4.0159 & 4.5104 & 5.0740\\
\hline
$\Omega_p [10^3{\rm s}^{-1}]$
 & 9.7975 & 9.7329 & 9.6510 & 9.5304 & 9.4096 & 9.2539 & 9.0579 & 8.7849 & 8.4531 & 7.9384\\
\hline
$T/W [10^{-2}]$ & 0.0679 & 0.2368 & 0.5533 & 1.0555 & 1.6387 & 2.3739 & 3.3823 & 4.7272 & 6.4613 & 9.2817\\
\hline
$cJ/GM_{\odot}^2$ & 0.1641 & 0.3065 & 0.4702 & 0.6512 & 0.8160 & 0.9849 & 1.1851 & 1.4065 & 1.6653 & 2.0742\\
\hline
$I [10^{45}{\rm g}\;{\rm cm}^2]$ & 2.7037 & 2.7147 & 2.7411 & 2.7782 & 2.8268 & 2.8824 & 2.9691 & 3.0780 & 3.2448 & 3.5927\\
\hline
$\Phi_2 [10^{42}{\rm g}\;{\rm cm}^2]$ & 2.8764 & 10.1367 & 23.4709 & 44.8791 & 70.2820 & 102.8550 & 148.8100 & 212.4910 & 299.7310 & 458.3680\\
\hline
$h_+ [{\rm km}]$ & 0.0000 & 0.0000 & 0.0000 & 0.0000 & 0.0000 & 0.0000 & 0.0000 & 0.0000 & 0.0000 & 0.0000\\
\hline
$h_- [{\rm km}]$ & 0.0000 & 0.0000 & 0.0000 & $-\infty$ & $-\infty$ & $-\infty$ & $-\infty$ & $-\infty$ & $-\infty$ & $-\infty$\\
\hline
$Z_p$ & 0.2852 & 0.2852 & 0.2861 & 0.2867 & 0.2880 & 0.2884 & 0.2898 & 0.2896 & 0.2905 & 0.2945\\
\hline
$Z_f$ & 0.2456 & 0.2118 & 0.1746 & 0.1342 & 0.0994 & 0.0641 & 0.0252 & -0.0170 & -0.0609 & -0.1178\\
\hline
$Z_b$ & 0.3251 & 0.3600 & 0.4008 & 0.4453 & 0.4859 & 0.5260 & 0.5726 & 0.6205 & 0.6734 & 0.7492\\
\hline
$\omega _c/\Omega$  & 0.4748  & 0.4741  & 0.4740  & 0.4727  & 0.4720  & 0.4697  & 0.4676  & 0.4628  & 0.4584  & 0.4549\\
\hline
$r_e  [{\rm km}]$ & 10.6958 & 10.7316 & 10.7913 & 10.8947 & 11.0128 & 11.1748 & 11.3977 & 11.7252 & 12.1652 & 12.9435\\
\hline
$r_p/r_e$ & 0.9977 & 0.9920 & 0.9815 & 0.9650 & 0.9460 & 0.9225 & 0.8910 & 0.8500 & 0.7990 & 0.7200\\
\hline
\end{tabular}
\caption{The physical parameters for a neutron star model with RMF stiff EOS, a total mass of $1.8M_{\odot}$ and angular velocities from $5\times10^2 {\rm s}^{-1}$ to $5\times10^3 {\rm s}^{-1}$.}
\label{table5}
\end{table*}
\begin{table*}
\centering
\begin{tabular}{|l|l|l|l|l|l|l|l|l|l|l|l|}
\hline
$\rho_c\;[10^{15}{\rm g}/{\rm cm}^{3}]$ & 0.4800 & 0.4770 & 3.2000 & 0.4750 & 0.4700 & 0.4620 & 0.4540 & 0.4400 & 0.4230 & 0.3980 & 0.3690\\
\hline
$M\;[M_{\odot}]$ & 1.8061 & 1.8031 & 1.7925 & 1.8076 & 1.8086 & 1.8017 & 1.8088 & 1.8002 & 1.8039 & 1.8033 & 1.8541\\
\hline
$M_0\; [M_{\odot}]$ & 2.0118 & 2.0070 & 2.1073 & 2.0113 & 2.0103 & 1.9990 & 2.0034 & 1.9876 & 1.9842 & 1.9725 & 2.0155\\
\hline
$R_e [{\rm km}]$ & 15.2003 & 15.2651 & 9.4389 & 15.3585 & 15.5171 & 15.7140 & 16.0281 & 16.4469 & 17.1311 & 18.2747 & 21.0394\\
\hline
$\Omega [10^3{\rm s}^{-1}]$ & 0.5081 & 1.0560 & 1.0536 & 1.5041 & 2.0308 & 2.5000 & 3.0348 & 3.5301 & 4.0582 & 4.5293 & 4.9020\\
\hline
$\Omega_p [10^3{\rm s}^{-1}]$ & 8.2289 & 8.1525 & 16.6910 & 8.0845 & 7.9708 & 7.8258 & 7.6457 & 7.3877 & 7.0275 & 6.4676 & 5.3725\\
\hline
$T/W [10^{-2}]$ & 0.0865 & 0.3780 & 0.0764 & 0.7757 & 1.4473 & 2.2664 & 3.4814 & 5.0168 & 7.2439 & 10.3451 & 14.9104\\
\hline
$cJ/GM_{\odot}^2$ & 0.1891 & 0.3949 & 0.1702 & 0.5701 & 0.7835 & 0.9799 & 1.2345 & 1.4877 & 1.8274 & 2.2429 & 2.9524\\
\hline
$I [10^{45}{\rm g}\;{\rm cm}^2]$ & 3.2706 & 3.2868 & 1.4195 & 3.3309 & 3.3905 & 3.4448 & 3.5750 & 3.7037 & 3.9573 & 4.3520 & 5.2932\\
\hline
$\Phi_2 [10^{42}{\rm g}\;{\rm cm}^2]$ & 4.8687 & 21.1326 & 1.1571 & 43.4479 & 81.4042 & 127.9780 & 199.1870 & 290.8950 & 432.8830 & 650.3940 & 1061.9200\\
\hline
$h_+ [{\rm km}]$ & $-\infty$ & 0.0000 & 5.9751 & 0.0000 & 0.0000 & 0.0000 & 0.0000 & 0.0000 & 0.0000 & 0.0000 & 0.0000\\
\hline
$h_- [{\rm km}]$ & 1.3339 & 1.9240 & 6.8894 & 2.4997 & 3.1777 & 3.7552 & 4.5759 & 5.2774 & 6.2012 & 7.1106 & 7.8501\\
\hline
$Z_p$ & 0.2415 & 0.2416 & 0.5094 & 0.2432 & 0.2446 & 0.2448 & 0.2480 & 0.2486 & 0.2515 & 0.2529 & 0.2601\\
\hline
$Z_f$ & 0.2020 & 0.1596 & 0.4348 & 0.1260 & 0.0856 & 0.0481 & 0.0056 & -0.0380 & -0.0885 & -0.1461 & -0.2296\\
\hline
$Z_b$ & 0.2814 & 0.3251 & 0.5859 & 0.3634 & 0.4094 & 0.4504 & 0.5038 & 0.5534 & 0.6165 & 0.6842 & 0.7891\\
\hline
$\omega _c/\Omega$  & 0.4201  & 0.4194  & 0.7328  & 0.4203  & 0.4202  & 0.4183  & 0.4191  & 0.4160  & 0.4145  & 0.4099  & 0.4102\\
\hline
$r_e  [{\rm km}]$ & 12.3910 & 12.4567 & 6.5254 & 12.5372 & 12.6848 & 12.8822 & 13.1688 & 13.5832 & 14.2357 & 15.3506 & 18.0144\\
\hline
$r_p/r_e$ & 0.9970 & 0.9870 & 0.9975 & 0.9735 & 0.9510 & 0.9240 & 0.8850 & 0.8370 & 0.7700 & 0.6800 & 0.5400\\
\hline
\end{tabular}
\caption{The physical parameters for a neutron star model with STOS0 EOS, a total mass of $1.8M_{\odot}$ and angular velocities from $5\times10^2 {\rm s}^{-1}$ to $5\times10^3 {\rm s}^{-1}$.}
\label{table6}
\end{table*}
\begin{table*}
\centering
\begin{tabular}{|l|l|l|l|l|l|l|l|l|}
\hline
$\rho_c\;[10^{15}{\rm g}/{\rm cm}^{3}]$ & 1.0800 & 1.0700 & 1.0500 & 1.0300 & 1.0000 & 0.9800 & 0.9490 & 0.9310\\
\hline
$M\;[M_{\odot}]$ & 1.8039 & 1.8065 & 1.8032 & 1.8029 & 1.8000 & 1.8014 & 1.7999 & 1.8071\\
\hline
$M_0\; [M_{\odot}]$ & 2.1221 & 2.1241 & 2.1176 & 2.1146 & 2.1071 & 2.1057 & 2.0989 & 2.1045\\
\hline
$R_e [{\rm km}]$ & 11.2268 & 11.2717 & 11.3243 & 11.3914 & 11.4900 & 11.5804 & 11.7097 & 11.8329\\
\hline
$\Omega [10^3{\rm s}^{-1}]$ & 1.4909 & 2.0426 & 2.5022 & 2.9918 & 3.5684 & 3.9966 & 4.5072 & 4.9097\\
\hline
$\Omega_p [10^3{\rm s}^{-1}]$ & 12.8593 & 12.7703 & 12.6630 & 12.5507 & 12.3985 & 12.2828 & 12.1180 & 11.9932\\
\hline
$T/W [10^{-2}]$ & 0.3678 & 0.6975 & 1.0654 & 1.5517 & 2.2722 & 2.9149 & 3.8377 & 4.6703\\
\hline
$cJ/GM_{\odot}^2$ & 0.3650 & 0.5047 & 0.6224 & 0.7523 & 0.9103 & 1.0352 & 1.1904 & 1.3280\\
\hline
$I [10^{45}{\rm g}\;{\rm cm}^2]$ & 2.1519 & 2.1717 & 2.1862 & 2.2100 & 2.2420 & 2.2765 & 2.3212 & 2.3772\\
\hline
$\Phi_2 [10^{42}{\rm g}\;{\rm cm}^2]$ & 10.3179 & 19.1974 & 30.1157 & 43.9566 & 64.5770 & 83.4277 & 110.8010 & 136.1140\\
\hline
$h_+ [{\rm km}]$ & 0.0000 & 0.0000 & 0.0000 & 0.0000 & 0.0000 & 0.0000 & 0.0000 & 0.0000\\
\hline
$h_- [{\rm km}]$ & 0.0000 & 0.0000 & 0.0000 & 0.0000 & 0.0000 & 0.0000 & 0.0000 & 0.0000\\
\hline
$Z_p$ & 0.3822 & 0.3837 & 0.3832 & 0.3839 & 0.3843 & 0.3859 & 0.3870 & 0.3906\\
\hline
$Z_f$ & 0.2775 & 0.2405 & 0.2084 & 0.1752 & 0.1355 & 0.1066 & 0.0712 & 0.0443\\
\hline
$Z_b$ & 0.4905 & 0.5339 & 0.5685 & 0.6080 & 0.6550 & 0.6930 & 0.7386 & 0.7805\\
\hline
$\omega _c/\Omega$  & 0.5400  & 0.5399  & 0.5374  & 0.5359  & 0.5329  & 0.5317  & 0.5289  & 0.5292\\
\hline
$r_e  [{\rm km}]$ & 8.3558 & 8.3912 & 8.4438 & 8.5041 & 8.5967 & 8.6746 & 8.7925 & 8.8900\\
\hline
$r_p/r_e$ & 0.9893 & 0.9800 & 0.9690 & 0.9549 & 0.9350 & 0.9170 & 0.8920 & 0.8700\\
\hline
\end{tabular}
\caption{The physical parameters for a quark star model with Q EOS, a total mass of $1.8M_{\odot}$ and angular velocities from $1.5\times10^3 {\rm s}^{-1}$ to $5\times10^3 {\rm s}^{-1}$.}
\label{table7}
\end{table*}
\begin{table*}
\centering
\begin{tabular}{|l|l|l|l|l|l|l|l|l|l|}
\hline
$\rho_c\;[10^{15}{\rm g}/{\rm cm}^{3}]$ & 0.7830 & 0.7790 & 0.7750 & 0.7670 & 0.7580 & 0.7470 & 0.7350 & 0.7200 & 0.7050\\
\hline
$M\;[M_{\odot}]$ & 1.8003 & 1.8000 & 1.8036 & 1.8015 & 1.8012 & 1.7994 & 1.8022 & 1.7997 & 1.8069\\
\hline
$M_0\; [M_{\odot}]$ & 2.1162 & 2.1149 & 2.1179 & 2.1129 & 2.1098 & 2.1042 & 2.1035 & 2.0954 & 2.0986\\
\hline
$R_e [{\rm km}]$ & 11.7280 & 11.7561 & 11.8075 & 11.8591 & 11.9321 & 12.0175 & 12.1406 & 12.2707 & 12.4636\\
\hline
$\Omega [10^3{\rm s}^{-1}]$ & 1.0163 & 1.4994 & 2.0421 & 2.5147 & 3.0197 & 3.5035 & 4.0415 & 4.5235 & 5.0640\\
\hline
$\Omega_p [10^3{\rm s}^{-1}]$ & 12.0724 & 12.0028 & 11.9281 & 11.8403 & 11.7438 & 11.6396 & 11.5143 & 11.3772 & 11.2109\\
\hline
$T/W [10^{-2}]$ & 0.1987 & 0.4351 & 0.8147 & 1.2517 & 1.8350 & 2.5197 & 3.4382 & 4.4366 & 5.7661\\
\hline
$cJ/GM_{\odot}^2$ & 0.2749 & 0.4071 & 0.5603 & 0.6944 & 0.8430 & 0.9894 & 1.1645 & 1.3263 & 1.5338\\
\hline
$I [10^{45}{\rm g}\;{\rm cm}^2]$ & 2.3770 & 2.3865 & 2.4112 & 2.4269 & 2.4533 & 2.4820 & 2.5323 & 2.5767 & 2.6619\\
\hline
$\Phi_2 [10^{42}{\rm g}\;{\rm cm}^2]$ & 7.1160 & 14.9170 & 27.8321 & 42.6462 & 62.2833 & 85.5102 & 117.0170 & 151.5570 & 198.8980\\
\hline
$h_+ [{\rm km}]$ & 0.0000 & 0.0000 & 0.0000 & 0.0000 & 0.0000 & 0.0000 & 0.0000 & 0.0000 & 0.0000\\
\hline
$h_- [{\rm km}]$ & 0.0000 & 0.0000 & 0.0000 & 0.0000 & 0.0000 & 0.0000 & 0.0000 & 0.0000 & 0.0000\\
\hline
$Z_p$ & 0.3539 & 0.3545 & 0.3566 & 0.3573 & 0.3589 & 0.3604 & 0.3635 & 0.3654 & 0.3703\\
\hline
$Z_f$ & 0.2819 & 0.2487 & 0.2125 & 0.1803 & 0.1463 & 0.1135 & 0.0775 & 0.0441 & 0.0070\\
\hline
$Z_b$ & 0.4276 & 0.4641 & 0.5077 & 0.5451 & 0.5872 & 0.6285 & 0.6784 & 0.7234 & 0.7813\\
\hline
$\omega _c/\Omega$  & 0.5010  & 0.5008  & 0.5016  & 0.5008  & 0.5005  & 0.4997  & 0.4999  & 0.4988  & 0.4997\\
\hline
$r_e  [{\rm km}]$ & 8.8688 & 8.8932 & 8.9318 & 8.9796 & 9.0431 & 9.1197 & 9.2223 & 9.3399 & 9.4977\\
\hline
$r_p/r_e$ & 0.9935 & 0.9864 & 0.9750 & 0.9620 & 0.9450 & 0.9250 & 0.8990 & 0.8710 & 0.8350\\
\hline
\end{tabular}
\caption{The physical parameters for a neutron star model with CFL150 EOS, a total mass of $1.8M_{\odot}$ and angular velocities from $10^3 {\rm s}^{-1}$ to $5\times10^3 {\rm s}^{-1}$.}
\label{table8}
\end{table*}

\bsp

\label{lastpage}


\begin{thebibliography}{99}

\bibitem[Akmal et al. 1998]{Ak98} Akmal, A., Pandharipande, V. R., \&  Ravenhall, D. G. 1998, Phys. Rev. C, 58, 1804


\bibitem[Alford et al. 1999]{cfl1} Alford, M. G., Rajagopal, K., \& Wilczek, F. 1999, Nucl. Phys. B, 537, 433

\bibitem[Alford et al. 2007]{cfl3} Alford, M. G., Rajagopal, K., Schaefer, T., \&  Schmitt, A. 2008, Rev. Mod. Phys., 80, 1455
\bibitem[Asano \& Fukuyama 2000]{AsFu00} Asano, K. \& Fukuyama, T. 2000, \apj, 531, 949

\bibitem[Asano \& Fukuyama 2001]{AsFu01} Asano, K. \& Fukuyama, T. 2001, \apj, 546, 1019

\bibitem[Asano \& Iwamoto 2002]{AsIw02} Asano, K. \& Iwamoto 2002, \apj, 581, 381

\bibitem[Baldo et al. 1997]{baldo} Baldo, M., Bombaci, I., \& Burgio, G. F. 1997, Astron. Astrophys., 328, 274

\bibitem[Bardeen et al. 1972]{BPT72} Bardeen, J. M., Press, J. H., \& Teukolsky, S. A. 1972, \apj, 178, 347

\bibitem[Baym et al. 1971a]{Ba71a} Baym, G., Bethe, H. A., \& Pethick, C. J. 1971, Nucl. Phys. A, 175, 225

\bibitem[Baym et al. 1971b]{Ba71b} Baym, G., Pethick, C. J., \& Sutherland, P. 1971, \apj, 170, 299

\bibitem[Bhattacharyya et al. 2009]{Ba09} Bhattacharyya, A., Ghosh, S. K., Mallick, R., \& Raha, S. 2009, arXiv0905.3605


\bibitem[Bethe 1990]{Be90} Bethe, H. A. 1990, Rev. Mod. Phys., 62, 801






\bibitem[Bethe \& Wilson 1985]{BeWi85} Bethe, H. A. \& Wilson, J. R., 1985, \apj, 295, 14

\bibitem[Birkl et al. 2007]{Bi07} Birkl, R., Aloy, M. A., Janka, H.-Th., \& Muller, E. 2007, Astron. Astrophys. 463, 51

\bibitem[Bodmer 1971]{Bo71} Bodmer, A. R., 1971, Phys. Rev. D, 4, 1601


\bibitem[Cadeu et al. 2005]{Ca05} Cadeu, C., Leahy D., \& Morsink S. 2005, \apj, 618, 451

\bibitem[Cadeu et al. 2007]{Ca07} Cadeu, C., Morsink S. \& Leahy D. 2007, \apj, 654, 458

\bibitem[Chan et al. 2009]{Cha09}  Chan, T. C., Cheng, K. S., Harko, T., Lau, H. K., Lin, L. M., Suen, W. M., \& Tian,  X. L. 2009, \apj, 695, 732


\bibitem[Cheng et al. 1998]{Ch98} Cheng,  K. S., Dai, Z. G., \& Lu, T. 1998a, Int. J. Mod. Phys. D, 7, 139




\bibitem[Cheng et al. 2009] {Ch09} Cheng, K. S., Harko, T., Huang, Y. F., Lin, L. M., Suen, W. M., \& Tian, X. L. 2009, JCAP, 09, 007


\bibitem[Cooperstein et al. 1986]{Co86} Cooperstein, J., van den Horn, L. J., \& Baron, E. A. 1986, \apj, 309, 653

\bibitem[Cooperstein et al. 1987]{Co87} Cooperstein, J., van den Horn, L. J., \& Baron, E. A. 1987, \apj, 321, L129

\bibitem[Dai et al. 1995]{Da} Dai,  Z. G.,  Peng, Q. H., \& Lu, T. 1995, \apj,  440, 815


\bibitem[Dey et al. 1998]{De98}  Dey, M.,  Bombacci, I.,  Dey, J., Ray, S., \& Samanta, B. C. 1998, Phys.
Lett. B, 438, 123

\bibitem[Douchin \& Haensel 2001]{DoHa01} Douchin, F. \& Haensel, P. 2001, Astron. Astrophys., 380, 151

\bibitem[Feynman et al. 1949]{Fe49} Feynman, R. P.,  Metropolis, N., \& Teller, E. 1949, Phys. Rev., 75, 1561


\bibitem[Goodman et al. 1987]{Go87} Goodman, J., Dar, A., \& Nussinov S., 1986 \apj, 314, L7









\bibitem[Horvath \& Lugones 2004]{HoLu04} Horvath, J. E., \& Lugones, G. 2004, Astron. Astrophys., 422, L1

\bibitem[Itoh 1970]{It70} Itoh, N. 1970, Prog. Theor. Phys., 44, 291




\bibitem[Kov\'acs et al. 2010]{Ko09} Kov\'acs, Z., Cheng, K. S., \& Harko, T. 2010, \mnras, 402, 1714

\bibitem[Kubis \& Kutschera 1997]{Kubis} Kubis, S., \& Kutschera, M. 1997, Phys. Lett. B, 399, 191

\bibitem[Lugones \& Horvath 2002]{LuHo02} Lugones, G., \& Horvath, J. E. 2002, Phys. Rev. D, 66, 074017

\bibitem[Mallick \& Majumder 2009]{MaMa09} Mallick, R., \& Majumder, S. 2009, Phys. Rev. D. 79, 023001

\bibitem[M\'esz\'aros \& Rees 1992]{MeRe92} M\'esz\'aros, P. \& Rees, M. J. 1992, \mnras, 257, 29

\bibitem[M\'esz\'aros 2006]{Me06} M\'esz\'aros, P. 2006, Rept. Prog. Phys., 69, 2259

\bibitem[Miller et al. 2003]{Mi03} Miller, W. A., George, N. D., Kheyfets, A., \& McGhee, J. M. 2003, \apj, 583, 833




\bibitem[Novikov \& Thorne 1973]{NoTh73} Novikov, I. D., \& Thorne, K. S. 1973, in Black Holes, ed. C. DeWitt and B. DeWitt, New York: Gordon and Breach

\bibitem[Nozawa et al. 1998]{No98} 	
	Nozawa, T., Stergioulas, N., Gourgoulhon, E., \& Eriguchi, Y. 1998, Astron. Astrophys., 132, 431

\bibitem[Page \& Thorne 1974]{PaTh74} Page, D. N., \& Thorne, K. S. 1974, \apj, 191, 499


\bibitem[Paczynski 1990]{Pa90} Paczynski, B. 1990, \apj, 363, 218

\bibitem[Pandharipande 1971]{Pa71} Pandharipande, V. R. 1971, Nucl. Phys. A, 178, 123

\bibitem[Perez-Giz \& Levin 2009]{PGL09} Perez-Giz, G., Levin, J., 2009, Phys. Rev. D, 79 124014



\bibitem[Rapp et al. 2000]{cfl2} Rapp, R., Schaffer, T., Shuryak, E. V., \& Velkovsky, M. 2000, Ann. Phys. (N. Y.), 280, 35

\bibitem[Ruffert \& Janka 1998]{RuJa98} Ruffert, M. \& Janka, H.-T. 1998, Astron. Astrophys., 338, 535

\bibitem[Ruffert \& Janka 1999]{RuJa99} Ruffert, M. \& Janka, H.-T. 1999, Astron. Astrophys., 344, 573


\bibitem[Salmonson \& Wilson 1999]{SaWi99} Salmonson, J. D. \& Wilson, J. R., 1999, \apj, 517, 895

\bibitem[Salmonson \& Wilson 2001]{SaWi01} Salmonson, J. D. \& Wilson, J. R., 2001, \apj, 561, 950

\bibitem[Shen et al. 1998]{Shen} Shen, H., Toki, H., Oyamatsu, K., \&  Sumiyoshi, K. 1998, Nucl. Phys. A, 637, 435

\bibitem[Shibata \& Sasaki 1998]{ShSa98} Shibata, M., \& Sasaki, M. 1998, \prd, 58, 104011


\bibitem[Stergioulas \& Friedman 1995]{SteFr95} Stergioulas, N., \& Friedman, J. L. 1995, \apj, 444, 306

\bibitem[Stergioulas et al. 1999]{Ste99} Stergioulas, N., Kluzniak, W., \& Bulik, T. 1999, Astron. Astrophys., 352, L116
\bibitem[Stergioulas 2003]{Sterev} Stergioulas, N. 2003, Living Rev. Rel., 6, 3










\bibitem[Witten 1984]{Wi84} Witten, E. 1984, Phys. Rev. D,  30, 272


\bibitem[Zhang 2007]{Zh07} Zhang, B. 2007, Chin. J. Astron. Astrophys., 7, 1

\bibitem[Zhang \& Dai 2009]{ZhDa09} Zhang, D. \& Dai, Z. G. 2009, \apj, 703, 461

\end{thebibliography}
\end{document}